\documentclass[12pt,a4paper]{article}

\usepackage{amsmath,amssymb,amsthm,color,bm}
\usepackage{verbatim,textcomp}
\usepackage{graphicx,epstopdf}
\usepackage{subfigure}
\usepackage{amscd}
\usepackage{subfigure}

\usepackage{rotating}

\usepackage{float}

\usepackage{bbm,ascii,hyperref}
\usepackage{mysec,times1}

\usepackage{cancel}
\usepackage[normalem]{ulem}

\newcommand{\myp}{\mbox{$\:\!$}}
\newcommand{\mypp}{\mbox{$\;\!$}}

\newcommand{\myn}{\mbox{$\;\!\!$}}
\newcommand{\mynn}{\mbox{$\:\!\!$}}

\def\MR#1{\href{http://www.ams.org/mathscinet-getitem?mr=#1}{MR#1}}

\numberwithin{equation}{section}

\newtheorem{theorem}{Theorem}[section]
\newtheorem{proposition}[theorem]{Proposition}
\newtheorem{lemma}[theorem]{Lemma}

\theoremstyle{definition}
\newtheorem{assumption}{Assumption}[section]

\theoremstyle{remark}
\newtheorem{remark}{Remark}[section]

\newtheorem{example}{Example}[section]

\newcommand{\PP}{{\normalfont\text{P}}}
\newcommand{\EE}{{\normalfont\text{E}}}
\newcommand{\Var}{{\normalfont\text{Var}}}
\newcommand{\rd}{\mathrm{d}}
\newcommand{\re}{\mathrm{e}}
\newcommand{\RR}{\mathbb{R}}
\newcommand{\NN}{\mathbb{N}}

\newcommand{\const}{\mathrm{const}}

\newcommand{\eNPV}{{\normalfont\text{e\kern.04pc{}NPV}}}
\newcommand{\stheta}{\theta}

\numberwithin{equation}{section}

\makeatletter 
\renewcommand{\@listI}{\setlength{\topsep}{.4ex}
\setlength{\labelwidth}{15pt}\setlength{\labelsep}{5pt}
\setlength{\leftmargin}{26pt}} \makeatother



\begin{document}

\title{
Optimal Stopping and Utility in a Simple Model\\
of Unemployment Insurance}

\author{Jason S.\,Anquandah\thanks{\,Corresponding author. E-mail: {\tt
\href{mailto:mmjsa@leeds.ac.uk}{mmjsa@leeds.ac.uk}}}\ \ and Leonid
V.\,Bogachev\thanks{\,E-mail: {\tt
\href{mailto:L.V.Bogachev@leeds.ac.uk}{L.V.Bogachev@leeds.ac.uk}}}\\[.5pc]
{\it Department of Statistics, School of Mathematics, University of Leeds,}\\
{\it  Leeds LS2 9JT, UK}}

\date{}

\maketitle

\begin{abstract}
Managing unemployment is one of the key issues in social policies.
Unemployment insurance schemes are designed to cushion the financial
and morale blow of loss of job but also to encourage the unemployed
to seek new jobs more pro-actively due to the continuous reduction
of benefit payments. In the present paper, a simple model of
unemployment insurance is proposed with a focus on optimality of the
individual's entry to the scheme. The corresponding optimal stopping
problem is solved, and its similarity and differences with the
perpetual American call option are discussed. Beyond a purely
financial point of view, we argue that in the actuarial context the
optimal decisions should take into account other possible
preferences through a suitable utility function. Some examples in
this direction are worked out.



%
%
%

\end{abstract}

\section{Introduction}\label{sec:1}

Assessing the risk in financial industries often aims at finding
optimal choices in decision making. In the insurance sector,
optimality considerations are crucial primarily for the insurers,
who have to address monetary issues (such as how to price the
insurance policy so as not to run it at a loss but also to keep the
product competitive) and time issues (e.g., when to release the
product to the market). Less studied but also important are optimal
decisions on behalf of the insured individuals, related to monetary
issues (e.g., how profitable is taking up an insurance policy and
the right portion of wealth to invest), consumption decisions (e.g.,
whether to maximize or optimize own consumption), or time-related
decisions (such as when it is best to enter or exit an insurance
scheme).

In this paper we focus on the particular type of products related to
\emph{unemployment insurance (UI)}, whereby an employed individual
is covered against the risk of involuntary unemployment (e.g., due
 to redundancy). Various UI
systems are designed to help cushion the financial (as well as
morale) blow of loss of job and to encourage unemployed workers to
find a new job as early as possible in view of the continued
reduction of benefits. The protection is normally provided in the
form of regular financial benefits (usually tax free) payable after
the insured individual becomes unemployed and until a new job is
found, but often only up to a certain maximum duration and with
payments gradually decreasing over time. Many countries have UI
schemes in place
\cite{holmlund1998unemployment,kerr1996unemployment}, often run and
funded by the governments, with contributions from employers and
workers, but also by private insurance companies~\cite{GoCompare}.
For example, the governmental UI systems administered in France and
Belgium in the 1990s provided benefits decreasing with time
according to a certain schedule; the amount of the benefit was
determined by the age of the worker, their final wage/salary, the
number of qualifying years in employment, family circumstances, etc.

\begin{figure}[h!]
        \centering
        \includegraphics[width=0.6\textwidth]{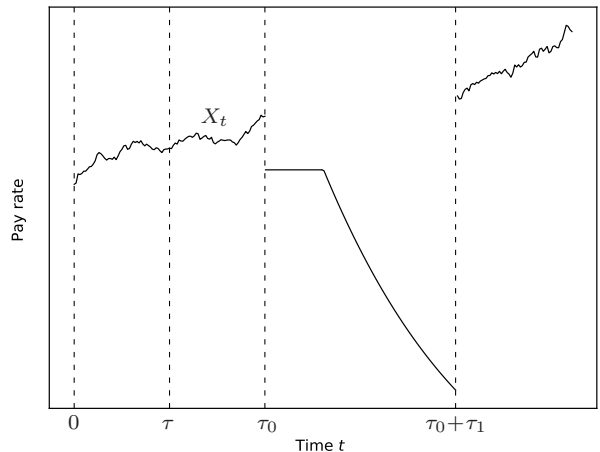}
        \put(-224.0,14.2){\mbox{\scriptsize$0$}}
        \put(-189.0,14.2){\mbox{\scriptsize$\tau$}}
        \put(-153.3,14.2){\mbox{\scriptsize$\tau_0$}}
        \put(-90.3,14.2){\mbox{\scriptsize$\tau_0{+}\myp\tau_1$}}
        \put(-174.4,129.5){\mbox{\scriptsize$X_t$}}
        \caption{A time chart of the unemployment insurance scheme.
            The horizontal axis shows (continuous) time; the vertical axis indicates the pay rate (i.e., income receivable per unit time).
            The origin $t=0$ indicates the start of employment. Two pieces of a random path $X_t$ depict the dynamics of the individual's wage
            whilst in employment.
            The individual joins the UI scheme at entry time $\tau$ (by paying a premium $P$).
            When the current job ends (at time
            $\tau_0>\tau$), a benefit proportional to the final wage $X_{\tau_0}$ is payable according to a predefined schedule
            (e.g., see Example~\ref{ex:2.1}), until a new job is found
            after the unemployment spell of duration~$\tau_1$.}
        \label{fig1}
\end{figure}

In this work we introduce and analyse a simple UI model focusing on
the optimal time for the individual to join the scheme. Before
setting out the model formally, let us describe the situation in
general terms. Consider an individual currently at work but who is
concerned about possible loss of job, which may be a genuine
potential threat due to the fluidity of the job market and the level
of demand in this employment sector. To mitigate this risk, the
employer or the social services have an unemployment insurance
scheme in place, available to this person (perhaps after a certain
qualifying period at work), which upon payment of a one-off entry
premium would guarantee to the insured a certain benefit payment
proportional to their final wage and determined by a specified
declining benefit schedule, until a new job is found (see
Fig.~\ref{fig1}).

The decision the individual is facing is \emph{when} (rather than
\emph{if}) to join the scheme. What are the considerations being
taken into account when contemplating such a decision? On the one
hand, delaying the entry may be a good idea in view of the monetary
inflation over time\,---\,since the entry premium is fixed, its
actual value is decreasing with time. Also, it may be reasonably
expected that the wage is likely to grow with time (e.g., due to
inflation but also as a reward for improved skills and experience),
which may have a potential to increase the total future benefit
(which depends on the final wage). Last but not least, some savings
may be needed before paying the entry premium becomes financially
affordable. On the other hand, delaying the decision to join the
insurance scheme is risky, as the individual remains unprotected
against loss of job, with its financial as well as morale impact.

Thus, there is a scope for optimizing the decision about the entry
time\,---\,probably not too early but also not too late. Apparently,
such a decision should be based on the information available to
date, which of course includes the inflation rate and also the
unemployment and redeployment rates, all of which should, in
principle, be available through the published statistical data.
Another crucial input for the decision-making is the individual's
wage as a function of time. We prefer to have the situation where
this is modelled as a random process, the values of which may go up
as well as down. This is the reason why we do not consider salaries
(which are in practice piecewise constant and unlikely to decrease),
and instead we are talking about \emph{wages}, which are more
responsive to supply and demand and are also subject to
``real-wage'' adjustments (e.g., through the consumer price index,
CPI). Besides, loss of job is more likely in wage-based employments
due to the fluidity of the job market. For simplicity, we model the
wage dynamics using a diffusion process called \emph{geometric
Brownian motion}.\footnote{For technical convenience, we choose to
work with continuous-time models, but our ideas can also be adapted
to discrete time (which may be somewhat more natural, since the wage
process is observed by the individual on a weekly time scale).}

To summarize, the optimization problem for our model aims to
maximize the expected net present value of the UI scheme by choosing
an optimal entry time $\tau^*$. We will show that this problem can
be solved exactly by using the well-developed \emph{optimal stopping
theory} \cite{peskir2006optimal,pham2009continuous,Shiryaev}.
It turns out that the answer is provided by the \emph{hitting time}
of a suitable threshold $b^*\mynn$, that is, the first time
$\tau_{b^*}$ when the wage process $X_t$ will reach this level.
Since the value of $b^*\mynn$ is not known in advance, this leads to
solving a \emph{free-boundary problem} for the differential operator
(generator) associated with the diffusion process $(X_t)$. In fact,
we first conjecture the aforementioned structure of the solution and
find the value $b^*\mynn$, and then verify that this is indeed
the true solution to the optimal stopping problem.

In the insurance literature, there has been much interest towards
using optimality considerations, including optimal stopping
problems. From the standpoint of insurer seeking to maximize their
expected returns, the optimal stopping time may be interpreted as
the time to suspend the current trading if the situation is
unfavourable, and to recalculate premiums
(see, e.g.,
\cite{jensen1997optimal,Karpowicz-Szajowski,muciek2002optimal} and
further references therein).
Insurance research has also focused on optimality from the
individual's perspective. One important direction relevant to the UI
context was the investigation of the job seeking processes,
especially when returning from the unemployed status
\cite{{Boshuizen1995},mccall1970economics,wang2016risk}. This was
complemented by a more general research exploring ways to optimize
and improve the efficacy of the UI systems (also in terms of
reducing government expenditure), using incentives such as a
decreasing benefit throughout the unemployment spell, in conjunction
with sanctions and workfare
(see
\cite{fredriksson2006optimal,sopraseuth2007optimal,hopenhayn1997optimal,
kolsrud2018optimal,landais2017risk}, to cite but a few). A related
strand of research is the study of optimal retirement strategies in
the presence of involuntary unemployment risks and borrowing
constraints
\cite{Choi2006,DeAngelis-Stabile,Gerrard,jang2013optimal,Stabile}.

To the best of our knowledge, optimal stopping problems in the UI
context (such as the optimal entry to\,/\,exit from a UI policy)
have not received sufficient research attention. This issue is
important, because knowing the optimal entry strategies is likely to
enhance the motivation for individuals to join the UI scheme, thus
ensuring better societal benefits through the UI policies (see
analysis and discussion in~\cite{rebollo2015unemployment}).
Knowledge of the optimal entry time for insured individuals, which
has impact on the amount and duration of benefits to be claimed,
will also help the insurers (both state and private) to optimize
their financial practices (see a discussion
in~\cite{landais2017value}).
Thus, our present work attempts
to fill in the gap by addressing the question of the optimal timing
to join the UI scheme.

It is interesting to point out that our optimal stopping problem and
its solution have a lot in common with (but are not identical to)
the well-known American call option in financial mathematics, where
the option holder has the right to exercise it at any time (i.e., to
buy a certain stock at an agreed price), and the problem is to
determine the best time to do that, aiming to maximize the expected
financial gain. However, unlike the American call option setting
based on purely financial objectives, the optimal stopping solution
obtained in our UI model is not entirely satisfactory from the
individual's point of view, because the (optimal) waiting time
$\tau_{b^*}$ may be infinite with positive probability (at least for
some values of the parameters), and even if it is finite with
probability one, the expected waiting time may be very long.

Motivated by this observation, we argue that certain elements of
\emph{utility} should be added to the analysis, aiming to quantify
the individual's ``impatience'' as a measure of purpose and
satisfaction. We suggest a few simple ideas of how utility might be
accommodated in the UI optimal stopping framework. Despite the
simplicity of such examples, in most cases they lead to much harder
optimal stopping problems. Not attempting to solve these problems in
full generality, we confine ourselves to exploring suboptimal
solutions in the class of hitting times, which nonetheless provide
useful insight into possible effects of inclusion of utility into
the optimal stopping context.

The general concept of utility in economics was strongly advocated
in the classical book by von Neumann and Morgenstern \cite{Neumann},
whose aim was in particular to overcome the idealistic assumption of
a strictly rational behaviour of market agents.\footnote{Impact of
individualistic (not always rational) perception in economics and
financial markets is the subject of the modern behavioural economics
(see, e.g., a recent monograph~\cite{Dhami}).}
These ideas were quickly adopted in insurance,
dating back to Borch~\cite{borch1961utility} and soon becoming part
of the insurance mainstream, culminating in the Expected Utility
Theory (see a recent book by Kaas et al.~\cite{Kaas}) routinely used
as a standard tool to price insurance products. In particular,
examples of use of utility in the UI analysis are ubiquitous (see,
e.g.,
\cite{acemoglu2000productivity,baily1978some,fredriksson2006optimal,
sopraseuth2007optimal,holmlund1998unemployment,hopenhayn1997optimal,
kolsrud2018optimal,landais2017risk,landais2017value}). There have
also been efforts to combine optimal stopping and utility
\cite{chen2019optimal,Choi2006,henderson2008explicit,
Karpowicz-Szajowski,muciek2002optimal,wang2016risk}. However, all
such examples were limited to using utility functions to
re-calculate wealth, while other important objectives and
preferences such as the desire to buy the policy or to reduce the
waiting times have not been considered as yet, as far as we can
tell.

%

\subsubsection*{Layout.} The rest of the paper is organized as follows. In
Section~\ref{sec:2}, our insurance model is specified and the
optimization problem is set up. In Section~\ref{sec:3}, the optimal
stopping problem is solved using a reduction to a suitable free
boundary problem, including the identification of the critical
threshold~$b^*\myn$. This is complemented in Section~\ref{sec:4} by
an elementary derivation using explicit information about the
distribution of the hitting times for the geometric Brownian motion.
Section~\ref{sec:5} addresses various statistical issues and also
provides a numerical example illustrating the optimality of the
critical threshold~$b^*\myn$. In Section~\ref{sec:6}, we carry out
the analysis of parametric dependence in our model upon two most
significant exogenous parameters, the unemployment rate and the wage
drift, and also give an economic interpretation thereof. In
Section~\ref{sec:7}, we make a useful comparison of our problem and
its solution with the classical American call option, which leads us
to the discussion of the necessity of utility-based considerations
in the optimal stopping context. Finally, Section~\ref{sec:8}
contains the summary discussion of our results, including
suggestions for further work.

\subsubsection*{Notation.} We use the standard notation
$a\wedge b:=\min\mynn\{a,b\}$, $a\vee b:=\max\mynn\{a,b\}$, and
$a^+\mynn:=a\vee 0$.

\section{Optimal stopping problem}\label{sec:2}

\subsection{The model of unemployment insurance}\label{sec:2.1}
Let us describe our model in more detail. Suppose that time $t\ge0$
is continuous and is measured (in the units of weeks) starting from
the beginning of the individual's employment We assume without loss
of generality that the unemployment insurance policy is available
immediately (although in practice, a qualifying period at work would
normally be required for eligibility). Let $X_t>0$ denote the
individual's wage (i.e., payment per week, paid in arrears) as a
function of time $t\ge0$, such that $X_0=x$. We treat
$X=(X_t,\,t\ge0)$ as a random process defined on a filtered
probability space $(\Omega, \mathcal{F}, (\mathcal{F}_t), \PP)$,
where $\Omega$ is a suitable sample space (e.g., consisting of all
possible paths of $(X_t)$), the filtration $(\mathcal{F}_t)$ is an
increasing sequence of $\sigma$-algebras $\mathcal{F}_t\subset
\mathcal{F}$, and $\PP$ is a probability measure on the measurable
space $(\Omega,\mathcal{F})$ which determines the distribution of
various random inputs in the model, including $(X_t)$. It is assumed
that the process $(X_t)$ is adapted to the filtration
$(\mathcal{F}_t)$, that is, $X_t$ is $\mathcal{F}_t$-measurable for
each $t\ge0$. Intuitively, $\mathcal{F}_t$ is interpreted as the
full information available up to time $t$, and measurability of
$X_t$ with respect to $\mathcal{F}_t$ means that this information
includes knowledge of the values of the process~$X_t$.

Furthermore, remembering that $X_t$ is positive valued, we use for
it a simple model of \emph{geometric Brownian motion} driven by the
stochastic differential equation
\begin{equation}\label{r1-0}
\frac{\rd X_t}{X_t} =  \mu \,\rd{t} + \sigma \mypp\rd B_t, \qquad
X_0=x,
\end{equation}
where $B_t$ is a standard Brownian motion (i.e., with mean zero,
$\EE(B_t)=0$, and variance $\Var(B_t)=t$, and with continuous sample
paths), and $\mu\in\RR$ and $\sigma>0$ are the drift and volatility
rates, respectively. The equation \eqref{r1-0} is well known to have
the explicit solution (see, e.g., \cite[Ch.\,III, \S\myp3a,
p.\,237]{Shiryaev})
\begin{equation}\label{r1}
X_t = x\exp\bigl\{(\mu - \tfrac{1}{2}\myp \sigma^2)\mypp t +\sigma
B_t\bigr\}\quad (t\ge0).
\end{equation}
Note that
\begin{equation}\label{eq:E(X)}
\EE_x(X_t)=x\mypp\re^{\myp\mu\myp t},\qquad
\Var_x(X_t)=x^2\myp\re^{2\mu\myp
t}\bigl(\re^{\myp\sigma^2t}-1\bigr),
\end{equation}
where $\EE_x$ and $\Var_x$ denote expectation and variance with
respect to the distribution of $X_t$ given the initial value
$X_0=x$.

Let us now specify the unemployment insurance scheme. An individual
who is currently employed may join the scheme by paying a fixed
one-off premium $P>0$ at the point of entry. If and when the current
employment ends (say, at time instant $\tau_0$), the benefit
proportional to the final wage $X_{\tau_0}$ is payable according to
the \emph{benefit schedule} $h(s)$; that is, the payout at time
$t\ge \tau_0$ is given by $X_{\tau_0}\myp h(t-\tau_0)$. However, the
payment stops when a new job is found after the unemployment spell
of duration~$\tau_1$. For simplicity, we assume that both $\tau_0$
and $\tau_1$ have \emph{exponential distribution} (with parameters
$\lambda_0$ and $\lambda_1$, respectively); as mentioned in the
Introduction, this guarantees a Markovian nature of the
corresponding transitions. These random times are also assumed to be
statistically independent of the process~$(X_t)$.

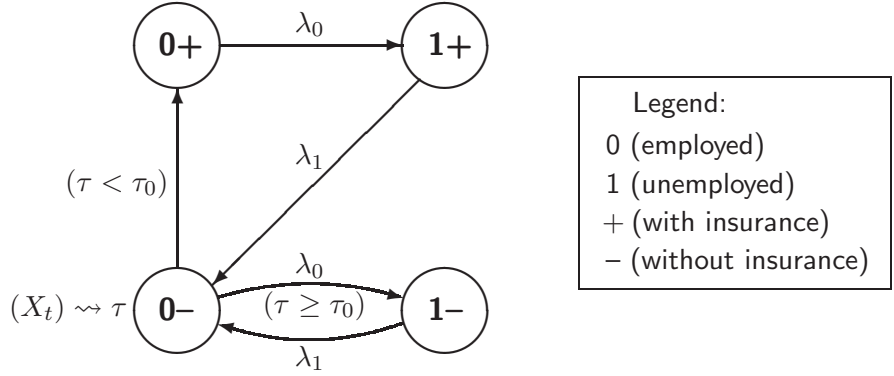
\begin{figure}[th]
 \thicklines\centering
    \hspace{-4pc}\begin{picture}(0,150)
    \put(-70,130){\circle{30}}
    \put(-77,126.5){\mbox{\bf 0+}}
    \put(23.7,126.5){\mbox{\bf 1+}}
   \put(30,130){\circle{30}}
    \put(-77,26.5){\mbox{\bf 0--}}
    \put(23.7,26.5){\mbox{\bf 1--}}
    \put(-70,30){\circle{30}}
   \put(30,30){\circle{30}}
   \put(-53.7,130){\vector(1,0){68.1}}
   \put(20.7,117){\vector(-1,-1){77.5}}
   \put(-70,46){\vector(0,1){68}}
   \qbezier(-54.5,35)(-20.25,45)(13,35.2)
   \qbezier(-52.7,24.3)(-20.25,13)(14,25)
   \put(10.4,35.9){\vector(4,-1){4.5}}
  \put(-50.4,23.6){\vector(-4,1){4.5}}
     \put(-112,75){\mbox{\small $(\tau<\tau_{0})$}}
     \put(-133,28){\mbox{\small $(X_t)\rightsquigarrow\tau$}}
   \put(-26,135){\mbox{\small $\lambda_{0}$}}
   \put(-26,85){\mbox{\small $\lambda_{1}$}}
   \put(-26,44){\mbox{\small $\lambda_{0}$}}
   \put(-38,29){\mbox{\small $(\tau\ge \tau_{0})$}}
   \put(-26,9){\mbox{\small $\lambda_{1}$}}
    \put(80,75){\fbox{\mbox{$\begin{array}{l} \hspace{1pc}\text{\small Legend:}\\[.1pc]
    \,\text{\small 0 (employed)}\\
    \,\text{\small 1 (unemployed)}\\
    \myp\text{\small + \mynn(with insurance)}\\
    \mypp\myp\text{\small -- (without insurance)}
    \end{array}$}}}\end{picture}
\caption{Schematic diagram of possible transitions in the
unemployment insurance scheme. Here, $\tau_0$ and $\tau_1$ are the
(exponential) holding times in states 0 and 1, with parameters
$\lambda_0$ and $\lambda_1$, respectively, whereas $\tau$ is the
entry time (i.e., from state 0-- to state 0+), which is subject to
optimal control based on observations over the wage process $(X_t)$.
}\label{fig2}\end{figure}

Possible transitions in the state space of our insurance model are
shown in Fig.~\ref{fig2}, where symbols ``0'' and ``1'' encode the
states of being employed and unemployed, respectively, whereas
suffixes ``+'' and ``--'' indicate whether insurance is in place or
not. Note that all transitions, except from 0-- to 0+ (which is
subject to optimal control based of observations over the wage
process $(X_t)$), occur in a Markovian fashion; that is, the holding
times are exponentially distributed (with parameters $\lambda_0$ if
in states 0-- and 0+, or $\lambda_1$ if in states 1-- and 1+).

The individual's decision about a suitable time to join the scheme
is based on the information available to date. In our model, this
information encoded in the filtration $(\mathcal{F}_t)$ is provided
by ongoing observations over the wage process $(X_t)$. Thus,
\emph{admissible strategies} for choosing $\tau$ must be
\emph{adapted} to the filtration $(\mathcal{F}_t)$; namely, at any
time instant $t\ge0$ it should be possible to determine whether
$\tau$ has occurred or not yet, given all the information in
$\mathcal{F}_t$. In mathematical terms, this means that $\tau$ is a
\emph{stopping time}, whereby for any $t\ge0$ the event $\{\tau>t\}$
belongs to the $\sigma$-algebra $\mathcal{F}_t$ (see, e.g.,
\cite[Ch.\mypp1, \S\mypp3, p.\,25]{Yeh}).

\begin{remark}
In general, a stopping time $\tau$ is allowed to take values in
$[\myp0,\infty]$ including $\infty$, in which case waiting continues
indefinitely and the decision to join the scheme is never taken. In
practice, it is desirable that the stopping time $\tau$ be finite
almost surely (a.s.)\ (i.e., $\PP_x(\tau<\infty)=1$), but this may
not always be the case (see Section~\ref{sec:4.1}).
\end{remark}

\subsection{Setting the optimal stopping problem}\label{sec:2.2}
As was explained informally in the Introduction, there is a scope
for optimizing the choice of the entry time~$\tau$, where optimality
is measured by maximizing the expected financial gain from the
scheme. Our next goal is to obtain an expression for the expected
gain under the contract. First of all, conditional on the final wage
$X_{\tau_0}$, the expected future benefit to be received under this
insurance contract is given by
\begin{equation}\label{r2a}
X_{\tau_0}\, \EE\mynn\left( \int_{0}^{\tau_1}
\re^{-rs}\,h(s)\,\rd{s} \right) = \beta X_{\tau_0},
\end{equation}
where $r$ is the \emph{inflation rate} and
\begin{equation}\label{eq:c1}
\beta:=\int_0^\infty \lambda_1\mypp\re^{-\lambda_1 t}
H(t)\,\rd{t},\qquad H(t):=\int_0^t \re^{-rs}\,h(s)\,\rd{s}.
\end{equation}
Note that the expectation in formula \eqref{r2a} is taken with
respect to the (exponential) random waiting time $\tau_1$ (with
parameter~$\lambda_1$), and that the expression inside integration
involves discounting to the beginning of unemployment at
time~$\tau_0$.

\begin{example}\label{ex:2.1}
A specific example of the benefit schedule $h(s)$ may be as follows,
\begin{equation}\label{eq:h}
h(s)=\begin{cases} h_0\myp,& 0\le s\le s_0,\\
h_0\mypp\re^{-\delta\myp (s-s_0)},&s\ge s_0,
\end{cases}
\end{equation}
where $0<h_0\le 1$, \,$0\le s_0\le\infty$ and $\delta>0$. Thus, the
insured receives a certain fraction of their final wage (i.e.,
$h_0\myp X_{\tau_0}$) for a grace period $s_0$, after which the
benefit is falling down exponentially with rate $\delta$. This
example is motivated by the declining unemployment compensation
system in France \cite{kerr1996unemployment}.\footnote{More
specifically, according to the French UI system back in the 1990s
(see~\cite[p.\,8]{kerr1996unemployment}), a worker aged 50 or more,
with eight months of insurable employment in the last twelve months,
was entitled to full benefits equal to 57.4\% of the final wage
payable for the first eight months, thereafter declining by 15\%
every four months; however, the payments continued for no longer
than 21 months overall. This leads to choosing the following
numerical values in~\eqref{eq:h}: $h_0=0.574$,
$s_0=8\left(52/12\right)\doteq34.7$ (weeks) and
$\delta=-(3/52)\ln\left(1-0.15\right)\doteq 0.0094
=0.94\%$ (per week). The restriction of the benefit term by
$21\left(52/12\right)=91$ weeks can be taken into account in our
model by adjusting the parameter $\lambda_1$ from the condition
$\EE(\tau_1)=91$, giving $\lambda_1\doteq 0.0110$.
A more conservative choice is to use a tail probability condition,
for example, $\PP(\tau_1>91)=0.10$, yielding $\lambda_1=
-\ln\left(0.10\right)/91\doteq 0.0253$
(with $\EE(\tau_1)\doteq39.5$).}
Having specified the schedule function, all calculations can be done
explicitly. In particular, the constant $\beta$ in \eqref{r2a} is
calculated from \eqref{eq:c1} as
$$
\beta = \frac{h_0\bigl( 1- \re^{-(r+\lambda_1)\myp
s_0}\bigr)}{r+\lambda_1} + \frac{h_0\mypp\re^{-(r+\lambda_1)\myp
s_0}}{r+\lambda_1+ \delta}.
$$
In the extreme cases $s_0=0$ or $s_0=\infty$, this expression
simplifies to
$$
\beta=\begin{cases}\displaystyle
\frac{h_0}{\lambda_1}\left(1-\frac{r+\delta}{r+\lambda_1+\delta}\right),&
s_0=0,\\[.9pc]
\displaystyle
\frac{h_0}{\lambda_1}\left(1-\frac{r}{r+\lambda_1}\right),&s_0=\infty.
\end{cases}
$$
Here, the first factor has a clear meaning as the product of pay per
    week ($h_0$) and the mean duration of the benefit payment
    ($\EE(\tau_1)=1/\lambda_1$), whereas the second factor takes into
    account the discounting at rates~$r$ and~$\delta$.
\end{example}

\smallskip
Returning to the general case, if the contract is entered
immediately (subject to the payment of premium $P$), then the net
expected benefit discounted to the entry time $t=0$ is given by the
\emph{gain function}
\begin{equation}\label{r2d0}
g(x): =\EE_x\bigl(\re^{-r\tau_0}\myp \beta X_{\tau_0}\bigr)-P,
\end{equation}
where $x=X_0$ is the starting wage and the symbol $\EE_x$ now
indicates expectation with respect to both $\tau_0$ and
$X_{\tau_0}$. Recall that the random time $\tau_0$ is independent of
the process $(X_t)$ and has the exponential distribution with
parameter $\lambda_0$. Using the total expectation formula (see,
e.g., \cite[\S\mypp{}II.7.4, Definition~3, p.\,214, and Property~G*,
p.\,216]{Shiryaev-pr}) and substituting the expression
\eqref{eq:E(X)}, the expectation in \eqref{r2d0} is computed as
follows,
\begin{alignat}{2}
\notag
\EE_x\bigl(\re^{-r\tau_0}X_{\tau_0}\bigr)&=\lefteqn{\EE_x\bigl[\re^{-r\tau_0}\,\EE_x(X_{\tau_0}\myp|\mypp\tau_0)\bigr]}\\
\notag &=\EE_x\bigl[\re^{-r\tau_0} (x\mypp \re^{\myp\mu\myp
\tau_0})\bigr] &&=x\int_0^\infty \re^{(\mu-r)\myp t}\mypp
\lambda_0\mypp
\re^{-\lambda_0\myp t}\mypp\rd{t}\\
&&&=\frac{\lambda_0\mypp x}{r+\lambda_0-\mu}. \label{eq:g-new}
\end{alignat}
Thus, substituting \eqref{eq:g-new} into \eqref{r2d0} and denoting
\begin{equation}\label{eq:c1-tilde}
\tilde{r}:=r+\lambda_0,\qquad \beta_1:=\frac{\beta\myp
\lambda_0}{\tilde{r}-\mu},
\end{equation}
the gain function is represented explicitly as
\begin{equation}\label{r2d}
g(x)=\beta_1x-P.
\end{equation}

Of course, the computation in \eqref{eq:g-new} is only meaningful as
long as
\begin{equation}\label{eq:upper}
\mu<r+\lambda_0=\tilde{r}.
\end{equation}
\begin{assumption}\label{as:1}
In what follows, we always assume that the condition
\eqref{eq:upper} is satisfied.
\end{assumption}

\smallskip
\begin{remark} In real life applications, the wage growth
rate $\mu$ is rather small (but may be either positive or negative).
It is unlikely to exceed the inflation rate $r$, but even if it
does, then it is hardly possible economically that it is greater
than the combined inflation--unemployment rate
$\tilde{r}=r+\lambda_0$. Thus, the condition \eqref{eq:upper} is
absolutely realistic.
\end{remark}

To generalize the expression \eqref{r2d}, consider a delayed entry
time $\tau>0$ (tacitly assuming that $\tau<\infty$). Discounting
first to the entry time $\tau$ when the deduction of the premium~$P$
is activated, and then further down to the initial time moment
$t=0$, yields the \emph{expected net present value}
of the total gain as a function of the initial wage~$x$,
\begin{equation}\label{r2d-new}
\eNPV(x;\tau):=\EE_x \bigl[ \re^{-r \tau} \bigl(
\re^{-r\myp(\tau_0-\tau)} \beta X_{\tau_0} - P\bigr)\myp
\mathbbm{1}_{ \{ \tau < \tau_0\} } \bigr],
\end{equation}
where the expectation on the right now also includes averaging with
respect to $\tau$, which is a functional of the path $(X_t)$. Note
that the indicator function under the expectation specifies that the
entry time $\tau$ must occur prior to $\tau_0$, for otherwise there
will be no gain.

\begin{remark}
The notation \eqref{r2d-new} emphasizes that the expected net
present value depends on the specific entry time~$\tau$. As was
intuitively explained in the Introduction, there is a scope for
optimizing the choice of~$\tau$, where optimality is measured by
\emph{maximizing} $\eNPV(x;\tau)$.
\end{remark}

Formula \eqref{r2d-new} indicates that the decision time $\tau$ has
a finite (random) expiry date $\tau_0$ (using the terminology of
financial options). However, the expectation in \eqref{r2d-new}
involves averaging with respect to $\tau_0$. Moreover, taking
advantage of exponential distribution of $\tau_0$, the expression
\eqref{r2d-new} can be rewritten without any expiry date (i.e., as a
\emph{perpetual option}).

\begin{lemma}\label{lm:NPV}
The expected net present value defined by formula \eqref{r2d-new}
can be expressed in the form
\begin{equation}\label{eq:NPV}
\eNPV(x;\tau)=\EE_x \bigl[\re^{-\tilde{r}\tau}\myn
g(X_\tau)\mypp\mathbbm{1}_{\{\tau<\infty\}}\bigr],
\end{equation}
where the function $g(\cdot)$ is defined in~\eqref{r2d0} and\
$\tilde{r}=r+\lambda_0$ \textup{(}see \eqref{eq:c1-tilde}\textup{)}.
\end{lemma}
\proof Since the distribution of $\tau_0$ is exponential,
the excess time $\tilde{\tau}_0:=\tau_0-\tau$ conditioned on
$\{\tau<\tau_0\}$ is again exponentially distributed (with the same
parameter $\lambda_0$) and independent of~$\tau$. Hence,
conditioning on $\tau$ (restricted to the event $\{\tau<\infty\}$)
and using the total expectation formula as before
\cite[\S\mypp{}II.7, Property~G*, p.\,216]{Shiryaev-pr}), together
with the (strong) Markov property of the process $(X_t)$, we get
from \eqref{r2d-new}
\begin{align}
\notag \eNPV(x;\tau)&=\EE_x \bigl( \EE_x\bigl[ \re^{-r \tau}\myp
(\re^{-r\myp (\tau_0-\tau)}\myp \beta X_{\tau_0} - P)\mypp
\mathbbm{1}_{ \{\tau_0>\tau\} }\myp\bigr|\mypp\tau \bigr]\bigr)\\
\notag &=\EE_x \Bigl(\re^{-r \tau}\myp \EE_x\bigl[(\re^{-r\myp
\tilde{\tau}_0}\myp \beta\myp X_{\tau+\tilde{\tau}_0} -
P)\myp\bigr|\mypp\tau
\bigr]\cdot\EE_x\bigl[\mathbbm{1}_{\{\tau_0>\tau\}}\myp\bigr|\mypp\tau
\bigr]\Bigr)\\&=\EE_x \Bigl(\re^{-r \tau}\myp\myp
\EE_{X_{\tau}}\mynn\bigl[(\re^{-r\myp \tilde{\tau}_0}\myp \beta\myp
\widetilde{X}_{\tilde{\tau}_0} -
P)\bigr]\cdot\PP_x\bigl(\tau_0>\tau\myp|\mypp\tau\bigr)\Bigr),
\label{eq:NPV-mid}
\end{align}
where $\widetilde{X}_t:=X_{\tau+t}$ ($t\ge0$) is a shifted wage
process starting at $\widetilde{X}_0=X_{\tau}$. Substituting
$\PP_x\bigl(\tau_0>\tau\myp|\mypp\tau\bigr)=\re^{-\lambda_0\tau}$
and recalling notation \eqref{r2d0}, formula \eqref{eq:NPV-mid} is
reduced to~\eqref{eq:NPV}.
\endproof


Finally, without loss we can remove the indicator from the
expression~\eqref{eq:NPV} by defining the value of the random
variable under expectation to be zero on the event
$\{\tau=\infty\}$. This definition is consistent with the limit at
infinity. Indeed, observe using \eqref{r1} and \eqref{eq:g-new} that
\begin{align}
\notag \re^{-\tilde{r}\myp t} g(X_t) &=\re^{-\tilde{r}\myp
t}\left(\beta_1\myp x\,\re^{\myp(\mu -
\sigma^2\mynn/2)\myp t +\sigma B_t} -P\right)\\
& =\beta_1\myp
x\mypp\exp\left\{-t\myp\bigl(\tilde{r}-\mu+\tfrac12\myp\sigma^2+\sigma\mypp
t^{-1}B_t\bigr)\right\}- P\mypp\re^{-\tilde{r}\myp
t}.\label{eq:expr_tau=infty}
\end{align}
Due to the condition~\eqref{eq:upper},
$\tilde{r}-\mu+\tfrac12\myp\sigma^2>\tfrac12\myp\sigma^2>0$. In
addition, by the (strong) law of large numbers for the Brownian
motion
(see, e.g., \cite[Exercise~6.4, p.\,265]{Durrett} or \cite[Ch.\,III,
\S\myp3b, p.\mypp246]{Shiryaev}),
$$
\lim_{t\to\infty}t^{-1}\mynn B_t=0\quad(\text{$\PP$-a.s.}).
$$
Thus, the limit of \eqref{eq:expr_tau=infty} as $t\to\infty$ is zero
($\PP_x$-a.s.). Hence, the event $\{\tau=\infty\}$ does not
contribute to the expectation \eqref{eq:NPV}, so that, substituting
\eqref{eq:g-new}, we get
\begin{equation}\label{eq:NPV1+}
\eNPV(x;\tau) =\EE_x \bigl[\re^{-\tilde{r}\tau} g(X_\tau)\bigr].
\end{equation}

To summarize, identification of the optimal entry time
$\tau=\tau^*$, in the sense of maximizing the expected net present
value $\eNPV(x;\tau)$ as a function of strategy $\tau$
(see~\eqref{eq:NPV1+}), is reduced to solving the following
\emph{optimal stopping problem},
\begin{equation}\label{eq:r2e1}
v(x) =\sup_{\tau} \EE_x \bigl[\re^{-\tilde{r}\tau} g(X_\tau)\bigr],
\end{equation}
where the function $g(x)$ is given by \eqref{r2d} and the supremum
is taken over the class of all admissible stopping times $\tau$
(i.e., adapted to the filtration $(\mathcal{F}_t)$). The supremum
$v(x)$ in \eqref{eq:r2e1} is called the \emph{value function} of the
optimal stopping problem.

\subsection{Allowing for mortality}\label{sec:2.3}

The simple model of unemployment insurance set out in
Section~\ref{sec:2.1} can be easily extended to include mortality.
Following
\cite[pp.\,399--401]{Merton}, suppose that the individual who
contemplates taking out the unemployment insurance policy may die
(say, at a random time $\tau_2$ from zero), independently of
employment-related events and subject to a constant force of
mortality $\lambda_2$. That is to say, given that the individual is
alive at current age $t\ge0$, the residual lifetime $\tau_2-t$ is an
independent random variable exponentially distributed with parameter
$\lambda_2$,
$$
\PP\myn\left(\tau_2-t>s\mypp\myp|\mypp\myp\tau_2>t\right)=\re^{-\lambda_2\myp
s}\quad (s\ge0).
$$

The necessary modifications to the unemployment insurance model of
Section~\ref{sec:2.2} start by adjusting the formula for the
expected future benefit (see~\eqref{r2a}). Assuming that death does
not occur prior to the time $\tau_0$ of losing the job (i.e.,
$\tau_2>\tau_0$, so that $\tilde{\tau}_2:=\tau_2-\tau_0$ is
exponentially distributed with parameter $\lambda_2$), the benefit
payments cease at $\tau_1\myn\wedge\tilde{\tau}_2$ (i.e., when a new
job is found or at death, whichever occurs first). Since $\tau_1$
and $\tilde{\tau}_2$ are independent and both have exponential
distributions, the random variable $\tau_1\wedge\tilde{\tau}_2$ has
the exponential distribution with parameter
$\lambda_1\myn+\lambda_2$. Hence, the constant $\beta$ from
\eqref{eq:c1} is now written as
$$
\beta=\int_0^\infty
(\lambda_1+\lambda_2)\mypp\myp\re^{-(\lambda_1\mynn+\lambda_2)\mypp
t} \mypp H(t)\,\rd{t}.
$$
Next, we need to take into account the effect of death in service,
that is, if $\tau_2\le \tau_0$. To be specific, it is reasonable to
assume that the lump sum to be paid by the employer in this case is
proportional to the final wage, say $a^\dag X_{\tau_2}$. Then,
separating the cases where death occurs after or prior to loss of
job, it is easy to see that the definition \eqref{r2d0} of the gain
function (i.e., net expected benefit discounted to the policy entry
time) takes the form
\begin{equation}\label{r2d0-mort}
g(x)=\EE_x\bigl(\re^{-r\tau_0}\myp \beta\mypp
X_{\tau_0}\myp\mathbbm{1}_{\{\tau_0<\tau_2\}}\bigr)+\EE_x\bigl(\re^{-r\tau_0}\myp
a^\dag X_{\tau_2}\myp\mathbbm{1}_{\{\tau_2\le\tau_0\}}\bigr)-P.
\end{equation}

The first expectation in \eqref{r2d0-mort} is computed using
conditioning on $\tau_0$ and the total expectation formula
(cf.~\eqref{eq:g-new}),
\begin{alignat}{2}
\notag \EE_x\myn\bigl(\re^{-r\tau_0}\mypp
X_{\tau_0}\myp\mathbbm{1}_{\{\tau_0<\tau_2\}}\bigr)&=\lefteqn{\EE_x\mynn\bigl[\re^{-r\tau_0}\,
\EE_x\myn\bigl(X_{\tau_0}\myp\mathbbm{1}_{\{\tau_0<\tau_2\}}\mypp\big|\mypp\myp\tau_0\bigr)\bigr]}\\
\notag &=\lefteqn{\EE_x\mynn\bigl[\re^{-r\tau_0}\,
\EE_x(X_{\tau_0}\myp|\mypp\myp\tau_0)\cdot
\PP_x(\tau_2>\tau_0\mypp |\mypp\myp\tau_0)\bigr]}\\
\notag &=\EE_x\myn\bigl(\re^{-r\tau_0}\mypp x\mypp\re^{\myp\mu\myp
\tau_0}\mypp \re^{-\lambda_2\tau_0}\bigr)&&=x\int_0^\infty
\!\re^{(\mu-r-\lambda_2)\myp t}\mypp \lambda_0\mypp
\re^{-\lambda_0\myp t}\mypp\rd{t}\\
&&&=\frac{\lambda_0\mypp x}{r+\lambda_0+\lambda_2-\mu},
\label{eq:g-new-mort1}
\end{alignat}
where in the second line we used conditional independence of
$X_{\tau_0}$ and $\tau_2$ given $\tau_0$. Similarly, by conditioning
on $\tau_2$ the second expectation in \eqref{r2d0-mort} is
represented as
\begin{align}
\notag \EE_x\myn\bigl(\re^{-r\tau_0}\mypp
X_{\tau_2}\myp\mathbbm{1}_{\{\tau_2\le
\tau_0\}}\bigr)&=\lefteqn{\EE_x\mynn\bigl[X_{\tau_2}\,
\EE_x\mynn\bigl(\re^{-r\tau_0}\myp\mathbbm{1}_{\{\tau_0\ge \tau_2\}}\mypp\big|\mypp\myp\tau_2\bigr)\bigr]}\\
\notag &=\EE_x\mynn\biggl[X_{\tau_2}\!\int_{\tau_2}^\infty
\!\re^{-r\myp t}\mypp \lambda_0\mypp \re^{-\lambda_0\myp
t}\mypp\rd{t}\biggr]\\
&=\frac{\lambda_0}{r+\lambda_0}\mypp\EE_x\myn\bigl(X_{\tau_2}\mypp
\re^{-(r+\lambda_0)\myp \tau_2}\bigr). \label{eq:g-new-mort2}
\end{align}
Again conditioning on $\tau_2$, the last expectation is computed as
follows,
\begin{align}
\notag \EE_x\myn\bigl(X_{\tau_2}\mypp \re^{-(r+\lambda_0)\myp
\tau_2}\bigr)&=\EE_x\mynn\bigl[\re^{-(r+\lambda_0)\myp
\tau_2}\myp\EE_x(X_{\tau_2}\mypp|\mypp\tau_2)\bigr]\\
\notag &=\EE_x\myn\bigl(\re^{-(r+\lambda_0)\myp \tau_2}\myp x\mypp
\re^{\myp\mu\myp\tau_2}\bigr)\\
\notag &=x\int_0^\infty \!\re^{-(r+\lambda_0-\mu)\myp
t}\lambda_2\mypp\re^{-\lambda_2\myp t}\,\rd{t}\\
&=\frac{\lambda_2\mypp x}{r+\lambda_0+\lambda_2-\mu}.
 \label{eq:g-new-mort3}
\end{align}

Finally, substituting the expressions \eqref{eq:g-new-mort1},
\eqref{eq:g-new-mort2} and \eqref{eq:g-new-mort3} into the
definition \eqref{r2d0-mort}, we obtain explicitly
$$
g(x)=\frac{\lambda_0\mypp
x}{r+\lambda_0+\lambda_2-\mu}\left(\beta+\frac{\lambda_2 \myp
a^\dag}{r+\lambda_0}\right) - P.
$$
This expression has the same form as \eqref{r2d} but with the
parameters $\tilde{r}$ and $\beta_1$ redefined as follows
(cf.~\eqref{eq:c1-tilde}),
$$
\tilde{r}:=r+\lambda_0+\lambda_2,\qquad
\beta_1:=\frac{\lambda_0}{\tilde{r}-\mu}\left(\beta+\frac{\lambda_2
\myp a^\dag}{r+\lambda_0}\right).
$$
In addition, the inequality \eqref{eq:upper} of Assumption
\ref{as:1} is updated accordingly. Subject to this
reparameterization, all subsequent calculations leading to the
optimal stopping problem \eqref{eq:r2e1} remain unchanged.

For the sake of clarity and in order not to distract the reader by
unnecessary technicalities, in the rest of the paper we adhere to
the original version of the model \textup{(}i.e., with no mortality,
$\lambda_2=0$\textup{)}; however, see the discussion at the end of
Section~\ref{sec:6.4} indicating an important regularizing role of
mortality, helping to avoid undesirable inconsistencies of the model
at small unemployment rates~$\lambda_0$.

\subsection{A priori properties of the value function}\label{sec:2.4}
The next lemma shows that the optimal stopping problem
\eqref{eq:r2e1} is well posed.
\begin{lemma}\label{lm:v(x)} The value function $x\mapsto v(x)$ of the optimal stopping
problem \eqref{eq:r2e1} has the following properties:
\begin{itemize}
\item[{\rm (i)}]
$v(0)=0$ and, moreover, $v(x)\ge 0$ for all $x\ge0$\textup{;}
\item[\rm (ii)]
$v(x)<\infty$ for all $x\ge0$.
\end{itemize}
\end{lemma}
\proof (i) If $x=0$ then, due to \eqref{r1}, $X_t\equiv 0$
($\PP_0$-a.s.)\ and the stopping problem \eqref{eq:r2e1} is reduced
to
\begin{equation*}
v(0)=\sup_{\tau} \EE_0 (-P\mypp\re^{-\tilde{r}\tau}),
\end{equation*}
which has the obvious solution $\tau=\infty$ ($\PP_0$-a.s.), with
the corresponding supremum value $v(0)=0$. Furthermore, by
considering $\tau=\infty$ ($\PP_x$-a.s.)\ it readily follows from
\eqref{eq:r2e1} that $v(x)\ge 0$ for all $x\ge0$.

\smallskip
(ii) Recalling that $\mu<\tilde{r}$ (see Assumption~\ref{as:1}),
observe that the process $\re^{-\tilde{r}\myp t}X_t$ is a
\emph{supermartingale}; indeed, for $0\le s\le t$ we have, using
\eqref{r1} and \eqref{eq:E(X)},
\begin{align*}
\EE_x\bigl[\re^{-\tilde{r}\myp
t}X_t\mypp|\mypp\mathcal{F}_s\bigr]&=\re^{-\tilde{r}\myp
t}X_s\mypp\EE\bigl[\re^{\myp\sigma (B_t-B_s)+(\mu -
\frac{1}{2}\sigma^2)(t-s)}\bigr]\\
&=\re^{-\tilde{r}\myp t}X_s\mypp \re^{\myp\mu(t-s)}\\
&\le \re^{-\tilde{r}\myp s}X_s\quad (\PP_x\text{-a.s.}).
\end{align*}
In particular,
$$
\EE_x(\re^{-\tilde{r}\myp t}X_t)\le \EE_x(X_0)=x.
$$
Hence, by Doob's optional
sampling theorem for non-negative, right-continuous supermartingales
(see, e.g., \cite[Theorem~8.18, pp.\mypp140--141]{Yeh}),
for any stopping time $\tau$
we have
$$
\EE_x(\re^{-\tilde{r} \tau}X_{\tau})\le\EE_x(X_0)=x,
$$
and it follows that the supremum in \eqref{eq:r2e1} is finite.
\endproof

\subsection{The optimal stopping rule}\label{sec:2.5}
\begin{figure}[th!]
\vspace{-1.5pc}    \centering
   \includegraphics[%
height=0.32\textheight] {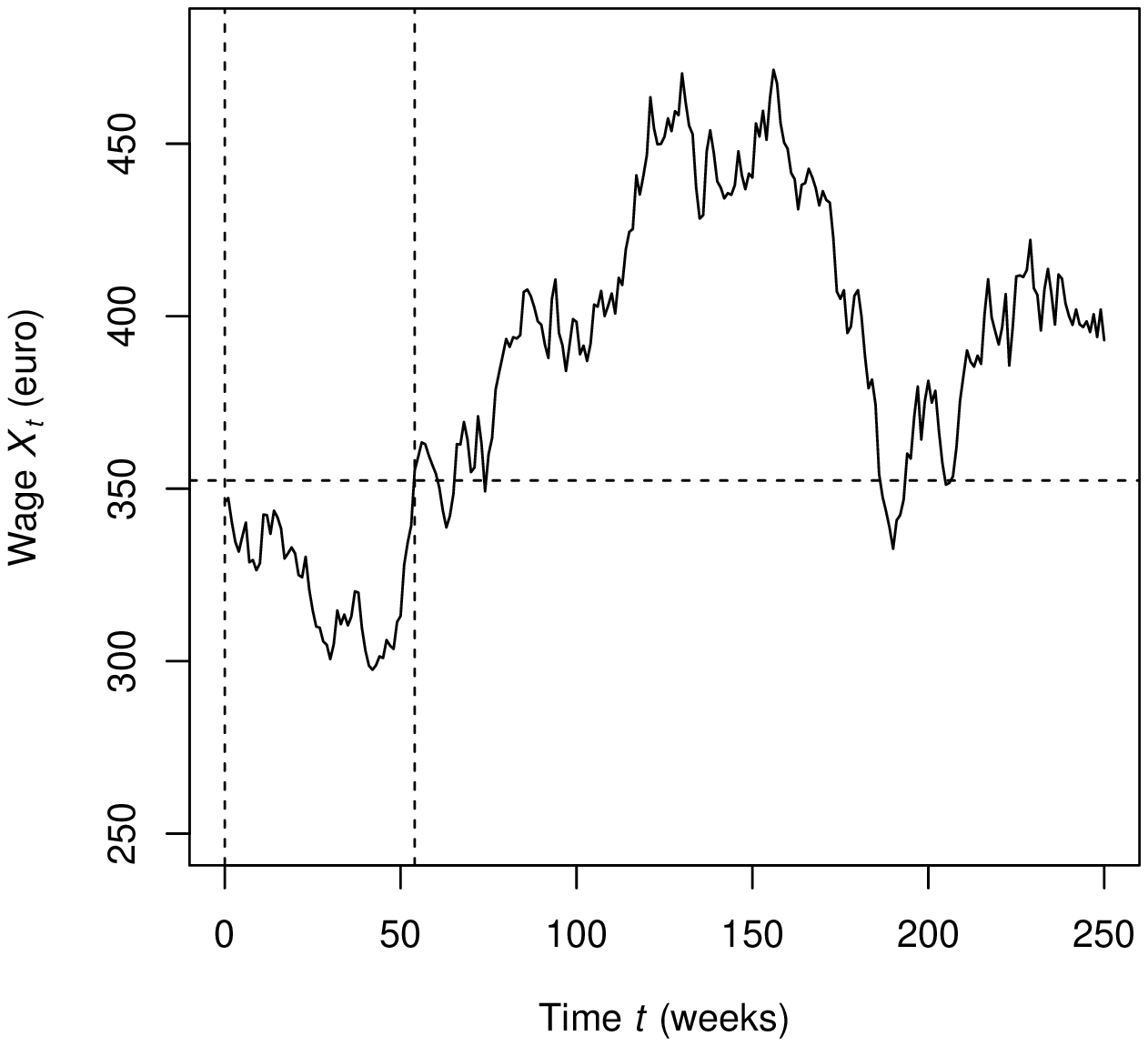}
\put(-163.0,110.0){\mbox{\scriptsize$b^*$}}
\put(-141.5,42.5){\mbox{\scriptsize$\tau^*$}}
   \hspace{0.5pc}
   \includegraphics[%
        height=0.32\textheight]
        {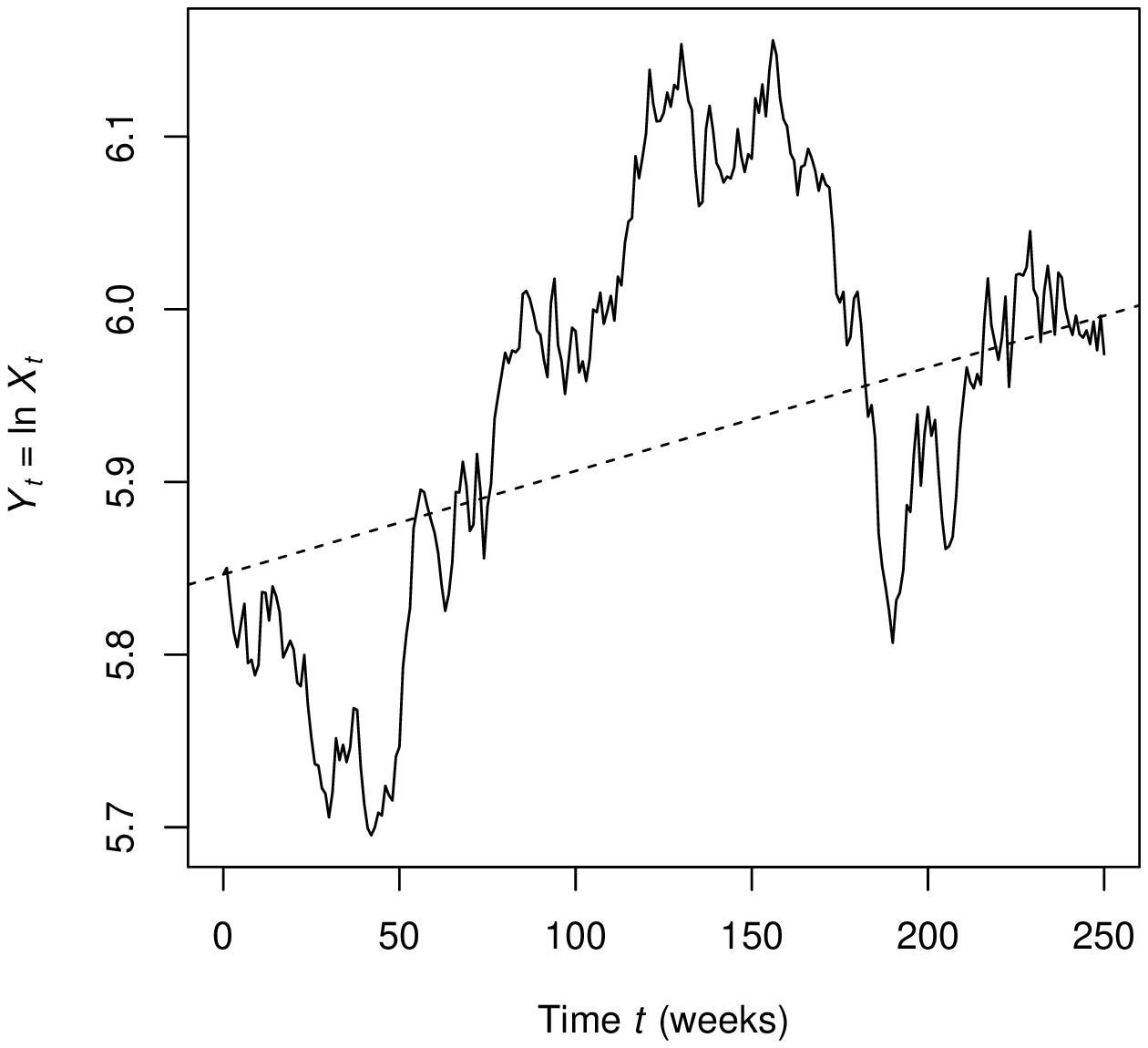}
    \caption{Simulated wage process $X_t$ (left) and
    $Y_t=\ln X_t$ (right) according to the geometric
    Brownian motion model \eqref{r1}, with $X_0=346$ (euros) and parameters $\mu=0.0004$ and $\sigma=0.02$ (see Example~\ref{ex:5.2}).
    The dashed horizontal line on the left plot indicates the optimal threshold $b^*\doteq
    352.37$ (euros)
    first attained in this simulation at $\tau^*\!=54$ (weeks).
    The dashed line on the right plot shows the estimated drift of the log-transformed data (see Section~\ref{sec:5.2}).
   }
    \label{fig3}
\end{figure}

For the wage process $(X_t)$, consider the \emph{hitting time}
$\tau_b$ of a threshold $b\in\RR$, defined by
\begin{equation*}
\tau_b:=\inf\{t\ge 0\colon X_t\ge b\}\in[\myp0,\infty].
\end{equation*}
(As usual, we make a convention that $\inf\varnothing =\infty$.)
Clearly, $\tau_b$ is a stopping time, that is, $\{\tau\le
t\}\in\mathcal{F}_t$ for all $t\ge0$. Since the process $X_t$ has
a.s.-continuous sample paths, on the event $\{\tau_b<\infty\}$ we
have $X_{\tau_b}=b$ ($\PP_x$-a.s.). As we will show, the optimal
strategy for the optimal stopping problem \eqref{eq:r2e1} is to wait
until the random process $X_t$ hits a certain threshold~$b^*\mynn$
(see Fig.~\ref{fig3}). More precisely, the solution to
\eqref{eq:r2e1} is provided by the following stopping rule,
\begin{equation}\label{eq:stopping_rule}
\tau^*=\begin{cases} \tau_{b^*}&\text{if}\ \,x\in[\myp0,b^*],\\
0&\text{if}\ \,x\in[\myp b^*,\infty).
\end{cases}
\end{equation}
That is to say, if $x\ge b^*\mynn$ then one must stop and buy the
policy immediately, or else wait until the hitting time
$\tau_{b^*}\!\ge 0$ occurs and buy the policy then. (Of course,
these two rules coincide when $x=b^*\myn$.) However, if it happens
so that $\tau_{b^*}=\infty$, then, according to the above rule, one
must wait indefinitely and, therefore, never buy the policy.

The specific value of the critical threshold $b^*\mynn$ is given by
\begin{equation}\label{eq:b*}
b^*=\frac{P q_*}{\beta_1\myp(q_*-1)},
\end{equation}
where
\begin{equation}\label{eq:q1*}
q_*= \frac{1}{\sigma^2}\mynn\left(-\bigl(\mu-\tfrac12\myp
\sigma^2\bigr) + \sqrt{\bigl(\mu-\tfrac12\myp \sigma^2\bigr)^2 +
2\myp\tilde{r}\sigma^2}\mypp\right).
\end{equation}
It is straightforward to check, using condition \eqref{eq:upper},
that $q_*>1$ (see also Section~\ref{sec:3.2}). Finally, the
corresponding value function \eqref{eq:r2e1} is specified as
\begin{equation}\label{eq:value-b*}
    v(x) = \begin{cases}
    \displaystyle (\beta_1 b^* - P) \left( \frac{x}{\displaystyle b^*}\right)^{\myn q_*}\!, &x\in[\myp0,b^*],\\
     \displaystyle\beta_1x - P,& x\in[\myp b^*,\infty).
    \end{cases}
\end{equation}
Equivalently, substituting the expression \eqref{eq:b*}, formula
\eqref{eq:value-b*} is explicitly rewritten as
\begin{equation}\label{eq:value-b*-exp}
v(x)=\begin{cases} \displaystyle\frac{P}{q_*-1}\left(\frac{\beta_1
(q_*-1)\myp x}{P q_*}\right)^{\myn q_*}\!,&\displaystyle \ 0\le x\le
\frac{P q_*}{\beta_1(q_*-1)},\\[.8pc]
\displaystyle \beta_1x-P,&\displaystyle \hphantom{\ 0\le{}}x\ge
\frac{P q_*}{\beta_1(q_*-1)}.
\end{cases}
\end{equation}
In particular, the function $x\mapsto v(x)$ is strictly increasing
for $x\ge 0$, with $v(0)=0$ (cf.\ Lemma~\ref{lm:v(x)}).

These results will be proved in Section~\ref{sec:3}.

\subsection{Deterministic case}\label{sec:2.6}

For orientation, it is useful to consider the simple baseline case
$\sigma=0$, where the random process $X_t$ (see~\eqref{r1})
degenerates to the deterministic function
$$
X_t=x\mypp\re^{\myp \mu t}\quad (t\ge0).
$$
Hence, any stopping time $\tau$ is non-random, say $\tau=t$, and the
optimal stopping problem \eqref{eq:r2e1} is reduced to
\begin{equation}\label{eq:r2e2-det}
v(x)=\sup_{t\ge0} \bigl[ \re^{-\tilde{r}\myp t} (\beta_1
x\mypp\re^{\myp\mu t}-P)\bigr].
\end{equation}
The problem \eqref{eq:r2e2-det} is easily solved, with the maximizer
$t^*$ given by
\begin{equation}\label{eq:t<t*}
t^*=\inf\bigl\{t\ge0\colon x\myp\re^{\myp\mu t}\ge
b_0^*\bigr\}\in[\myp0,\infty],
\end{equation}
where
\begin{equation}\label{eq:b*det}
b_0^*=\begin{cases} \displaystyle
\frac{P\myp\tilde{r}}{\beta_1(\tilde{r}-\mu)},
&\ \mu>0,\\[.8pc]
\displaystyle \frac{P}{\beta_1},
&\ \mu\le 0.
\end{cases}
\end{equation}
The expression \eqref{eq:b*det} is consistent with the general
formula \eqref{eq:b*}, noting that, in the limit as
$\sigma\downarrow0$, the quantity \eqref{eq:q1*} is reduced to
(cf.~\eqref{eq:upper})
$$
q_*=
\begin{cases}\displaystyle \frac{\tilde{r}}{\mu}>1,&\mu>0,\\
\myp\infty,&\mu\le0.
\end{cases}
$$
With this convention, it is easy to check that the value function
\eqref{eq:r2e2-det} is given by the general
formula~\eqref{eq:value-b*}. In particular, if $\mu\le 0$ and
$x<b_0^*$ then, according to \eqref{eq:t<t*}, $t^*\myn=\infty$ and
from \eqref{eq:r2e2-det} we get $v(x)=0$; indeed, the function
$t\mapsto x\myp\re^{\myp \mu t}$ is non-increasing, so it never
attains the required threshold $b_0^*>x$. In contrast, if $x\ge
b_0^*$ then by \eqref{eq:t<t*} $t^*\myn=0$ (for any~$\mu$), and
\eqref{eq:r2e2-det} readily yields $v(x)=\beta_1x-P$.

\section{Solving the optimal stopping problem}\label{sec:3}
The optimal stopping problem \eqref{eq:r2e1} involves two tasks: (i)
evaluating the value function $v(x)$, and (ii) identifying the
maximizer $\tau=\tau^*$. A standard approach is to try and
\emph{guess} the solution and then to \emph{verify} that it is
correct.

\subsection{Guessing the solution}\label{sec:3.1}
Let us look more closely at the nature of the value function $v(x)$
that we are trying to identify. Observe that by picking $\tau =0$ in
\eqref{eq:r2e1} yields the lower estimate
\begin{equation}\label{eq:v>g}
v(x) \ge g(x).
\end{equation}
Clearly, if $v(x)>g(x)$ then we have not yet achieved the maximum
payoff available, so we should continue to wait. On the other hand,
if $v(x)=g(x)$ then the maximum has been attained and we should
stop. This motivates the definition of the two regions, $C$
(\emph{continuation}) and $S$ (\emph{stopping}),
\begin{equation*}
C:=\{ x\ge0\colon v(x) > g(x)\},\qquad S:=\{ x\ge0\colon v(x) \le
g(x)\}.
\end{equation*}

By virtue of the Markov property of the process $X_t$, the same
argument can be propagated to any time $t\ge0$,
provided that stopping has not yet occurred. Namely, if $X_t=x'$
(and $\tau\ge t$) then the problem \eqref{eq:r2e1} is updated with
the new (residual) stopping time $\tau'=\tau-t$ and with the initial
value $x$ replaced by $x'$.

Thus, it is natural to expect that the optimal strategy prescribes
to continue as long as the current wage value $X_t$ belongs to the
region $C$ (i.e., $v(X_t)>g(X_t)$), but to stop when $X_t$ first
enters the region~$S$ (i.e., $v(X_t)\le g(X_t)$). That is to say,
the optimal stopping time should be given by\footnote{This
conclusion is in accord with the general optimal stopping theory
\cite[\S\mypp{}II.2.2]{peskir2006optimal}.}
\begin{equation}\label{eq:tau*}
\tau^*= \inf\{t\ge0\colon X_t\in S\}=\inf\{t\ge0\colon v(X_t)\le
g(X_t)\}\in[\myp0,\infty].
\end{equation}

To clarify the plausible structure of the stopping set $S$, recall
(see the proof of Lemma \ref{lm:v(x)}(i)) that a zero value of the
stopping problem \eqref{eq:r2e1} is achieved by simply using the
strategy $\tau\equiv\infty$, that is, by never joining the scheme.
Thus, if the initial wage $X_0=x$ is small (e.g., such that
$g(x)=\beta_1x-P<0$) then, in order to secure a positive payoff, we
should wait for a sufficiently high wage $X_t$. This suggests that
the stopping rule \eqref{eq:tau*} is reduced to the first hitting
time for a certain set on the plane $\{(t,x)\colon t\ge0,\,x\ge0\}$.
Furthermore, noting that the definition \eqref{eq:tau*} is time
homogeneous, in that it does not change in the course of time $t$,
we also hypothesize the simplest situation whereby the regions $C$
and $S$ are determined by a constant threshold $y=b^*>0$,
\begin{align}\label{def:b*}
C=[\mypp 0,b^*),\qquad S=[\myp b^*,\infty).
\end{align}
In other words, the conjectural hitting boundary does not depend on
time.

Hence, we are led to the reduced optimal stopping problem over the
subclass of hitting times,
\begin{equation}\label{eq:r2e**red}
u(x)=\sup_{b\ge 0} \EE_x \bigl[ \re^{-\tilde{r}\tau_b} \myp
g(X_{\tau_b})\bigr].
\end{equation}
In particular, formula \eqref{eq:tau*} specializes to
\begin{equation}\label{eq:tau*b}
\tau_{b^*}= \inf\{t\ge0\colon X_t\ge b^*\}=\inf\{t\ge0\colon
u(X_t)\le g(X_t)\}\in[\myp0,\infty].
\end{equation}
Our first task is to identify the value function $u(x)$ in
\eqref{eq:r2e**red} and the corresponding maximizer $b=b^*\mynn$ by
solving the corresponding free-boundary problem
(Section~\ref{sec:3.2}). After that, we will have to show that this
solution is optimal in the general class of stopping times, that is,
$u(x)=v(x)$ for all $x\ge 0$ (Section~\ref{sec:3.3}).

\subsection{Free-boundary problem}\label{sec:3.2}

According to general theory of optimal stopping (see, e.g.,
\cite[Ch.\,IV]{peskir2006optimal}), in the continuation region
$C=[\myp0,b)$ (see~\eqref{def:b*}) the value function $u(x)$ from
\eqref{eq:r2e**red} must be \emph{harmonic} with respect to the
underlying process $\widetilde{X}_t$ generated by $X_t$. More
precisely, \strut{}due to the discounting exponential factor in the
optimal stopping problem \eqref{eq:r2e**red}, the process
$\widetilde{X}_t$ is obtained from $X_t$ by independent
\emph{killing} (or discounting) with rate $\tilde{r}$ (see
\cite[\S\S\,5.4, 6.3]{peskir2006optimal}). Thus, if $b$ is a
suitable threshold
and $\tau_{b}$ is the corresponding hitting time,
then for any $x\ge 0$ the following condition must hold,
\begin{equation}\label{eq:harmonic1}
\EE_x\bigl[\re^{-\tilde{r}\myp(\tau_{b}\wedge t)}u(X_{\tau_{b}\wedge
t})\bigr]=u(x)\quad(t\ge 0).
\end{equation}

Note that the geometric Brownian motion $X_t$ determined by the
stochastic differential equation \eqref{r1-0} is a diffusion process
with the infinitesimal generator
\begin{equation}\label{lop}
L:=\mu\myp x \frac{\rd }{\rd x}+\tfrac12\myp \sigma^2 x^2
\frac{\rd^2 }{\rd x^2}\quad (x>0).
\end{equation}
The generator of the killed process $\widetilde{X}_t$ is then given
by (see \cite[\S\,6.3, p.\mypp127]{peskir2006optimal})
\begin{equation}\label{eq:tilde-L}
\tilde{L}=L-\tilde{r} I,
\end{equation}
where $I$ is the identity operator. Then the harmonicity condition
\eqref{eq:harmonic1} can be reduced to the differential equation
$\tilde{L} u =0$, that is, $Lu-\tilde{r}u=0$
(see~\eqref{eq:tilde-L}).

On the boundary $x=b$ of the set $C=[\myp0,b)$, due to the stopping
rule \eqref{eq:tau*b} we have $u(b)=g(b)$. Moreover, according to
the \emph{smooth fit principle} (see
\cite[\S\mypp9.1]{peskir2006optimal}), we must also satisfy the
condition $u'(b)=g'(b)$. Finally, in view of the equality $v(0)=0$
(see Lemma~\ref{lm:v(x)}(i)), we add a Dirichlet boundary condition
at zero, $u(0+)=\lim_{x\downarrow0}u(x)=0$. Thus, we arrive at the
following \emph{free-boundary problem},
\begin{equation}\label{eq:BC}
\begin{cases}
Lu(x)-\tilde{r}\myp u(x)=0,&\ x\in(0,b),\\
u(b)=g(b),\\
u'(b)=g'(b),\\
u(0+)=0,
\end{cases}
\end{equation}
where both $b>0$ and $u(x)$ are unknown.

Substituting \eqref{r2d} and~\eqref{lop}, the problem \eqref{eq:BC}
is rewritten explicitly as
\begin{align}\label{eq:ub}
\left\{ \!\myn\begin{array}{ll} \mu\myp x\mypp u'(x)+\tfrac12\myp
\sigma^2 x^2 u''(x)-\tilde{r}\myp
u(x)=0,&\ x\in(0,b),\\[.4pt]
u(b)=\beta_1 b-P,\\[.4pt]
u'(b)=\beta_1,\\[.4pt]
u(0+)=0.
\end{array}
\right.
\end{align}
Let us look for a solution of \eqref{eq:ub} in the form $u(x) = x^q$
($x>0$), with a suitable parameter $q\in\RR$. Then the differential
equation in \eqref{eq:ub} yields
\begin{equation}\label{e6}
\tfrac12\myp \sigma^2 q\mypp(q-1) + \mu\myp q -\tilde{r}=0.
\end{equation}
This quadratic equation has two distinct roots,
\begin{equation*}
q_{1,2}= \frac{1}{\sigma^2}\mynn\left(-\bigl(\mu-\tfrac12\myp
\sigma^2\bigr) \pm \sqrt{\bigl(\mu-\tfrac12\myp \sigma^2\bigr)^2 +
2\myp\tilde{r} \sigma^2}\mypp\right),
\end{equation*}
where $q_2<0<q_1=q_*$ (see~\eqref{eq:q1*}). Also note that, due to
the condition \eqref{eq:upper}, the left-hand side of \eqref{e6} is
negative at $q=1$, therefore $q_1>1$.  Thus, the general solution of
the differential equation \eqref{eq:ub} is given~by
\begin{equation}\label{ef3}
u(x) = A\myp x^{q_1} + B\myp x^{q_2},\quad x\in(0,b),
\end{equation}
with arbitrary constants $A$ and $B$. But since $q_2<0$, the
condition $u(0+)=0$ implies that $B=0$. Hence, \eqref{ef3} is
reduced to $u(x) = A\myp x^{q_1}\equiv A\myp x^{q_*}$ ($0<x<b$).
Furthermore, the boundary conditions in \eqref{eq:ub} yield
\begin{equation*}
\begin{cases}
A\myp b^{\myp q_*} =\beta_1 b-P,\\
A\myp q_* b^{\myp q_*-1}= \beta_1,
\end{cases}
\end{equation*}
whence we find
\begin{equation}\label{e11}
A = \frac{\beta_1 b-P}{b^{\myp q_*}}, \qquad b= \frac{P
q_*}{\beta_1(q_*-1)}.
\end{equation}
Thus, the required solution to \eqref{eq:ub} is given by
\begin{equation}\label{a3}
    u(x) = \begin{cases}
   (\beta_1 b - P) \left( \dfrac{x}{b}\right)^{q_*}\!, &x\in[\myp0,b\myp],\\[.3pc]
    \beta_1x - P,& x\in[\myp b,\infty)
    \end{cases}
\end{equation}
where the threshold $b$ is defined in \eqref{e11} and $q_*>1$ is the
positive root of the equation~\eqref{e6}, given explicitly by
formula~\eqref{eq:q1*}.

\subsection{Verification of the found solution}\label{sec:3.3}

Using \eqref{e11} and \eqref{a3}, it is easy to see that
\begin{equation}\label{eq:u>g}
\begin{aligned}
u(x)&=g(x),\ &&x\in [\myp b,\infty),\\
u(x)&>g(x),\ &&x\in[\myp 0,b),
\end{aligned}
\end{equation} in accord with the heuristics outlined in
Section~\ref{sec:3.1} (see~\eqref{def:b*}). However, there is no
need to check that the function $u(x)$ defined in \eqref{a3} solves
the reduced optimal stopping problem \eqref{eq:r2e**red}, because we
can prove directly that $u(x)$ provides the solution to the original
optimal stopping problem~\eqref{eq:r2e1}, that is, $u(x)=v(x)$ for
all $x\ge0$.

\begin{remark}
Since $u(0)=0$ by formula \eqref{a3}, and $v(0)=0$ according to
Lemma~\ref{lm:v(x)}(i), in what follows it suffices to assume that
$x>0$.
\end{remark}

The proof of the claim above (commonly referred to as
\emph{verification theorem}) consists of two parts.

\begin{itemize}
\item[(i)] Let us first show that $u(x)\ge v(x)$ $(x>0$). If the map
$x\mapsto u(x)$ was a $C^2$-function (i.e., with continuous second
derivative), then the classical \emph{It\^{o} formula} (see, e.g.,
\cite[Theorem 4.1.2, p.\,44]{Oksendal})
applied to $\re^{-\tilde{r}\myp t}\myp u(X_t)$ would yield, on
account of \eqref{r1-0} and~\eqref{lop},
\begin{equation}\label{a4-t}
\re^{-\tilde{r}\myp t}\myp u(X_t) =  u(x) + \int_{0}^{t}
\re^{-\tilde{r} s}\bigl(Lu(X_s) -\tilde{r}\myp u(X_s)\bigr)\mypp
\rd{s} + M_t \quad (\PP_x\text{-a.s.}),
\end{equation}
where
\begin{equation}\label{eq:M}
M_t:=\int_{0}^{t} \re^{-\tilde{r} s} u'(X_{s}) \,\sigma {X_{s}}
\,\rd B_s\quad (t\ge0).
\end{equation}
However, for the function $u(x)$ given by \eqref{a3}, its
$C^2$-smoothness breaks down at the point $x=b$, where it is only
$C^1$. But $u(x)$ is strictly convex on $(0,b)$ (i.e., $u''(x)>0$)
and linear on $(b,\infty)$, and we can define the action $Lu(x)$ at
$x=b$ by using the one-sided second derivative, say,
\begin{equation}\label{eq:u''}
u''(b-)=Pq_*b^{-2}.
\end{equation}
In this situation, a generalization of the It\^{o} formula holds,
known as the \emph{It\^{o}--Meyer formula} (see \cite[Ch.\,VIII,
\S\mypp2a, p.\mypp757]{Shiryaev}), which ensures that the
representation \eqref{a4-t} is still valid.

\smallskip
Recall that by construction (see the differential equation
in~\eqref{eq:BC}), we have
\begin{equation}\label{eq:Lu=0}
Lu(x)-\tilde{r}\myp u(x)=0,\quad x\in(0,b).
\end{equation}
Moreover, it is easy to check using \eqref{eq:u''} that the equality
\eqref{eq:Lu=0} also extends to $x=b$. On the other hand, on account
of the condition \eqref{eq:upper} and the definition of $b$ in
\eqref{e11}, for $x>b$ we get
\begin{align}
\notag
Lu(x)-\tilde{r}\myp u(x)&=\mu\myp
\beta_1x-\tilde{r}\mypp(\beta_1x-P)\\
\notag
&=\beta_1x\myp (\mu-\tilde{r})+\tilde{r} P\\
\notag
&<\beta_1b\myp (\mu-\tilde{r})+\tilde{r} P\\
&=\frac{P\mypp(\mu\myp q_*-\tilde{r})}{q_*-1}<0, \label{eq:Lu<0*}
\end{align}
because, due to the equation \eqref{e6} and the inequality $q_*>1$,
$$
\mu\myp q_*-\tilde{r}=-\tfrac12\myp\sigma^2 q_*(q_*-1)<0.
$$

Thus, combining \eqref{eq:Lu=0} and \eqref{eq:Lu<0*} we obtain
\begin{equation}\label{eq:L<0}
Lu(x)-\tilde{r}\myp u(x)\le0\quad (x>0).
\end{equation}
Substituting the inequality \eqref{eq:L<0} into formula
\eqref{a4-t}, we conclude that, for any $x>0$ and all $t\ge0$,
\begin{equation}\label{21}
u(x)+ M_t\ge \re^{-\tilde{r}\myp t}\myp u(X_t) \quad
(\PP_x\text{-a.s.}).
\end{equation}
According to formula \eqref{eq:M}, $(M_t)$ is a continuous local
martingale (see, e.g., \cite[Ch.\,II, \S\myp1c,
p.\myp101]{Shiryaev}). Let $(\tau_{n})$ be a localizing sequence of
bounded stopping times, so that $\tau_n\uparrow\infty$
($\PP_x$-a.s.)\ and the stopped process $(M_{\tau_n\wedge\myp t})$
is a martingale, for each~\mbox{$n\in\NN$}.

\smallskip
Now, let $\tau$ be an arbitrary stopping time of ($X_t)$.
From \eqref{21} we get
\begin{align}
\notag u(x)+M_{\tau_n\wedge\tau}&\ge \re^{-\tilde{r}\myp
(\tau_n\wedge\tau)}\myp u(X_{\tau_n\wedge\tau})\\
&\ge \re^{-\tilde{r}\myp (\tau_n\wedge\tau)}\myp
g(X_{\tau_n\wedge\tau})\quad (\PP_x\text{-a.s.}), \label{23}
\end{align}
using that $u(x)\ge g(x)$ for all $x\ge0$ (see~\eqref{eq:u>g}).
Taking expectation on both sides of the inequality \eqref{23} gives
\begin{equation}\label{24}
u(x)\ge \EE_x \bigl[\re^{-\tilde{r}{(\tau_{n} \wedge \tau)}} \mypp
g( X_{\tau_{n} \wedge \tau})\bigr],
\end{equation}
since by Doob's optional sampling
theorem (see, e.g., \cite[Theorem~8.10, p.\mypp131]{Yeh})
$$
\EE_x[M_{\tau_n\wedge\tau}]=\EE_x[M_0]=0.
$$
By Fatou's lemma (see, e.g., \cite[\S\mypp{}II.6, Theorem~2(a),
p.\mypp187]{Shiryaev-pr}),
from \eqref{24} it follows
\begin{equation}\label{26}
u(x)\ge \EE_x \bigl[\liminf_{n\to\infty} \re^{-\tilde{r}{(\tau_{n}
\wedge \tau)}} \mypp  g(X_{\tau_{n} \wedge \tau})\bigr]=\EE_x\bigl[
\re^{-\tilde{r}{\tau }}\mypp g(X_{\tau}) \bigr].
\end{equation}
Finally, taking in \eqref{26} the supremum over all stopping times
$\tau$, we obtain
$$
u(x)\ge \sup_{\tau} \EE_x \bigl[\re^{-\tilde{r}{\tau }}\mypp
g(X_{\tau})\bigr] =v(x)\quad (x>0),
$$
as claimed.

\item[(ii)] Let us now prove the opposite inequality, $u(x) \le v(x)$
($x>0$). According to \eqref{eq:v>g} and \eqref{eq:u>g}, we readily
have $u(x)=g(x)\le v(x)$ for $x\in[\myp b,+\infty)$. Next, fix
$x\in(0,b)$ and consider the representation \eqref{a4-t} with $t$
replaced by $\tau_n\wedge\tau_b$, where $(\tau_n)$ is the localizing
sequence of stopping times for ($M_t$) as before. Then, by virtue of
the identity~\eqref{eq:Lu=0} (which, as has been explained, is also
true for $x=b$), it follows that
\begin{equation}\label{23'}
u(x)+M_{\tau_n\wedge\tau}=\re^{-\tilde{r}\myp
(\tau_n\wedge\tau_b)}\myp u(X_{\tau_n\wedge\tau_b})\quad
(\PP_x\text{-a.s.}).
\end{equation}
Similarly as above, taking expectation on both sides of the equality
\eqref{23'} and again applying Doob's optional sampling theorem to
the martingale $(M_{\tau_{n}\wedge t})$, we obtain
\begin{equation}\label{28}
u(x)=\EE_x \bigl[\re^{-\tilde{r}\myp(\tau_n\wedge\tau_b)}\myp
u(X_{\tau_n\wedge\tau_b})\bigr].
\end{equation}
Note that, for $0<x<b$, we have $0\le u(x)\le u(b)$ and $0\le
X_{\tau_n\wedge\tau_b}\le b$ ($\PP_x$-a.s.), hence
$$
0\le \re^{-\tilde{r}\myp(\tau_n\wedge\tau_b)}\myp
u(X_{\tau_n\wedge\tau_b})\le u(b)\quad(\PP_x\text{-a.s.}).
$$
Using that $\tau_n\uparrow\infty$, observe that, $\PP_x$-a.s.,
\begin{align}
\notag \lim_{n\to\infty}
\re^{-\tilde{r}\myp(\tau_n\wedge\tau_b)}\myp
u(X_{\tau_n\wedge\tau_b})&= \re^{-\tilde{r}\tau_b}\myp
u(X_{\tau_b})\myp\mathbbm{1}_{\{\tau_b<\infty\}}+ \lim_{n\to\infty}
\re^{-\tilde{r}\tau_n}\myp
u(X_{\tau_n})\myp\mathbbm{1}_{\{\tau_b=\infty\}}\\
&=\re^{-\tilde{r}\tau_b}\myp
u(b)\myp\mathbbm{1}_{\{\tau_b<\infty\}}, \label{eq:lim-n}
\end{align}
because $X_{\tau_b}=b$ on the event $\{\tau<\infty\}$, while $0\le
u(X_{\tau_n})\le u(b)$ on the event $\{\tau=\infty\}$. Hence,
letting $n\to\infty$ in \eqref{28} and using the dominated
convergence theorem (see, e.g., \cite[\S\mypp{}II.6, Theorem~3,
p.\mypp187]{Shiryaev-pr}), we get, on account of~\eqref{eq:lim-n},
\begin{align*}
u(x)&=\EE_x \bigl[\re^{-\tilde{r}\tau_b}\myp
u(b)\myp\mathbbm{1}_{\{\tau_b<\infty\}}\bigr]\\
&=\EE_x \bigl[\re^{-\tilde{r}\tau_b}\myp
g(b)\myp\mathbbm{1}_{\{\tau_b<\infty\}}\bigr]\\
&=\EE_x \bigl[\re^{-\tilde{r}\tau_b}\myp
g(X_{\tau_b})\myp\mathbbm{1}_{\{\tau_b<\infty\}}\bigr]\\
&\le v(x),
\end{align*}
according to~\eqref{eq:r2e1}. That is, we have proved that $u(x)\le
v(x)$ for all $0<x<b$, as required.
\end{itemize}

Thus, the proof of the verification theorem is complete.

\section{Elementary solution of the reduced problem}\label{sec:4}

\subsection{Distribution of the hitting time}\label{sec:4.1}

In view of the formula \eqref{r1}, the hitting problem for the
process $X_t$ is reduced to that for the Brownian motion with drift,
\begin{equation}\label{eq:tau-b-drifted}
\tau_b:=\inf\{t\ge 0\colon X_t=b\}\equiv \inf\{t\ge 0\colon
B_t+\tilde{\mu}\myp t=\tilde{b}\},
\end{equation}
where
\begin{equation}\label{eq:tilde}
\tilde{\mu}=\frac{\mu-\frac12\myp\sigma^2}{\sigma},\qquad
\tilde{b}=\frac{1}{\sigma}\ln\frac{b}{x}.
\end{equation}
Suppose that $x\le b$, so that $\tilde{b}\ge 0$. The explicit
expression for the Laplace transform of the hitting time
\eqref{eq:tau-b-drifted} is well known (see, e.g.,
\cite[Exercises 6.29 and~6.31, p.\,268]{Durrett} or
\cite[Proposition~3.3.5, p.\,61]{Etheridge}).
\begin{proposition} For $x\le b$ and any $\stheta>0$, set
\begin{equation}\label{eq:LT}
\Phi_{x,\myp b}(\stheta):=\EE_x(\re^{-\stheta\tau_b})\equiv
\EE_x\bigl(\re^{-\stheta\tau_b}\mypp\mathbbm{1}_{\{\tau_b<\infty\}}\bigr).
\end{equation} Then
\begin{equation}\label{eq:LT-tau}
\Phi_{x,\myp
b}(\stheta)=\exp\!\left\{-\tilde{b}\left(\sqrt{\tilde{\mu}^2+2\stheta}-\tilde{\mu}\right)\right\},\quad
\stheta>0,
\end{equation}
where $\tilde{\mu}$ and $\tilde{b}$ are defined in~\eqref{eq:tilde}.
\end{proposition}
Substituting the expressions \eqref{eq:tilde}, the formula
\eqref{eq:LT-tau} is rewritten as
\begin{equation}\label{eq:LT-tau-orig}
\Phi_{x,\myp
b}(\stheta)=\left(\frac{x}{b}\right)^{q_1(\stheta)},\quad \stheta>0,
\end{equation}
where $q_1(\stheta)$ is given by (cf.~\eqref{eq:q1*})
\begin{equation}\label{eq:q1-theta}
q_{1}(\stheta) =
\frac{1}{\sigma^2}\mynn\left(-\bigl(\mu-\tfrac12\myp \sigma^2\bigr)
+ \sqrt{\bigl(\mu-\tfrac12\myp \sigma^2\bigr)^2 +
2\myp\stheta\sigma^2}\mypp\right).
\end{equation}

As usual, it is straightforward to extract from the Laplace
transform \eqref{eq:LT} some explicit information about the
distribution of the hitting time $\tau_b$. First, by the monotone
convergence theorem (see, e.g., \cite[\S\mypp{}II.6, Theorem~1(a),
p.\mypp186]{Shiryaev-pr} we have
$$
\lim_{\stheta \downarrow 0} \Phi_{x,\myp b}(\stheta)
=\EE_x(\mathbbm{1}_{\{\tau_b<\infty\}}) =\PP_x(\tau_b<\infty).
$$
Hence, noting from \eqref{eq:q1-theta} that
\begin{equation}\label{eq:q(0)}
q_1(0)=\begin{cases} \myp0& \text{if} \ \,\mu-\frac12\myp\sigma^2\ge0,\\
1-\dfrac{2\mu}{\sigma^2}& \text{if} \ \,\mu-\frac12\myp\sigma^2<0,
\end{cases}
\end{equation}
we obtain
\begin{equation}\label{eq:P(tau<infty)}
\PP_x(\tau_b<\infty)=\left(\frac{x}{b}\right)^{q_1(0)}
=\begin{cases} \,1,&\mu-\tfrac12\myp\sigma^2\ge0,\\
\left(\dfrac{x}{b}\right)^{1-2\mu/\sigma^2}\!,&
\mu-\tfrac12\myp\sigma^2<0.
\end{cases}
\end{equation}

\smallskip
\begin{remark}\label{rm:<1} The result
\eqref{eq:P(tau<infty)} shows that hitting the critical threshold
$b=b^*\mynn$, as required by the stopping rule, is only certain when
the wage growth rate is large enough, $\mu\ge\frac12\myp\sigma^2$.
Thus, the ``dangerous'' case is when $\mu<\frac12\myp\sigma^2$,
whereby relying only on the optimal stopping recipe may not be
practical. This observation may serve as a germ of the idea to
connect the optimality problem in the insurance context with the
notion of \emph{utility} (cf.\ the discussion in
Section~\ref{sec:7.1} below).
\end{remark}

Via the Laplace transform $\Phi_{x,\myp b}(\stheta)$, we can also
obtain the mean hitting time $\EE_x(\tau_b)$ in the case
$\mu\ge\frac12\myp\sigma^2$, where $\tau_b<\infty$ ($\PP_x$-a.s.).
Namely, again using the monotone convergence theorem we have
$$
\lim_{\stheta\downarrow0}\frac{\partial\myp \Phi_{x,\myp
b}(\stheta)}{\partial\stheta}=-\lim_{\stheta\downarrow0}\EE_x\bigl(\tau_b\,\re^{-\stheta\tau_b}\bigr)=-\EE_x(\tau_b).
$$
Hence, differentiating formula \eqref{eq:LT-tau-orig} at $\stheta=0$
and noting from \eqref{eq:q1-theta} that $q_1(0)=0$
(cf.~\eqref{eq:q(0)}) and
$$
q_1'(0)=\begin{cases} \myp \infty,&\mu=\tfrac12\myp\sigma^2,\\[.2pc]
\dfrac{1}{\mu-\frac12\myp\sigma^2},&\mu>\tfrac12\myp\sigma^2,
\end{cases}
$$
we get
\begin{equation}\label{eq:E(tau)}
\EE_x(\tau_b) =-\ln\left(\frac{x}{b}\right)
\left(\frac{x}{b}\right)^{q_1(0)}\!q_1'(0)=\begin{cases}
\, \infty,&\mu=\tfrac12\myp\sigma^2,\\[.2pc]
\dfrac{\ln(b/x)}{\mu-\frac12\myp\sigma^2},&\mu>\tfrac12\myp\sigma^2.
\end{cases}
\end{equation}

\subsection{Alternative derivation}\label{sec:4.2}
An alternative (and more direct) method to derive the formulas
\eqref{eq:P(tau<infty)} and~\eqref{eq:E(tau)} is based on general
theory of Markov processes by solving the suitable boundary value
problems (see, e.g.,
\cite[\S\mypp9]{Oksendal}). Namely, the hitting probability
$\pi(x):=\PP_x(\tau_b<\infty)$ as a function of $x>0$ satisfies the
Dirichlet problem \cite[\S\mypp9.2]{Oksendal}
\begin{equation}\label{eq:L=0}
\begin{cases}
\begin{aligned}
\!L\pi(x)&=0&(0<x<b),\\
\pi(b)&=1. \end{aligned}
\end{cases}
\end{equation}
The differential equation in \eqref{eq:L=0} reads
$$
\tfrac12\myp\sigma^2 x^2\pi''(x)+\mu\myp x\myp \pi'(x)=0\quad
(0<x<b),
$$
which is easily solved to give
$$
\pi(x)=c_1\myp x^{1-2\mu/\sigma^2}\!+c_2.
$$
If $1-2\mu/\sigma^2<0$ (i.e., $\mu-\frac12\myp\sigma^2>0$) then
$c_1=0$ (since $\pi(x)$ is bounded), and due to the boundary
condition $\pi(b)=1$ it follows that $c_2=1$ and $\pi(x)\equiv 1$. A
similar argument shows that $\pi(x)\equiv 1$ in the case
$1-2\mu/\sigma^2=0$. Finally, if $1-2\mu/\sigma^2>0$ then, noting
that $\pi(0)=0$, we conclude that $c_2=0$ and, due to the boundary
condition, $c_1=b^{-1+2\mu/\sigma^2}\mynn$. Thus, formula
\eqref{eq:P(tau<infty)} is proved.

\smallskip
Similarly, the mean hitting time $m(x):=\EE_x(\tau_b)$ (with
$\mu-\frac12\myp\sigma^2>0$) satisfies the Poisson problem
\cite[\S\mypp9.3]{Oksendal}
\begin{equation}\label{eq:L=-1}
\!\begin{cases}
\begin{aligned}
Lm(x)&=-1&\quad(0<x<b),\\
m(b)&=0. \end{aligned}\end{cases}
\end{equation}
As usual, to solve the problem~\eqref{eq:L=-1}, it is convenient to
approximate it with a two-sided boundary problem by adding an
auxiliary Neumann (reflection) condition at $\varepsilon>0$,
\begin{equation}\label{eq:L=-1-eps}
\begin{cases}
\begin{aligned}
Lm_\varepsilon(x)&=-1&\quad(\varepsilon<x<b),\\
m_\varepsilon(b)&=0,\\
m'_\varepsilon(\varepsilon)&=0, \end{aligned}\end{cases}
\end{equation}
and then taking the limit of $m_\varepsilon(x)$ as
$\varepsilon\downarrow0$. This procedure will produce the correct
solution $m(x)$ since
$\lim_{\varepsilon\downarrow0}\PP_x(\tau_\varepsilon<\infty)=\PP_x(\tau_0<\infty)=0$
(for any $x>0$).

A particular solution to the inhomogeneous differential equation
$$
\tfrac12\myp\sigma^2 x^2 m_\varepsilon''(x)+\mu\myp x\myp
m_\varepsilon'(x)=-1\quad (\varepsilon<x<b)
$$
can be sought in the form $m_0(x)=c_0\ln x$, which gives
$c_0=-1/(\mu-\frac12\myp\sigma^2)$. Thus, the general solution
of~\eqref{eq:L=-1-eps} can be expressed as
\begin{equation}\label{eq:m-eps}
m_\varepsilon(x)=-\frac{\ln x}{\mu-\frac12\myp\sigma^2}+c_1\myp
x^{1-2\mu/\sigma^2}+c_2.
\end{equation}
Now, using the boundary conditions in \eqref{eq:L=-1-eps} it is
straightforward to check that
$$
\lim_{\varepsilon\downarrow0} c_1=0,\qquad
\lim_{\varepsilon\downarrow0} c_2=\frac{\ln
b}{\mu-\frac12\myp\sigma^2}.
$$
Hence, from \eqref{eq:m-eps} we get
$$
m(x)=\lim_{\varepsilon\downarrow0} m_{\varepsilon}(x)=\frac{\ln
\left(b/x\right)}{\mu-\frac12\myp\sigma^2},
$$
which retrieves the result~\eqref{eq:E(tau)}.

\begin{remark}
The same method applied to the killed process $\widetilde{X}_t$ with
generator $\tilde{L}=L-\tilde{r} I$ (see~\eqref{eq:tilde-L})
provides a neat interpretation of the value function $u(x)$ as given
by~\eqref{a3}. Namely, rewrite the expectation in
\eqref{eq:r2e**red} (i.e., $\eNPV(x;\tau_b)$) in the form
$\tilde{\EE}_x \bigl[g(\widetilde{X}_{\tau_b})\bigr]$, where
$\tilde{\EE}_x$ denotes expectation with respect to the killed
process $(\widetilde{X}_t)$, and note that, for $b\ge0$,
$$
\tilde{\EE}_x \bigl[g(\widetilde{X}_{\tau_b})\bigr]=
\begin{cases}
g(b)\,\tilde{\PP}_x(\tau_b<\infty),\ \ &x\in[\myp0,b\myp],\\
g(x),& x\in[\myp b,\infty).
\end{cases}
$$
In turn, the hitting probability
$\tilde{\pi}(x):=\tilde{\PP}_x(\tau_b<\infty)$ can be easily found
by solving the corresponding Dirichlet problem (cf.~\eqref{eq:L=0}),
\begin{equation*}
\begin{cases}
\!\begin{aligned}
\tilde{L}\tilde{\pi}(x)&=0&(0<x<b),\\
\tilde{\pi}(b)&=1.
\end{aligned}
\end{cases}
\end{equation*}
Indeed, repeating the calculations in Section~\ref{sec:3.2}, it is
straightforward to get $\tilde{\pi}(x)=(x/b)^{q^*}$.
\end{remark}

\subsection{Direct maximization}\label{sec:4.3}

Using the results of the previous sections, we can easily solve the
optimal stopping problem \eqref{eq:r2e1}, at least in the subclass
of hitting times $\tau=\tau_b$ (see~\eqref{eq:r2e**red}),
\begin{equation}\label{eq:r2e**red-again}
u(x)=\sup_{b\ge0}\eNPV(x;\tau_b)=\sup_{b\ge 0} \EE_x \bigl[
\re^{-\tilde{r}\tau_b}(\beta_1 X_{\tau_b}-P)\bigr].
\end{equation}

Observe that if $x\ge b$ then $\tau_{b}=0$ and $X_{\tau_b}=x$
($\PP_x$-a.s.), so that $\eNPV(x;\tau_b)\equiv \beta_1x-P$ for all
$b\in[\myp0,x]$. Let now $b\in[x,\infty)$. As already noted, on the
event $\{\tau_b<\infty\}$ we have $X_{\tau_b}=b$ ($\PP_x$-a.s.),
hence, according to \eqref{eq:r2e1} and~\eqref{eq:LT-tau-orig},
\begin{equation}\label{eq:NPV-explicit}
\eNPV(x;\tau_b)=(\beta_1 b-P)\,\EE_x\bigl(\re^{-\tilde{r}
\tau_b}\bigr) =(\beta_1 b-P)\left(\frac{x}{b }\right)^{q_*}\quad
(b\ge x),
\end{equation}
where $q_*=q_1(\stheta)|_{\stheta=\tilde{r}}$ (cf.\ \eqref{eq:q1*}
and~\eqref{eq:q1-theta}). It is straightforward to find the
maximizer for the function~\eqref{eq:NPV-explicit}. Indeed, the
condition $(\partial/\partial\myp b)\mypp \eNPV(x;\tau_b)\ge0$,
equivalent to
$$
\beta_1\myp b^{-q_*}\!-q_*(\beta_1 b-P)\mypp b^{-q_*-1}\ge0,
$$
holds for all $b\in[\myp0,b^*]$, where
\begin{equation}\label{eq:b*new}
b^*=\frac{Pq_*}{\beta_1(q_*-1)},
\end{equation}
which is the same optimal threshold as before (cf.~\eqref{eq:b*}).
Thus, the supremum of $\eNPV(x;\tau_b)$ over $b\ge x$ is attained at
$b=b^*\mynn$ if $x\le b^*\mynn$ or
at $b=x$ if $x\ge b^*\myn$.

The corresponding value function $u(x)$ is then calculated as
(cf.~\eqref{eq:value-b*})
\begin{equation}\label{eq:v-new}
u(x)=
\begin{cases} \displaystyle (\beta_1
b^*-P)\left(\frac{x}{b^*}\right)^{q_*}\!,&x\in[\myp0,b^*],\\
\displaystyle \beta_1x-P,&x\in[\myp b^*,\infty) .\end{cases}
\end{equation}
Finally, substituting \eqref{eq:b*new} into \eqref{eq:v-new}, we
obtain explicitly (cf.~\eqref{eq:value-b*-exp})
\begin{equation}\label{eq:v-new1}
u(x)=\begin{cases} \displaystyle\frac{P}{q_*-1}\left(\frac{\beta_1
(q_*-1)\myp x}{Pq_*}\right)^{q_*}\!,&\displaystyle 0\le x\le
\frac{P q_*}{\beta_1(q_*-1)},\\[.8pc]
\displaystyle \beta_1x-P,&\displaystyle \hphantom{0\le{}}x\ge
\frac{Pq_*}{\beta_1(q_*-1)}.
\end{cases}
\end{equation}

\section{Statistical issues and numerical illustration}\label{sec:5}

\subsection{Specifying the model parameters}\label{sec:5.1}
From the practical point of view, in order to exercise the stopping
rule \eqref{eq:stopping_rule} the individual concerned needs to be
able to compute the critical threshold $b^*\mynn$ expressed in
\eqref{eq:b*}, for which the knowledge is required about $\beta_1$
(defined in~\eqref{eq:c1-tilde}) and therefore about the parameters
$r$, $\lambda_0$, $\mu$ and $\beta$ (see~\eqref{eq:c1});
furthermore, to evaluate the quantity $q_*$ defined in
\eqref{eq:q1*}, one needs to estimate $\mu-\frac12\myp\sigma^2$ and
$\sigma^2$ itself. Specifically:
\begin{itemize}
\item The loss-of-job rate $\lambda_0$ can be extracted from the publicly available
data about the mean length at work, which is theoretically given by
$\EE(\tau_0)=1/\lambda_0$.

\item Likewise, the inflation rate $r$ is also in the public domain.

\item
To specify the wage growth rate $\mu$, a simple approach is just to
set $\mu=r$ as a crude version of a ``tracking'' rule. However, it
may be possible that the individual's wage growth rate $\mu$ is, to
some extent, stipulated by the job contract\,---\,for example, that
it must not exceed the inflation rate $r$ by more than 1\% per annum
(applicable, e.g., to civil servants) or, by contrast, that it must
be no less than $r$ minus 0.5\% per annum (more realistic in the
private sector). In practical terms, this would often mean that the
actual growth rate $\mu$ is kept on the lowest predefined level.

\item
More generally, the wage growth rate $\mu$ can be estimated by
observing the wage process $X_t$. This can be implemented by first
using regression analysis on $Y_t=\ln X_t$ and estimating the
regression line slope $\mu-\frac12\myp\sigma^2$ (see~\eqref{r1}). In
addition, the volatility $\sigma^2$ can be estimated by using a
suitable quadratic functional of the sample paths $Y_t$.

\item Finally, knowing the benefit schedule (which should be
available through the insurance policy's terms and conditions), it
is in principle possible to calculate, or at least estimate the
value~$\beta$.
\end{itemize}
To summarize, certain
estimation procedures need to be carried out along with the on-line
observation of the sample path $(X_t)$. More details (most of which
are quite standard) are provided in the next two subsections.

\subsection{Estimating the drift and volatility}\label{sec:5.2}
Denote for short $a:=\mu - \tfrac{1}{2}\myp \sigma^2$. According to
the geometric Brownian motion model \eqref{r1}, we have
\begin{equation*}
Y_t:=\ln X_t = \ln x + \sigma B_t + a\myp t,\qquad Y_0 = \ln x.
\end{equation*}
Suppose the process $X_t $ is observed over the time interval
$t\in[\myp0,T\myp]$ on a discrete-time grid $t_i=iT/n$
($i=0,\dots,n$), and consider the consecutive increments
\begin{equation}\label{r6}
Z_i:= Y_{t_i} - Y_{t_{i-1}} = \sigma (B_{t_i} - B_{t_{i-1}}) +
a\mypp (t_i-t_{i-1})\quad (i=1,\dots,n).
\end{equation}
Note that the increments of the Brownian motion in \eqref{r6} are
mutually independent and have normal distribution with zero mean and
variance $t_i - t_{i-1}=T/n$, respectively. Therefore, $(Z_i)$ is an
independent random sample with normal marginal distributions,
$$
Z_i \sim \mathcal{N}\!\left(\frac{a\myp T}{n}, \frac{\sigma^2\myp
T}{n} \right) \quad (i=1,\dots,n).
$$
Then, it is standard to estimate the parameters via the sample mean
and sample variance,
\begin{align}
\label{eq:a-hat} \hat{a}_n &:= \frac{n}{T}\cdot \bar{Z} = \frac{Z_1
+
\cdots + Z_n}{T}=\frac{Y_T - Y_0}{T},\\
\label{eq:sigma-hat} \hat{\sigma}^2_n &:=
\frac{n}{T}\cdot\frac{1}{n-1} \sum_{i=1}^{n} (Z_i - \bar{Z})^2.
\end{align}
These estimators are unbiased,
$$
\EE(\hat{a}_n)=a=\mu-\tfrac12\myp\sigma^2,\qquad
\EE(\hat{\sigma_n}^2)=\sigma^2,
$$
with mean square errors
$$
\Var(\hat{a}_n)=\frac{\sigma^2}{T},\qquad
\Var(\hat{\sigma}^2_n)=\frac{2\myp\sigma^4}{n-1}.
$$
In turn, the parameter $\mu$ is estimated by
$$
\hat{\mu}_n=\hat{a}_n+\tfrac12\myp\hat{\sigma}_n^2,
$$
with mean
$\EE(\hat{\mu}_n)=\EE(\hat{a}_n)+\tfrac12\mypp\EE(\hat{\sigma}_n^2)
=a+\tfrac12\myp\sigma^2= \mu$ and mean square error
$$
\Var(\hat{\mu}_n)=\Var(\hat{a}_n)+\tfrac14\mypp\Var(\hat{\sigma}_n^2)=\frac{\sigma^2}{T}+\frac{\sigma^4}{2\myp(n-1)}
$$
(due to independence of the estimators $\hat{a}_n$ and
$\hat{\sigma}_n^2$).

Note that the estimator $\hat{a}_n$ in \eqref{eq:a-hat} only employs
the last observed value, $Y_T$; in particular, its mean square error
is not sensitive to the grid size $\Delta t_i=T/n$, and only tends
to zero with increasing observational horizon, $T\to\infty$. This
makes the estimation of the drift parameter $a$ difficult in the
sense that very long observations over $Y_t$ are required to achieve
an acceptable precision (see, e.g., \cite[Example~2.1,
p.\,3]{Ekstrom}). For instance, let $\mu=0.004$ and $\sigma=0.02$
(per week), then $a=0.0038$; if $T=25$ (weeks) then the
95\%-confidence bounds for $a$ are given by $\hat{a}\pm
1.96\,\sigma/\sqrt{T}=\hat{a}\pm 0.00784$, so the margin of error is
about twice as big as the value of $a$ itself. To reduce it, say to
$0.5\myp a$, one needs $T\approx 425$ (weeks), which exemplifies
slow convergence.

In contrast, the mean square error of the estimator
$\hat{\sigma}_n^2$ in \eqref{eq:sigma-hat} tends to zero as
$n\to\infty$, with $T$ fixed. Thus, estimation of the parameter
$\sigma^2$ can be made asymptotically precise.

A numerical example illustrating the estimation of $\mu$ and
$\sigma^2$ using simulated data will be given at the end of
Section~\ref{sec:5.4}. A brief discussion of practical choices of
$\mu$, based on sensitivity analysis, is provided at the end of
Section~\ref{sec:6.3}.

\subsection{Hypothesis testing}\label{sec:5.3}

In view of the drawback in the general solution of the optimal
stopping problem in that the stopping time $\tau_{b^*}$ may be
infinite, that is, $\PP_x(\tau_{b^*}=\infty)>0$ (which occurs when
$a=\mu-\frac12\myp\sigma^2<0$, see Section~\ref{sec:4.1}), a
reasonable pragmatic approach to decision making in our model may be
based on testing the null hypothesis $H_0\colon a\ge 0$ versus the
alternative $H_1\colon a<0$ (at some intuitively acceptable
significance level, e.g.\ $\alpha=0.05$). Namely, as long as $H_0$
remains tenable, one keeps waiting for the hitting time $\tau_{b^*}$
to occur, but once $H_0$ has been rejected, it is reasonable to
terminate waiting and buy the policy immediately.

The corresponding test is specified as follows. Again, suppose that
the process $Y_t$ is observed on a discrete time grid $t_i=i\myp
T/n$, and set $Z_i=Y_{t_i}-Y_{t_{i-1}}$ ($i=1,\dots,n$). Let
$z(\alpha)$ be the upper $\alpha$-quantile of the \strut{}standard
normal distribution $\mathcal{N}(0,1)$, that is,
$1-\Phi(z(\alpha))=\alpha$, where
$\Phi(x)=\frac{1}{\sqrt{2\pi}}\int_{-\infty}^x
\re^{-u^2/2}\,\rd{u}$. Then the null hypothesis $H_0\colon a\ge 0$
is to be rejected at significance level $\alpha$ \strut{}whenever
\begin{equation*}
Z_1+\dots+Z_n\le
\inf_{a\ge0}\left\{aT-z(\alpha)\mypp\sigma\sqrt{T}\right\},
\end{equation*}
that is,
\begin{equation}\label{eq:normal-test}
Y_T-Y_0\le -z(\alpha)\mypp\sigma\sqrt{T}.
\end{equation}
This test is uniformly most powerful among all tests with
probability of error of type I not exceeding $\alpha$, that is,
$\PP(\text{reject $H_0$}\,|\,\text{$H_0$ true})\le\alpha$.

The normal test \eqref{eq:normal-test} assumes that the variance
$\sigma^2$ is known. As mentioned before, this presents no real
restriction if the process $Y_t$ is observable continuously (i.e.,
if the grid $(t_i)$ can be refined indefinitely). If this is not the
case (e.g., because the wage process can only be observed on the
weekly basis) then the test \eqref{eq:normal-test} is replaced by
the $t$-test,
\begin{equation*}
Y_T-Y_0\le -t_{n-1}(\alpha)\mypp \hat{\sigma}\mypp\sqrt{T},
\end{equation*}
where $\hat{\sigma}^2$ is the sample variance
(see~\eqref{eq:sigma-hat}) and $t_{n-1}(\alpha)$ is the upper
$\alpha$-quantile of the $t$-distribution with $n-1$ degrees of
freedom.

In practice, the hypothesis testing is carried out sequentially
(e.g., weekly) as the observational horizon $T$ increases. The
advantage of this approach is that the resulting stopping time is
finite with probability one (i.e., $\PP_x$-a.s.); indeed, it is the
minimum between the optimal stopping time $\tau_{b^*}$ (which is
finite $\PP_x$-a.s.\ under the null hypothesis $H_0\colon a\ge0$)
and the first time of rejecting $H_0$ (which is finite $\PP_x$-a.s.\
if $H_0$ is false).

\subsection{Numerical examples}\label{sec:5.4}

To be specific, we use euro as the monetary unit. First of all, the
value of the constant $\beta$, which encapsulates information about
the benefit schedule as well as the rate $\lambda_1$ of finding new
job (see~\eqref{eq:c1}), is chosen to be
\begin{equation*}
\beta=30.
\end{equation*}
Thus, the overall expected benefit payable over the lifetime of the
policy (and projected to the beginning of unemployment) is taken to
be equal to 30 weekly wages; that is, if the final wage is 400 (euro
per week) then the total to be received is
$$
400.00\times 30 = 12\,000.00\ \text{(euro)}.
$$
Further, we set
\begin{equation*}
\lambda_0=0.01,\qquad r=0.0004.
\end{equation*}
This means that the expected time until loss of job is
$1/\lambda_0=100$ (weeks), that is, about 1 year and 11 months,
whereas the annual inflation rate is
$$
\re^{(365/7)\cdot 0.0004}-1=
0.02107617 \approx 2.11\%,
$$
which is quite realistic.

Next, we need to specify the premium $P$ and the parameters of the
wage process $X_t$, First, choose the initial value $x=X_0$ as
\begin{equation*}
x=346.00\ \text{(euro)}.
\end{equation*}
This is motivated by the French labour legislation, whereby the
current minimum pay rate is set as 9.88 euro per hour
\cite{minwage}, with a 35-hour workweek \cite{the35w,35week}, giving
$$
9.88 \times 35 = 345.80\ \text{(euro per week)}.
$$
As for the premium, it is set at the value
\begin{equation*}
P=9\,000.00\ \text{(euro)},
\end{equation*}
which equates to about 26 minimum weekly wages (i.e., income over
about half a year). For simplicity, we also choose
\begin{equation}\label{eq:mu=r}
\mu=r=0.0004,
\end{equation}
so that the wage growth rate is the same as inflation $r$ (in
reality, it could be slightly less). Then from \eqref{eq:c1-tilde},
using \eqref{eq:mu=r}, we get
$$
\beta_1=\frac{\lambda_0\myp \beta}{\tilde{r}-\mu}=\beta=30.
$$
For the volatility $\sigma$, we will illustrate two opposite cases,
$\mu<\tfrac12\myp\sigma^2$ and $\mu>\tfrac12\myp\sigma^2$.
\begin{example}\label{ex:5.1}
Set $\sigma=0.04$, then $\mu-\tfrac12\myp\sigma^2=-0.0004<0$. From
\eqref{eq:q1*} we calculate $q_*=3.864208$, then \eqref{e11} yields
$$
b^*=404.7410=404.74\ \text{(euro)}.
$$
Using~\eqref{eq:P(tau<infty)}, the hitting probability is calculated
as
$$
\PP_x(\tau_{b^*}<\infty)= 0.9245906.
$$
Finally, using \eqref{a3}, we obtain the value of this contract,
$$
v(346)=1\mypp714.2780
 = 1\mypp714.28 \ \text{(euro)}.
$$
\end{example}

\begin{example}\label{ex:5.2}
Now, set $\sigma=0.02$, then $\mu-\tfrac12\myp\sigma^2=0.0002>0$.
Furthermore, using \eqref{eq:q1*} we calculate $q_*=6.728416$, and
from~\eqref{e11}
$$
b^*=352.3705= 352.37\ \text{(euro)}.
$$
Hence, using \eqref{eq:E(tau)}, the expected hitting time is found
to be
$$
\EE(\tau_{b^*})=91.22197=91.2\ \text{(weeks)}.
$$
Finally, according to formula \eqref{eq:value-b*}, the value of this
contract is calculated as
$$
v(346)= 1\mypp389.6190=1\mypp389.62\ \text{(euro)}.
$$

In the simulation of the process $X_t$ shown in Fig.~\ref{fig3}, the
drift $a=\mu-\frac12\myp\sigma^2$ is estimated using
formula~\eqref{eq:a-hat} as $\hat{a}\doteq0.0005994$. Estimation of
the variance $\sigma^2$ according to formula \eqref{eq:sigma-hat}
(on a weekly time grid) gives $\hat{\sigma}^2\doteq0.0003723$, while
the true value is $\sigma^2=0.0004$. Hence, the parameter $\mu$ is
estimated by $\hat{\mu}\doteq0.0007855$; recall that the true value
is $\mu=0.0004$.
\end{example}

\section{Parametric dependencies}\label{sec:6}

In this section, we aim to explore the parametric dependencies of
the solution of our insurance problem, that is, of the optimal
threshold $b^*\mynn$ given by~\eqref{eq:b*} and the value function
$v=v(x)$ given  by~\eqref{eq:value-b*}. In particular, it is helpful
to analyse different asymptotic regimes as well as (the sign of)
appropriate partial derivatives, so as to ascertain the direction of
changes under small perturbations and to understand their economic
meaning. This is a key ingredient of sensitivity analysis and of the
so-called \emph{comparative statics} \cite[Section~VII]{Merton-CS}.

In what follows, we confine ourselves to a discussion of the two
most important exogenous parameters\,---\,the wage drift $\mu$ and
the unemployment rate~$\lambda_0$. The constraint \eqref{eq:upper}
implies that the range of the parameters $\mu$ and $\lambda_0$ is
specified as follows,
$$
-\infty< \mu <\tilde{r}=r+\lambda_0,\qquad 0\vee
(\mu-r)<\lambda_0<\infty.
$$

\begin{remark}
     The next two technical subsections are elementary but rather
     tedious, and the reader wishing to grasp the results quickly may
     just inspect the plots in Figs.\ \textup{\ref{fig:5}} and~\textup{\ref{fig:5e}}.
\end{remark}

\subsection{Monotonicity}\label{sec:6.1}

\begin{figure}[bt!]
    \centering
    \subfigure[$\mu \mapsto b^*$]{\includegraphics[%
       height=0.30\textheight]{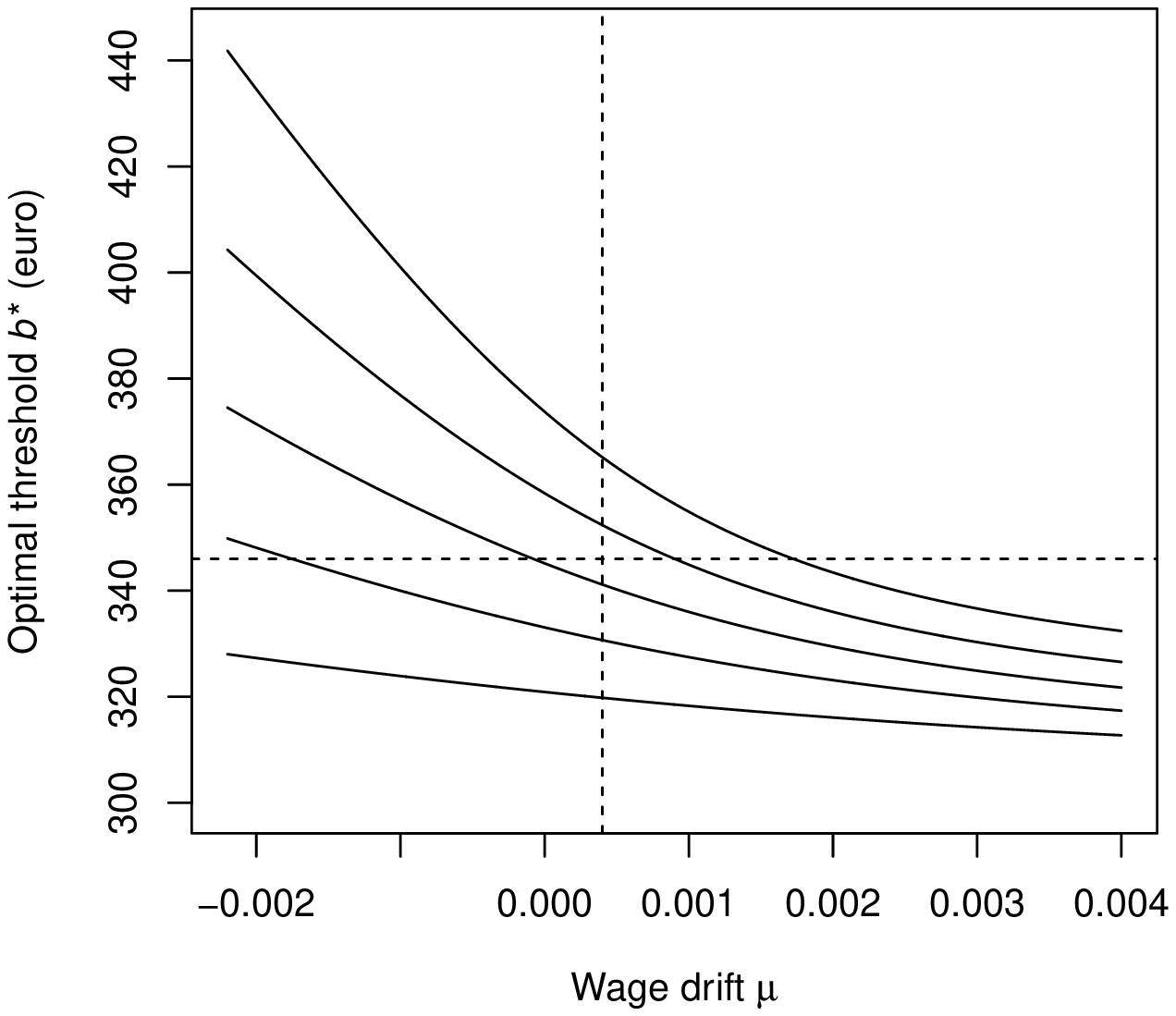}
       \put(-110.5,33.8){\mbox{\scriptsize $r$}}
       \put(-183.0,83.5){\mbox{\scriptsize $x$}}
       \put(-165.2,164.5){\mbox{\myp\tiny {\sc I}}}
       \put(-165.2,133.0){\mbox{\tiny {\sc II}}}
       \put(-165.2,108.4){\mbox{\tiny {\sc III}}}
       \put(-165.4,88.0){\mbox{\tiny {\sc IV}}}
       \put(-165.2,70.0){\mbox{\tiny {\sc V}}}
            \put(-90.7,160.2){\mbox{\scriptsize {\sc \hphantom{II}I\mypp:} \,$\lambda_0=0.007$}}
            \put(-90.7,152.2){\mbox{\scriptsize {\sc \hphantom{I}II\mypp:} \,$\lambda_0=0.01$}}
            \put(-90.7,144.2){\mbox{\scriptsize {\sc III\mypp:} \,$\lambda_0=0.015$}}
            \put(-90.7,136.2){\mbox{\scriptsize {\sc IV\myp:} \,$\lambda_0=0.025$}}
            \put(-90.7,128.2){\mbox{\scriptsize {\sc \hphantom{I}V\myp:} \,$\lambda_0=0.055$}}
}
    \hspace{0.0pc}
    \subfigure[$\lambda_0 \mapsto b^*$]{\includegraphics[%
        height=0.30\textheight]{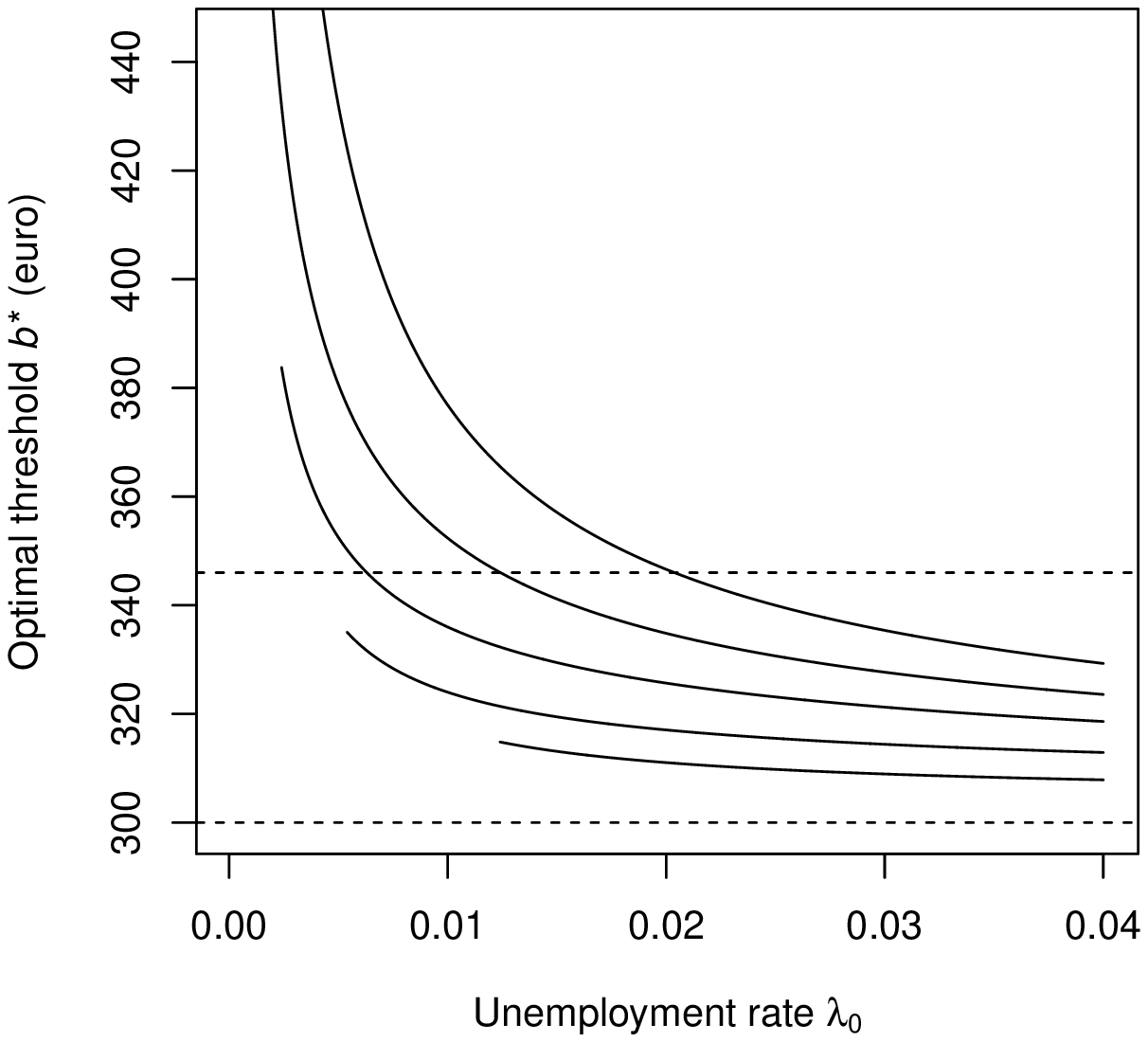}
    \put(-175.5,83.5){\mbox{\scriptsize $x$}}
       \put(-14.4,42.5){\mbox{\scriptsize $P/\beta$}}
       \put(-115.0,49.5){\mbox{\mypp\tiny {\sc V}}}
       \put(-142.0,62.0){\mbox{\myp\tiny {\sc IV}}}
       \put(-161.0,100.0){\mbox{\tiny {\sc III}}}
       \put(-160.0,140.0){\mbox{\tiny {\sc II}}}
       \put(-145.0,164.0){\mbox{\myp\tiny {\sc I}}}
            \put(-104.7,160.2){\mbox{\scriptsize {\sc \hphantom{II}I\mypp:} \,$\mu=-0.001$}}
            \put(-104.7,152.2){\mbox{\scriptsize {\sc \hphantom{I}II\mypp:} \,$\mu=0.0004\ (=r)$}}
            \put(-104.7,144.2){\mbox{\scriptsize {\sc III\mypp:} \,$\mu=0.002$}}
            \put(-104.7,136.2){\mbox{\scriptsize {\sc IV\myp:} \,$\mu=0.005$}}
            \put(-104.7,128.2){\mbox{\scriptsize {\sc \hphantom{I}V\myp:} \,$\mu=0.012$}}
}\\[-.5pc]
    \caption{Graphs illustrating parametric dependencies of the optimal threshold~\eqref{eq:b*}: (a) on the
    wage drift~$\mu<\tilde{r}$ and (b) on the unemployment rate
    $\lambda_0>0\vee\mynn (\mu-r)$, for selected values of $\lambda_0$ and $\mu$,
    respectively.
    The values of other model parameters used throughout are as in Example~\ref{ex:5.2}:
    $r=0.0004$, $P=9\,000$, $\beta=30$, and $\sigma
    =0.02$. The dashed horizontal line in both plots indicates the initial wage~$x=346$.
    The dashed vertical line in (a) indicates $\mu=r$.
    The lower dashed horizontal line in (b) shows the asymptote $P/\beta=300$ (see~\eqref{eq:b*|lambda1}).}
    \label{fig:5}
\end{figure}

By virtue of the quadratic equation \eqref{e6}, the formula
\eqref{eq:b*} can be conveniently rewritten as
\begin{equation}\label{eq:b*1}
b^*=\frac{P\myp(\frac12\myp\sigma^2q_*\myn+\tilde{r})}{\beta\myp\lambda_0}.
\end{equation}
First, fix $\lambda_0$ and consider the function $\mu\mapsto
b^*\myn$. Differentiating the equation \eqref{e6} and then again
using \eqref{e6} to eliminate $\mu$, we obtain
\begin{equation}\label{eq:q'mu}
\frac{\partial q_*}{\partial \mu}=
-\frac{q_*}{\frac12\myp\sigma^2(2q_*-1)+\mu}=-\frac{q_*^2}{\frac12\myp\sigma^2\myp
q_*^2+\tilde{r}}<0.
\end{equation}
Hence, using \eqref{eq:b*1} and \eqref{eq:q'mu},
\begin{equation}\label{eq:b'mu}
\frac{\rd b^*}{\rd\mu}=\frac{\partial b^*}{\partial
\mu}+\frac{\partial b^*}{\partial q_*}\cdot \frac{\partial
q_*}{\partial\mu}
=-\frac{P\mypp(\frac12\myp\sigma^2q_*^2)}{\beta\myp\lambda_0\mypp(\frac12\myp\sigma^2q_*^2+\tilde{r})}<0,
\end{equation}
and, therefore, $b^*\mynn$ is a decreasing function of $\mu$ (see
Fig.~\ref{fig:5}(a)).

Similarly, the equation \eqref{e6} yields
\begin{equation}\label{eq:q'lambda}
\frac{\partial q_*}{\partial\lambda_0}=
\frac{1}{\frac12\myp\sigma^2(2q_*-1)+\mu}=\frac{q_*}{\frac12\myp\sigma^2\myp
q_*^2+r+\lambda_0}>0.
\end{equation}
From \eqref{eq:b*1} and \eqref{eq:q'lambda}, after some
rearrangements we obtain
\begin{align}
\notag \frac{\rd b^*}{\rd \lambda_0}&=\frac{\partial
b^*}{\partial\lambda_0}+\frac{\partial
b^*}{\partial q_*}\cdot \frac{\partial q_*}{\partial\lambda_0}\\
\notag
&=-\frac{P\myp(\frac12\myp\sigma^2q_*\myn+r)}{\beta\myp\lambda^2_0}+
\frac{P\mypp(\frac12\myp\sigma^2q_*)}{\beta\myp\lambda_0\mypp(\frac12\myp\sigma^2q_*^2+r+\lambda_0)}\\
&=-\frac{P\left[(\frac12\myp\sigma^2q_*\myn+r)(\frac12\myp\sigma^2q_*^2+r)+\lambda_0\myp
r\right]}{\beta\myp\lambda_0^2\mypp(\frac12\myp\sigma^2q_*^2+r+\lambda_0)}<0,
\label{eq:b'lambda}
\end{align}
and it follows that the function $\lambda_0\mapsto b^*\mynn$ is
decreasing (see Fig.~\ref{fig:5}(b)).

\begin{figure}[htb]
    \centering
    \subfigure[$\mu\mapsto v(x)$]{\includegraphics[%
       height=0.30\textheight]{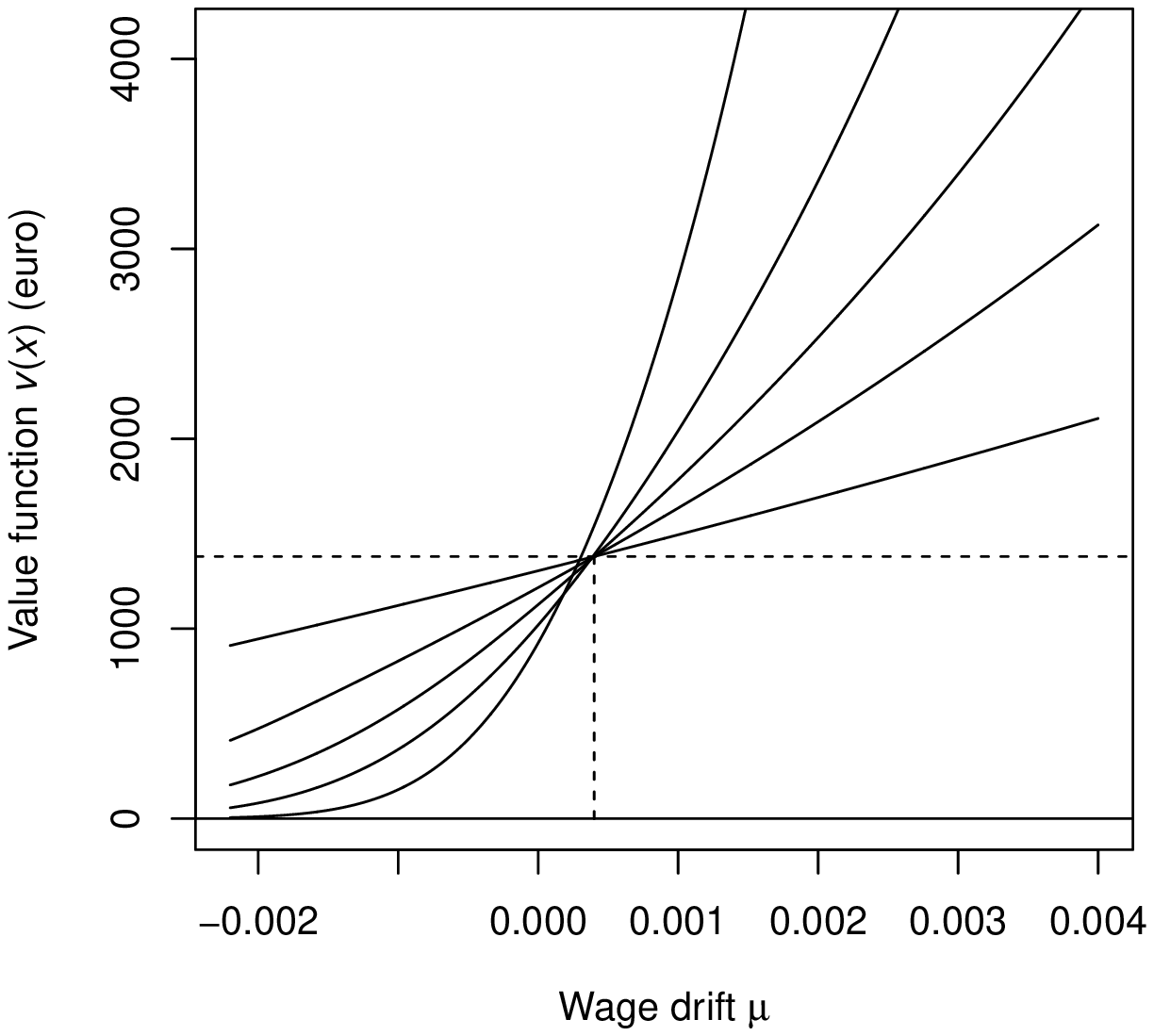}}
       \put(-178.5,85.5){\mbox{\scriptsize $v_{*}$}}
       \put(-102.4,46.0){\mbox{\scriptsize $r$}}
       \put(-87.5,164.5){\mbox{\myp\tiny {\sc I}}}
       \put(-67.5,161.0){\mbox{\tiny {\sc II}}}
       \put(-53.0,150.0){\mbox{\tiny {\sc III}}}
       \put(-40.5,133.0){\mbox{\tiny {\sc IV}}}
       \put(-32.5,109.0){\mbox{\tiny {\sc V}}}
            \put(-158.7,153.2){\mbox{\scriptsize {\sc \hphantom{II}I\mypp:} \,$\lambda_0=0.005$}}
            \put(-158.7,145.2){\mbox{\scriptsize {\sc \hphantom{I}II\mypp:} \,$\lambda_0=0.01$}}
            \put(-158.7,137.2){\mbox{\scriptsize {\sc III\mypp:} \,$\lambda_0=0.016$}}
            \put(-158.7,129.2){\mbox{\scriptsize {\sc IV\myp:} \,$\lambda_0=0.025$}}
            \put(-158.7,121.2){\mbox{\scriptsize {\sc \hphantom{I}V\myp:} \,$\lambda_0=0.055$}}
        \hspace{1.0pc}
    \subfigure[$\lambda_0 \mapsto v(x)$]{\includegraphics[%
        height=0.30\textheight]{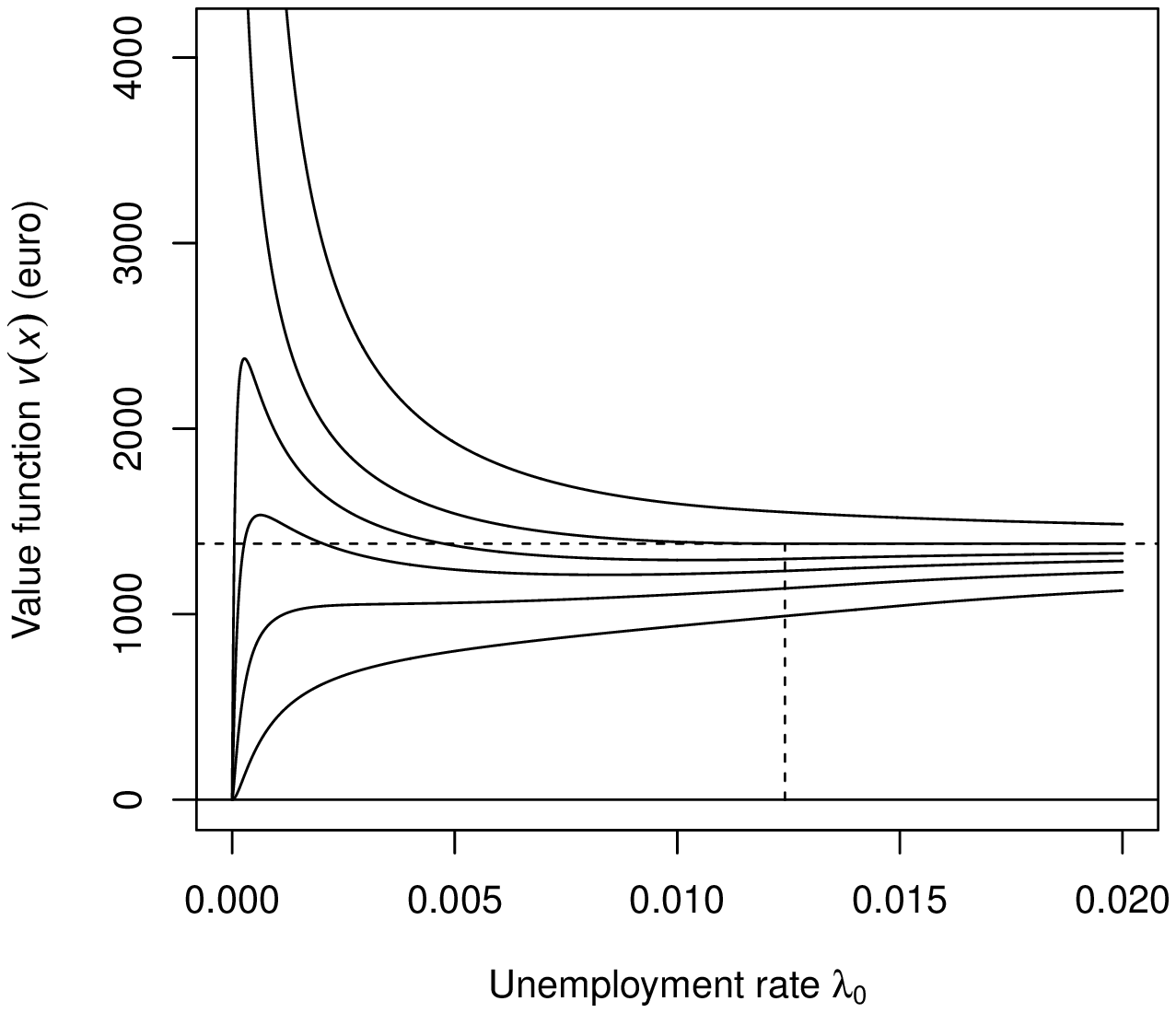}}
       \put(-186.0,85.5){\mbox{\scriptsize $v_{*}$}}
       \put(-77.2,46.0){\mbox{\scriptsize $\lambda_{*}$}}
       \put(-158.8,164.0){\mbox{\myp\tiny {\sc I}}}
       \put(-172.5,142.5){\mbox{\tiny {\sc II}}}
       \put(-172.0,119.8){\mbox{\tiny {\sc III}}}
       \put(-168.5,93.5){\mbox{\tiny {\sc IV}}}
       \put(-164.0,77.0){\mbox{\tiny {\sc V}}}
       \put(-160.0,55.5){\mbox{\tiny {\sc VI}}}
            \put(-101.7,158.2){\mbox{\scriptsize {\sc \hphantom{II}I\mypp:} \,$\mu=0.0006$}}
            \put(-101.7,150.2){\mbox{\scriptsize {\sc \hphantom{I}II\mypp:} \,$\mu=0.0004\ (=r)$}}
            \put(-101.7,142.2){\mbox{\scriptsize {\sc III\mypp:} \,$\mu=0.0003$}}
            \put(-101.7,134.2){\mbox{\scriptsize {\sc IV\myp:} \,$\mu=0.00022$}}
            \put(-101.7,126.2){\mbox{\scriptsize {\sc \hphantom{I}V\myp:} \,$\mu=0.0001$}}
            \put(-101.7,118.2){\mbox{\scriptsize {\sc VI\myp:} \,$\mu=-0.0001$}}
\\[-.5pc]
    \caption{Graphs illustrating parametric dependencies of the value function~\eqref{eq:value-b*}: (a) on the
    wage drift $\mu<\tilde{r}$ and (b) on the unemployment rate
    $\lambda_0>0\vee\mynn(\mu-r)$, for selected values of $\lambda_0$ and $\mu$,
    respectively.
    The values of other model parameters used throughout are as in Example~\ref{ex:5.2}: $r=0.0004$,
    $P=9\,000$, $\beta=30$, $\sigma
    =0.02$, and $x = 346$. The dashed horizontal lines in both plots correspond to the value
    $v_*:=\beta\myp x-P =
    1380$. The dashed vertical line in (a) indicates
    $\mu=r$; in this case, shown as  curve {\large {\sc ii}} in plot (b), $v(x)\equiv v_*$
    for all $\lambda_0\ge \lambda_{*}\doteq0.012420$ (see~\eqref{eq:lambda*}).
    That is why curves {\large {\sc iii}}, {\large {\sc iv}} and {\large {\sc v}} in plot (a)
    all intersect at $\mu=r$.}
    \label{fig:5e}
\end{figure}

Let us now turn to the value function $v=v(x)$. First, consider $v$
as a function of $\mu$, thus keeping $\lambda_0$ fixed. Using the
expression \eqref{eq:b*}, we can rewrite the first line of the
formula \eqref{eq:value-b*} (i.e., for $x\le b^*\myn$) as
\begin{equation}\label{eq:v*}
v=\frac{P}{q_*-1}\left(\frac{x}{b^*}\right)^{q_*}.
\end{equation}
Differentiating \eqref{eq:v*}, we get
\begin{align}
\label{eq:dv/dq} \frac{\partial v}{\partial
q_*}&=-\frac{P}{(q_*-1)^2}\left(\frac{x}{b^*}\right)^{q_*}\left(1+(q_*-1)
\ln\left(\frac{b^*}{x}\right)\right)<0,\\
\frac{\partial v}{\partial b^*}&=-\frac{P\myp q_*}{(q_*-1)\mypp
b^*}\left(\frac{x}{b^*}\right)^{q_*}<0. \label{eq:dv/db}
\end{align}
Hence, on account of the inequalities \eqref{eq:q'mu},
\eqref{eq:q'lambda}, \eqref{eq:dv/dq} and \eqref{eq:dv/db},
\begin{equation}\label{eq:v'|mu1}
\frac{\rd v}{\rd \mu}=\frac{\partial v}{\partial \mu}+\frac{\partial
v}{\partial q_*}\cdot \frac{\partial
q_*}{\partial\mu}+\frac{\partial v}{\partial b^*}\cdot \frac{\rd
b^*}{\rd\mu}>0.
\end{equation}
If $x\ge b^*\mynn$, then from the second line of \eqref{eq:value-b*}
we readily obtain
\begin{equation}\label{eq:v'|mu2}
\frac{\rd v}{\rd \mu}=\frac{\beta\myp\lambda_0\myp
x}{(\tilde{r}-\mu)^2}>0.
\end{equation}
Thus, in all cases $\rd v/\rd \mu>0$, which implies that the
function $\mu\mapsto v$ is increasing (see Fig.~\ref{fig:5e}(a)).

Finally, fix $\mu$ and consider the function $\lambda_0\mapsto v$.
If $x\ge b^*\myn$ then $v$ is given by the second line of
\eqref{eq:value-b*}, that is,
\begin{equation}\label{eq:v2nd}
v=\frac{\beta\myp\lambda_0\myp x}{\lambda_0+r-\mu}-P.
\end{equation}
In particular, if $\mu=r$ then \eqref{eq:v2nd} is reduced to
$v\equiv v_*:=\beta\myp x -P$. From \eqref{eq:v2nd} it follows that
\begin{equation*}
\frac{\rd v}{\rd\lambda_0}=\frac{\beta\myp
x\myp(r-\mu)}{(\lambda_0+r-\mu)^2} \left\{\begin{aligned}\!{}<0,&\ \
\quad\mu>r,\\
\!{}=0,&\ \ \quad\mu=r,\\
\!{}>0,&\ \ \quad\mu<r.
\end{aligned}
\right.
\end{equation*}
Due to monotonicity of the function $\lambda_0\mapsto b^*\mynn$
(see~\eqref{eq:b'lambda}), $v$ is given by \eqref{eq:v2nd} as long
as $\lambda_0\ge\lambda_*$, for some critical value $\lambda_*\equiv
\lambda_*(\mu)\le \infty$. It will be shown below
(see~\eqref{eq:b*|lambda1}) that
$\lim_{\lambda_0\to\infty}b_*=P/\beta$, so $\lambda_*<\infty$ if and
only $x>P/\beta$. Clearly, $\lambda_*$ is determined by the
condition $b^*=x$ (see~\eqref{eq:b*}) together with the
equation~\eqref{e6}. In the special case $\mu=r$ (assuming that
$x>P/\beta$), these equations can be solved to yield
\begin{equation}\label{eq:lambda*}
\lambda_*=\frac{P}{\beta\myp x}
\left(\frac{\tfrac12\myp\sigma^2\beta\myp x}{\beta\myp
x-P}+r\right).
\end{equation}
In particular, in Example~\ref{ex:5.2} this gives
$\lambda_*\doteq0.012420$. From the consideration above, it also
follows that if $x>P/\beta$ then (see~\eqref{eq:v2nd})
\begin{equation}\label{eq:lim-v-2}
\lim_{\lambda_0\to\infty} v=v_*=\beta\myp x- P.
\end{equation}

In the case $x\le b^*\mynn$, we use formula \eqref{eq:v*}. Similarly
to \eqref{eq:v'|mu1},
\begin{equation}\label{eq:v'|lambda1}
\frac{\rd v}{\rd \lambda_0}=\frac{\partial v}{\partial\lambda_0}
+\frac{\partial v}{\partial q_*}\cdot \frac{\partial
q_*}{\partial\lambda_0} +\frac{\partial v}{\partial b^*}\cdot
\frac{\rd b^*}{\rd\lambda_0}.
\end{equation}
Substituting the expressions \eqref{eq:q'mu}, \eqref{eq:q'lambda},
\eqref{eq:dv/dq} and \eqref{eq:dv/db} into \eqref{eq:v'|lambda1},
cancelling immaterial factors and recalling formula~\eqref{eq:b*1},
the condition $\rd v/\rd \lambda_0<0$ is reduced to
\begin{equation}\label{eq:ineq-l0}
\left(\frac12\myp\sigma^2q_*+r\right)\left(\frac12\myp\sigma^2q_*^2+r\right)+\lambda_0\myp
r<\left(\frac{1}{q_*-1}+\ln\left(\frac{b^*}{x}\right)\right)\left(\frac12\myp\sigma^2q_*^2+r+\lambda_0\right).
\end{equation}

It can be proved that if $\mu\ge r$ then the inequality
\eqref{eq:ineq-l0} holds for all $\lambda_0<\lambda_*$, but the
analysis becomes difficult for $\mu<r$. Numerical plots (see
Fig.~\ref{fig:5e}(b)) suggest that in the latter case the function
$\lambda_0\mapsto v$ may be non-monotonic, with the derivative $\rd
v/\rd\lambda_0$ possibly vanishing in up to two points, provided
that $r-\varepsilon<\mu<r$ with $\varepsilon>0$ small enough. To be
more specific, the plots in Fig.~\ref{fig:5e}(b) illustrate the case
$x>P/\beta$, with the common asymptote~\eqref{eq:lim-v-2}. For $x\le
P/\beta$, the plots look similar (not shown here) but with
$\lim_{\lambda_0\to\infty} v= 0$ (see \eqref{eq:lim-v-1} below), so
the derivative $\rd v/\rd\lambda_0$ may vanish in at most one point.

\subsection{Limiting values}\label{sec:6.2}

Let us investigate the functions $b^*\mynn$ and $v$ in the limits
(i) $\mu\to-\infty$ or $\mu\uparrow \tilde{r}$, and (ii)
$\lambda_0\to\infty$ or $\lambda_0\downarrow 0$ ($\mu<r$),
$\lambda_0\downarrow \mu-r$ ($\mu\ge r$). Start by observing, using
equation \eqref{e6}, that
\begin{equation}\label{eq:mu->}
\lim_{\mu\to-\infty} q_*=\infty,\qquad \lim_{\mu\myp\uparrow\myp
\tilde{r}} q_*=1,
\end{equation}
and moreover,
\begin{equation}\label{eq:q*|mu1}
q_*-1\sim \frac{\tilde{r}-\mu}{\frac12\myp\sigma^2\mynn+\tilde{r}}\
\ \quad (\mu\uparrow \tilde{r}).
\end{equation}
Similarly, $\lim_{\lambda_0\to\infty} q_*=\infty$; on the other
hand, if $\mu<r$ then $\lim_{\lambda_0\downarrow 0}
q_*=q_*|_{\lambda_0=0}>1$, while if $\mu\ge r$ then
\begin{equation}\label{eq:q*|mu2}
q_*-1\sim \frac{\lambda_0-(\mu-r)}{\frac12\myp\sigma^2\mynn+\mu}\ \
\quad (\lambda_0\downarrow \mu-r).
\end{equation}
Hence, from \eqref{eq:b*1} and \eqref{eq:mu->} it readily follows
that $b^*\to\infty$ \,($\mu\to-\infty$) and
\begin{equation*}
b^*\to
\frac{P\myp(\frac12\myp\sigma^2\mynn+\tilde{r})}{\beta\myp\lambda_0}
\quad \ \ (\mu\uparrow\tilde{r}).
\end{equation*}
Also, using that $q_*\to\infty$ \,($\lambda_0\to\infty$), from
\eqref{eq:b*} we get
\begin{equation}\label{eq:b*|lambda1}
b^*\to\frac{P}{\beta}\ \ \quad (\lambda_0\to\infty).
\end{equation}
In the opposite limit, if $\mu>r$ then, according to \eqref{eq:b*1}
and~\eqref{eq:q*|mu2},
\begin{equation}\label{eq:b*|lambda2}
b^* \to
\frac{P\mypp(\frac12\myp\sigma^2\mynn+\mu)}{\beta\myp(\mu-r)}\ \
\quad (\lambda_0\downarrow \mu-r),
\end{equation}
while if $\mu\le r$ then $\lim_{\lambda_0\myp\downarrow\myp
0}b^*=\infty$; in particular, for $\mu=r$
\begin{equation}\label{eq:b|m=r}
b^*\sim \frac{P\myp(\frac12\myp\sigma^2+r)}{\beta\myp\lambda_0} \ \
\quad (\lambda_0\downarrow0).
\end{equation}

For the value function $v=v(x)$, from formula \eqref{eq:v*} we get,
using \eqref{eq:mu->} and~\eqref{eq:q*|mu1},
\begin{equation*}
\lim_{\mu\to-\infty} v=0,\qquad  \lim_{\mu\myp\uparrow\myp
\tilde{r}} v=\infty.
\end{equation*}
Furthermore, according to \eqref{eq:lim-v-2}, if $x>P/\beta$ then
$v\to v_*=\beta\myp x-P$ as $\lambda_0\to\infty$. In the opposite
case, due to monotonicity of $b^*\mynn$ (see~\eqref{eq:b'lambda})
and the limit \eqref{eq:b*|lambda1} we have $b^*>P/\beta\ge x$, so
using formula \eqref{eq:v*} and recalling that $q_*\to \infty$, we
get
\begin{equation}\label{eq:lim-v-1}
v \le \frac{P}{q_*-1}\to0\ \ \quad (\lambda_0\to\infty).
\end{equation}
Now, consider the limit of $v$ as $\lambda_0$ approaches the lower
edge of its range. If $\mu<r$ then \eqref{eq:v*} implies that
$\lim_{\lambda\myp\downarrow\myp0}v=0$, since $b^*\to\infty$ and
$q_*\to q_*|_{\lambda_0=0}>1$. If $\mu=r$ then, using
\eqref{eq:q*|mu2} and \eqref{eq:b|m=r} (with $\mu=r$), we obtain
\begin{equation}\label{eq:v|m=r}
v\sim \beta\myp x\myp \lambda_0^{q_*-1}=\beta\myp x\myp
\exp\myn\bigl\{(q_*-1)\ln\lambda_0\bigr\}\to \beta\myp x\ \ \quad
(\lambda_0\downarrow 0).
\end{equation}
Finally, if $\mu>r$ then from \eqref{eq:v*} it readily follows,
according to \eqref{eq:q*|mu2} and \eqref{eq:b*|lambda2},
\begin{equation}\label{eq:lim-v-3}
v\sim\frac{\beta\myp x\myp(\mu-r)}{\lambda-(\mu-r)}\to\infty\ \
\quad (\lambda_0\downarrow \mu-r).
\end{equation}

\subsection{Comparative statics and sensitivity analysis}\label{sec:6.3}

\begin{figure}[bt]
    \centering
   \subfigure[Isolines of $b^*$]{\includegraphics[width=0.45\linewidth]
   {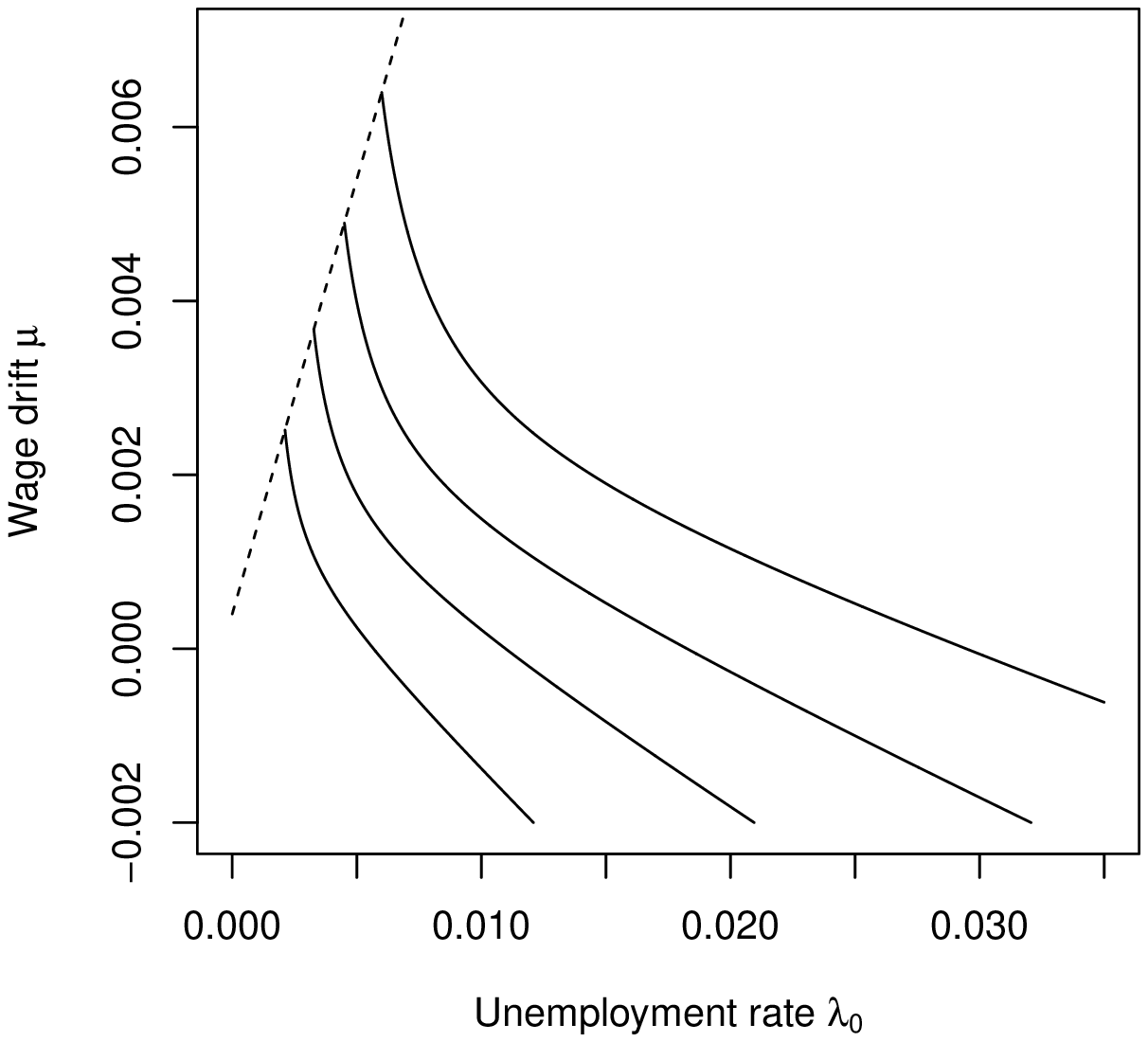}}
       \put(-30.0,65.9){\mbox{\myp\tiny {\sc I}}}
       \put(-54.0,54.5){\mbox{\tiny {\sc II}}}
       \put(-81.9,48.0){\mbox{\tiny {\sc III}}}
       \put(-111.5,46.0){\mbox{\tiny {\sc IV}}}
            \put(-82.7,149.2){\mbox{\scriptsize {\sc \hphantom{II}I\mypp:} \,$b^*=330$}}
            \put(-82.7,141.2){\mbox{\scriptsize {\sc \hphantom{I}II\mypp:} \,$b^*=340$}}
            \put(-82.7,133.2){\mbox{\scriptsize {\sc III\mypp:} \,$b^*=355$}}
            \put(-82.7,125.2){\mbox{\scriptsize {\sc IV\myp:} \,$b^*=385$}}
        \hspace{0.5pc}
\subfigure[Isolines of
$v=v(x)$]{\includegraphics[width=0.45\linewidth]
{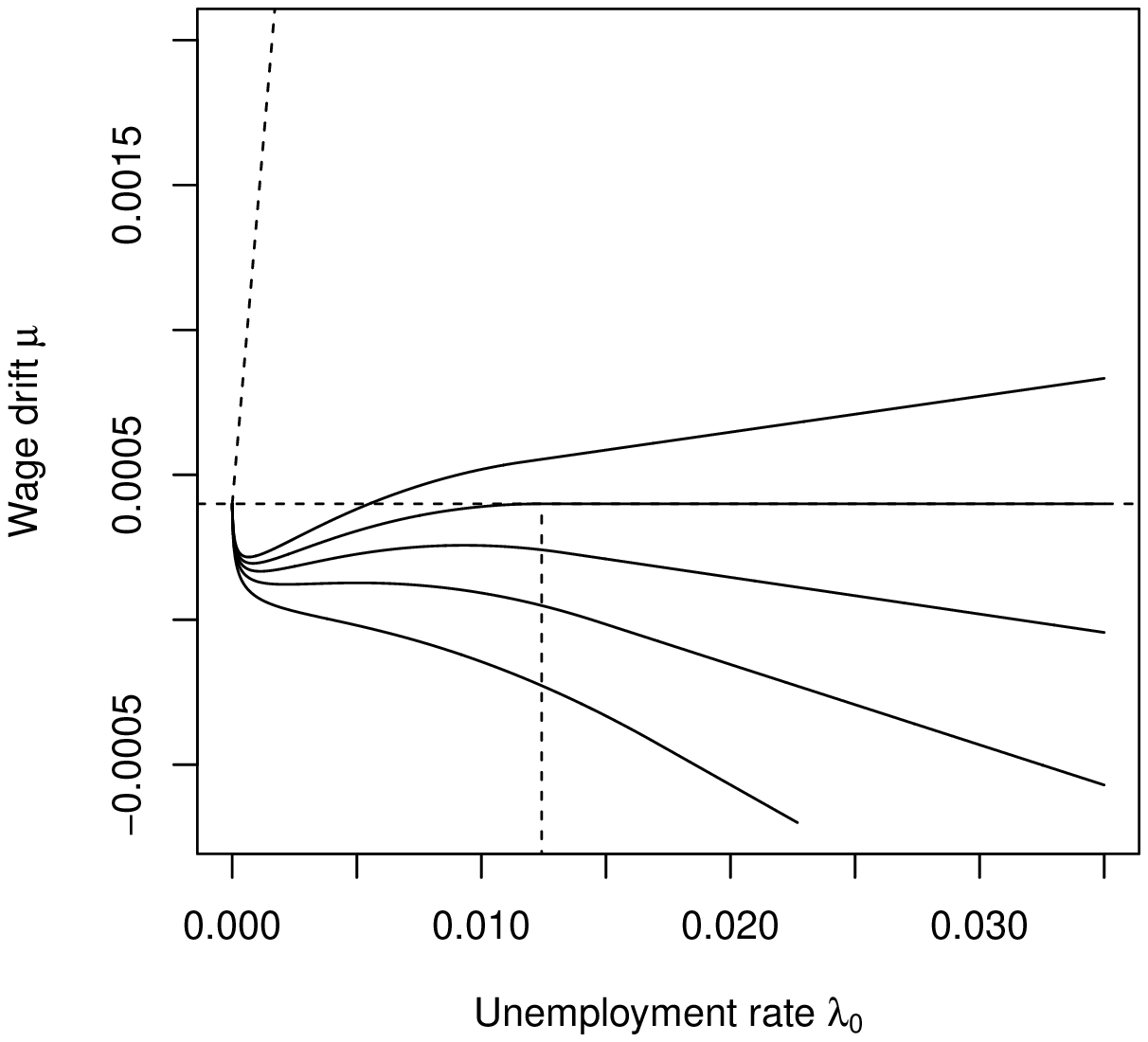}}
       \put(-166.5,90.5){\mbox{\scriptsize $r$}}
       \put(-107.0,39.1){\mbox{\scriptsize $\lambda_{*}$}}
       \put(-73.6,46.0){\mbox{\myp\tiny {\sc V}}}
       \put(-46.0,58.0){\mbox{\tiny {\sc IV}}}
       \put(-37.0,75.8){\mbox{\tiny {\sc III}}}
       \put(-34.2,94.1){\mbox{\tiny {\sc II}}}
       \put(-32.2,111.7){\mbox{\tiny {\sc I}}}
            \put(-102.7,153.2){\mbox{\scriptsize {\sc \hphantom{II}I\mypp:} \,$v=1510$}}
            \put(-102.7,145.2){\mbox{\scriptsize {\sc \hphantom{I}II\mypp:} \,$v=1380$}}
            \put(-102.7,137.2){\mbox{\scriptsize {\sc III\mypp:} \,$v=1250$}}
            \put(-102.7,129.2){\mbox{\scriptsize {\sc IV\myp:} \,$v=1100
            $}}
            \put(-102.7,121.2){\mbox{\scriptsize {\sc \hphantom{I}V\myp:} \,$v=900$}}
 \\[-.5pc]
   \caption%
   {Isolines (level curves) of the optimal stopping problem solution
   on the $(\lambda_0,\mu)$-plane: (a) $b^*\myn(\lambda_0,\mu)=\const$ (optimal
   threshold~\eqref{eq:b*}); (b) $v(\lambda_0,\mu)=\const$ (value function~\eqref{eq:value-b*}).
        The values of other parameters used throughout are as in
        Example~\ref{ex:5.2}: $r=0.0004$, $P=9\,000$, $\beta=30$, $\sigma =0.02$, and $x = 346$.
        The slanted dashed lines in both plots show the boundary $\mu=\lambda_0+r$ (see~\eqref{eq:upper}).
        In plot (b), the horizontal dashed line indicates $\mu=r$ and the vertical dashed line shows
        $\lambda_{*}\doteq0.012420$ (cf.\ Fig.~\ref{fig:5e}(b)).}
\label{fig:6}
\end{figure}

The goal of comparative statics is to understand how varying values
of exogenous parameters affect a target function of interest. For
instance, consider the optimal threshold $b^*\mynn$ as a function of
both unemployment rate $\lambda_0$ and wage drift $\mu$. Rather then
fixing one of these parameters and then plotting $b^*\mynn$ against
the remaining parameter (as was done in Figs.\ \ref{fig:5}(a)
and~\ref{fig:5}(b)), it is useful to plot a family of comparative
statics plots showing the \emph{isolines} (or \emph{level curves})
for different values (levels) of the function, that is,
$b^*(\lambda_0,\mu)=\const$ (see Fig.~\ref{fig:6}(a)). As may be
expected from Figs.\ \ref{fig:5}(a) and~\ref{fig:5}(b), the plots of
the function $\lambda_0=\lambda_0(\mu)$ (determined implicitly by
the level condition) behave as monotone decreasing graphs. Analogous
plots for the value function are presented in Fig.~\ref{fig:6}(b);
the plots become non-monotonic for $v$ large enough. If $\lambda_0$
is fixed then the value $v$ grows with $\mu$, in agreement with
\eqref{eq:v'|mu1} and~\eqref{eq:v'|mu2}. Similarly, if $\mu>r$ is
fixed then $v$ decreases with $\lambda_0$, converging to the limit
$v_*=\beta\myp x-P$ as $\lambda_0\to\infty$
(see~\eqref{eq:lim-v-2}), represented by curve {\sc II} in
Fig.~\ref{fig:5e}(b). If $v>v_*$ then there are up to two different
values of $\lambda_0$ (and common $\mu$) producing the same
value~$v$, while for $v$ smaller than but close enough to~$v_*$, the
number of such roots may increase to three (see the discussion in
Section~\ref{sec:6.4}).

Let us also comment on the sensitivity of our numerical examples
presented in Section~\ref{sec:5.4}. The question here is, how much
the output values (say, the optimal threshold $b^*\mynn$ and the
value $v$) would change under a small variation of one of the
background parameters. In the linear approximation, the change
factor is given by the corresponding partial derivative. As in the
previous sections, we address the sensitivity with regard to the
wage drift $\mu$ (around the set value $\mu=0.0004$) and the
unemployment rate $\lambda_0$ (around $\lambda_0=0.01$). Other model
parameters are fixed as in Section~\ref{sec:5.4}, that is, $r =
0.0004$, $P = 9\,000$, $\beta = 30$, and $x = 346$. As for the
volatility parameter $\sigma$, it is set to be $\sigma=0.04$ as in
Example~\ref{ex:5.1} or $\sigma=0.02$ as in Example~\ref{ex:5.2}.
The required partial derivatives of $b^*\mynn$ and $v$ can be
computed using the formulas derived in Section~\ref{sec:6.1}; the
results are presented in Table~\ref{table:1}(a).

\begin{table}[h]
\centering \caption{Sensitivity check of numerical results for the
functions $b^*\mynn$ and $v$ in Examples \ref{ex:5.1}
and~\ref{ex:5.2}: (a) parametric derivatives; (b) increments in
response to a $1\%$-change in the background parameters.}
\subtable[Derivatives]{
\begin{tabular}{|c|c|c|}
    \hline
       &&\\[-.9pc]
    Derivative & Example~\ref{ex:5.1} & Example~\ref{ex:5.2} \\[.1pc]
    \hline
    &&\\[-0.9pc]
 $\rd b^*\mynn/\rd \mu$ & $-16\,037.57$ & $-13\,962.43$ \\
     $\rd v/\rd\mu$ & \hspace{2.6pt}$842\,062.30$ & \hspace{2.5pt}$993\,991.20$ \\[.1pc]
    \hline
       &&\\[-.9pc]
 $\rd b^*\mynn/\rd\lambda_0$ & $\hphantom{0}{-}6\,323.813$ & $-3\mypp161.906$ \\
    $\rd v/\rd\lambda_0$ & \hspace{-.3pt}$-46\,485.530$ & \hspace{-.6pt}$-8\mypp768.435$ \\[.1pc]
    \hline
\end{tabular}}
\hspace{1.5pc}\subtable[Increments (euro)]{
\begin{tabular}{|c|c|c|}
    \hline
       &&\\[-.9pc]
    Increment& Example~\ref{ex:5.1} & Example~\ref{ex:5.2} \\[.1pc]
    \hline
    &&\\[-.9pc]
 $\Delta b^*$ ($\mu$)& $-0.06415$ & $-0.05585$ \\
     $\Delta v$ \mypp\,($\mu$)& $\hphantom{-}3.36825$ & $\hphantom{-}3.97597$ \\[.1pc]
    \hline
       &&\\[-.9pc]
  $\Delta b^*$ ($\lambda_0$) & $-0.63238$ & $-0.31619$ \\
   $\Delta v$ \mypp\,($\lambda_0$) & $-4.64855$ & $-0.87684$ \\[.1pc]
    \hline
\end{tabular}}
\label{table:1}
\end{table}

Numerical values in Table~\ref{table:1}(a) may seem quite big, but
they should be offset by small background values of the parameters,
$\mu=0.0004$ and $\lambda_0=0.01$. If we increase them by a small
amount, say by $1\%$, then the absolute increments would be
$$
\Delta\mu=0.0004/100=4\cdot 10^{-6},\qquad
\Delta\lambda_0=0.01/100=10^{-4}.
$$
Hence, using Table~\ref{table:1}(a), we obtain the corresponding
approximate increments of the target functions $b^*\mynn$ and $v$
(see Table~\ref{table:1}(b)), which look more palatable. One
interesting observation is that the value $v$ reacts about $5$ times
stronger to the change of the unemployment rate $\lambda_0$ when the
volatility $\sigma$ gets $2$ times bigger (from $\sigma=0.02$ in
Example~\ref{ex:5.2} to $\sigma=0.04$ in Example~\ref{ex:5.1}); in
contrast, the change of $v$ in response to an increase of the wage
drift is much less pronounced. This highlights the primary
significance of the unemployment rate, which is of course only
natural.

Sensitivity analysis with regard to the wage drift $\mu$ is also
useful in the light of the difficulty in estimation of $\mu$ from
the data, mentioned in Section~\ref{sec:5.2}. The results in
Table~\ref{table:1}(b) suggest that a reasonably small error in
selecting $\mu$ has only a minor effect on the identification of the
optimal threshold $b^*\mynn$ and the value $v$; for instance,
overestimating it by $1\%$ will decrease $b^*\mynn$ by just $0.01$
euro, while the value $v$ will be up by about $0.60$ euro. Thus, an
individual using a  moderately inflated value of their wage rate
would take a slightly over-optimistic view about the timing of
joining the insurance scheme and its expected benefit. On the other
hand, a risk-averse individual may take a more conservative view and
prefer to underestimate their wage drift $\mu$, which will raise the
threshold $b^*\mynn$ resulting in a longer waiting time. For the
insurance company though, it may be reasonable to try and avoid
underestimation of the wage drift of potential customers, so as to
reduce the risk of overpaying the benefits.

\subsection{Economic interpretation}\label{sec:6.4}

Monotonic decay of the optimal threshold $b^*\mynn$ with an increase
of the unemployment rate $\lambda_0$ (see~\eqref{eq:b'lambda} and
Fig.~\ref{fig:5}(b)) has a clear economic appeal: a bigger
unemployment rate $\lambda_0$ means a higher risk of losing the job,
which demands a lower target threshold $b^*\mynn$ in order to
expedite joining the insurance scheme. The economic rationale for
the monotonicity of $b^*\mynn$ as a function of $\mu$
(see~\eqref{eq:b'mu} and Fig.~\ref{fig:5}(a)) is different\,---\,a
bigger wage drift $\mu$ makes it more likely to reach a higher final
wage $X_{\tau_0}$ by the time of loss of job, so lowering the
threshold $b^*\mynn$ adds incentive to an earlier entry.

Monotonic growth of the value $v$ as a function of the wage drift
$\mu$ (see \eqref{eq:v'|mu1}, \eqref{eq:v'|mu2}, and
Fig.~\ref{fig:5e}(a)) is also meaningful\,---\,indeed, when the wage
drift $\mu$ gets bigger, there is more potential to reach a higher
final wage $X_{\tau_0}$ by the time of loss of job, which increases
the expected benefit $\beta_1$ (see~\eqref{eq:c1-tilde}) and,
therefore, the value $v=v(x)$ of the insurance policy.

The behaviour of the value function $v=v(x)$ in response to a
varying unemployment rate $\lambda_0$ is more interesting, as
indicated by the plots in Fig.~\ref{fig:5e}(b). In the case $\mu<r$,
it is satisfactory to see that the value $v$, vanishing in the limit
as $\lambda_0\downarrow 0$, starts growing with $\lambda_0$, thus
reflecting a good efficiency of the insurance policy against an
increasing risk of unemployment. On the other side of the policy,
this may present a growing risk for the insurance company which will
have to finance an increasing number of claims. But with the
unemployment rate $\lambda_0$ getting ever bigger, the value $v$
should stay bounded, so must converge to a limit as
$\lambda_0\to\infty$, given by $v_*=\beta\myp x-P$ if $x>P/\beta$
(see~\eqref{eq:lim-v-2}) or $v_*=0$ otherwise
(see~\eqref{eq:lim-v-1}). In particular, Fig.~\ref{fig:5e}(b) shows
that, for a certain range of $\mu$, the value $v$ achieves its
maximum at some $\lambda_0$. However, the graphs also reveal that if
$\mu$ keeps increasing then the value plots may have a more
complicated non-monotonic behaviour, which is harder to interpret
economically.

On the other hand, as is evident from Fig.~\ref{fig:5e}(b), in the
case $\mu\ge r$ our model produces a counter-intuitive increase of
the value $v$ as $\lambda_0$ approaches the left edge of its
range\,---\,it is hard to believe that the value may grow as the
risk of unemployment falls. Moreover, as was computed in
\eqref{eq:lim-v-3}, for $\mu>r$ the corresponding limit of $v$ is
infinite! But perhaps the most striking example emerges in the
borderline case $\mu=r$, whereby formally setting $\lambda_0=0$ we
would get, according to \eqref{eq:b|m=r}, that the threshold
$b^*\mynn$ is infinite (unlike the case $\mu>r$,
see~\eqref{eq:b|m=r}), so that the wage process $(X_t)$ never
reaches it; therefore, we never buy the insurance policy
(understandably so, as there is no risk of losing the job), and
nonetheless its value is positive in this limit
(see~\eqref{eq:lim-v-2}). The explanation of this paradox lies in
the way how the optimal stopping is exercised for small
$\lambda_0>0$: here, the threshold $b^*\mynn$ is high and there is
only a very small probability that it is ever reached; before this
happens, we stay idle, but if and when the threshold is hit then the
expected payoff is rather big, which contributes enough to the
expected net present value to keep it positive in the limit
$\lambda_0\downarrow 0$ (see~\eqref{eq:v|m=r}).

Thus, the artefacts in our model as indicated above are caused by
not putting any constraint on the waiting times. This can be
rectified, for example, by introducing \emph{mortality}, as was
sketched in Section~\ref{sec:2.3}; in particular, such a
regularization should restore a zero limit of $v$ at the lower edge
of~$\lambda_0$.

\section{Including utility considerations}\label{sec:7}

\subsection{Perpetual American call option}\label{sec:7.1}
Our model (and its solution) resembles that of the optimal stopping
problem for the \emph{(perpetual) American call option} (see a
detailed discussion in
\cite[Ch.\,VIII, \S\mypp2a]{Shiryaev}). More specifically, the
holder of a call option may exercise the right to buy an asset
(e.g., one unit of stock) at any time for a pre-determined strike
price $K$, where the decision is based on observations over the
random process of stock prices $(S_t)$, assumed to follow a
geometric Brownian motion model. The term \emph{perpetual} is used
to indicate that there is no expiration date, so the right to buy
extends indefinitely.

The optimal time instant $\tau=\tau^*$ to buy, bearing in mind a
purely financial target of maximizing the profit $S_{\tau}-K$, is
the solution of the following optimal stopping problem,
\begin{equation}\label{eq:ACO}
V(x)=\sup_\tau \EE_x\bigl(\re^{-r\tau}(S_\tau-{K})^{+}\bigr),
\end{equation}
where $S_t$ is a geometric Brownian motion with parameters $\mu<r$
and $\sigma>0$, the supremum is taken over all stopping times $\tau$
adapted to the filtration associated with $(S_t)$. The positive
truncation $(\cdot)^+$ corresponds to the constraint that the option
holder is not in a position to buy at the price $K$ higher than the
current spot price $S_t$. The solution to \eqref{eq:ACO} is well
known (see, e.g.,
\cite[Ch.\,VIII, \S\mypp2a]{Shiryaev}) to be given by the hitting
time $\tau^*=\tau_{b^*}$, with the optimal threshold
$$
b^*=\frac{Kq^*}{q^*-1},
$$
where $q^*$ is given by formula \eqref{eq:q1*} but with
$\tilde{r}=r+\lambda_0$ replaced by $r$. The corresponding value
function is given by
$$
V(x)=
\begin{cases}
   (b^* - K) \left( \dfrac{x}{b^*}\right)^{q_*}\!, &x\in[\myp0,b^*],\\[.3pc]
    x - K,& x\in[\myp b^*,\infty).
    \end{cases}
$$

Observe that our optimal stopping problem \eqref{eq:r2e1} can be
rewritten as
\begin{equation}\label{eq:eq:r2e1}
v(x)=\beta_1 \sup_{\tau} \EE_x \bigl[ \re^{-\tilde{r}
\tau}\myp(X_\tau-\tilde{K})\bigr], \qquad \tilde{K}:=P/\beta_1,
\end{equation}
which makes it look very similar to the perpetual American call
option problem~\eqref{eq:ACO}. However, there are several important
differences. Firstly, unlike the gain function in the American call
option problem \eqref{eq:ACO}, no truncation is applied
in~\eqref{eq:eq:r2e1}, because the financial gain is not the sole
priority in this context and therefore the individual is prepared to
tolerate negative values of $\beta_1X_\tau - P$ (despite the fact
that, under the optimal strategy, the value function $v(x)$ is
always non-negative, see Lemma~\ref{lm:v(x)}(i) and
formula~\eqref{eq:value-b*}).\footnote{The equivalence of the
problems \eqref{eq:ACO} and \eqref{eq:eq:r2e1}, which we have
established directly, is not a coincidence: it is known
\cite[Proposition~3.1, p.\mypp185]{Villeneuve} that, under mild
assumptions, the solution of the general optimal stopping problem
$v(x)=\sup_{\tau}\EE_x\mynn\left(\re^{-r\tau} g(X_\tau)\right)$ does
not change with the positive truncation of $g(\cdot)$.} In addition,
as was mentioned in Remark~\ref{rm:<1} and in Section~\ref{sec:5.3},
the hitting time $\tau_{b^*}$ may be infinite with a positive
probability (i.e., when $\mu<\frac12\myp\sigma^2$), which may be
deemed impractical
in the insurance context, but is considered to be acceptable for
exercising the American call option. This simple observation helps
to realize the fundamental conceptual difference between the two
problems; indeed, the insurance optimal stopping does not focus only
on the financial gain, but also places an ultimate priority on
acquiring an insurance cover \emph{per~se}. Hence, a more realistic
formulation of the optimal stopping problem in the UI model should
involve a certain \emph{utility}, which specifies the individual's
weighted preferences for satisfaction\,---\,for example, impatience
against waiting for too long before joining the UI scheme.

\subsection{Heuristic optimal stopping models with utility}\label{sec:7.2}

Here, we present a few informal thoughts about the possible
inclusion of utility in the optimality analysis. As already
mentioned, in the case $\mu<\frac12\myp\sigma^2$ the probability of
hitting the critical threshold $b^*\mynn$ is less than 1, so there
is a probability that the individual will never join the insurance
scheme if the optimal stopping rule is strictly followed. This is of
course not desirable, as the individual puts high priority on
getting insured at some point in time (hopefully, prior to loss of
job).

One simple way to take these additional requirements into account is
to extend the optimal stopping problem \eqref{eq:r2e1} as follows:
\begin{align}
\notag v^\dag(x)&=\sup_{\tau}\bigl[\kappa\, \PP_x(\tau<\infty) +
\eNPV(x;\tau)\bigr]\\
&=\sup_{\tau}\EE_x\bigl[\kappa\mypp\mathbbm{1}_{\{\tau<\infty\}}\!
+\re^{-\tilde{r}\tau} g(X_\tau)\bigr], \label{eq:optimal+utility}
\end{align}
where the supremum is again taken over all stopping times $\tau$
adapted to the process $(X_t)$, and the coefficient $\kappa\ge0$ is
a predefined weight representing the individual's personal attitude
(preference) towards the two contributing terms. If
$\PP_x(\tau<\infty)=1$ then the first term in
\eqref{eq:optimal+utility} is reduced to a constant ($\kappa$),
leading to a pure optimal stopping problem as before; however, if
$\PP_x(\tau<\infty)<1$ then the first term enhances the role of
candidate stopping times $\tau$ that are less likely to be infinite.

The problem \eqref{eq:optimal+utility} can be rewritten in a more
standard form by pulling out the common discounting factor under
expectation,
\begin{equation}\label{eq:optimal+utility1}
v^\dag(x)=\sup_{\tau}\EE_x\bigl[\re^{-\tilde{r}\tau}\myp
G(\tau,X_\tau)\bigr],
\end{equation}
with
\begin{equation}\label{eq:G}
G(t,x):=\kappa\,\re^{\myp\tilde{r}\myp t\!}+g(x),\quad
(t,x)\in[\myp0,\infty\myp]\times [\myp0,\infty).
\end{equation}
Unfortunately, the optimal stopping problem
\eqref{eq:optimal+utility1} is not amenable to an exact solution as
before, because the gain function \eqref{eq:G} depends also on the
time variable (see
\cite[Ch.\mypp{}IV]{peskir2006optimal}). In this case, the problem
\eqref{eq:optimal+utility1} may again be reduced to a suitable (but
more complex) free-boundary problem, but the hitting boundary (of a
certain
set on the $(t,x)$-plane) is no longer a straight
line.

More generally, our optimal stopping problem can be modified by
replacing the indicator in \eqref{eq:optimal+utility} with the
expression $\re^{-\rho\myp\tau}$ ($\rho>0$),
\begin{equation}\label{eq:OSP-general}
v^\dag(x)=\sup_{\tau}\EE_x\bigl[\kappa\mypp\re^{-\rho\myp\tau\!}
+\re^{-\tilde{r}\tau} g(X_\tau)\bigr],
\end{equation}
which retains the flavour of progressively penalizing larger values
of $\tau$, including $\tau=\infty$. Here, the gain function
\eqref{eq:G} takes the form
$$
G(t,x)=\kappa\,\re^{\myp(\tilde{r}-\rho)\myp t\!}+g(x).
$$
In particular, by choosing $\rho=\tilde{r}$ the problem
\eqref{eq:OSP-general} is transformed into
$$
v^\dag=\sup_{\tau}\EE_x\bigl[\re^{-\tilde{r}\tau}(\beta_1X_\tau
+\kappa-P)\bigr],
$$
which is the same problem as \eqref{eq:r2e1} but with the premium
$P$ replaced by $P-\kappa$.

Another, more drastic approach to amending the standard optimal
stopping problem \eqref{eq:r2e1} stems from the observation that
even if $\tau<\infty$ ($\PP_x$-a.s.), it may take long to wait for
$\tau$ to happen\,---\,for instance, if $\EE_x(\tau)=\infty$. In
other words, it is reasonable to take into account the expected
value of $\tau$, leading to the combined optimal stopping problem
\begin{equation}\label{eq:optimal+utility+exp}
v^\dag(x)=\sup_{\tau} \bigl[\kappa\,
\PP_x(\tau<\infty)+\kappa\myp\exp\{-\EE_x(\tau)\} +
\eNPV(x;\tau)\bigr].
\end{equation}
If $\PP_x(\tau<\infty)<1$ then $\EE_x(\tau)=\infty$ and the problem
\eqref{eq:optimal+utility+exp} is reduced
to~\eqref{eq:optimal+utility}, whereas if $\PP_x(\tau<\infty)=1$
then, effectively, only the term with the expectation remains
in~\eqref{eq:optimal+utility+exp}. However, a disadvantage of the
formulation \eqref{eq:optimal+utility+exp} is that it cannot be
expressed in the form~\eqref{eq:optimal+utility1}. Trying to amend
this would take us back to the version~\eqref{eq:OSP-general}.

It is interesting to look at how the value function depends on the
preference parameter~$\kappa$. The next property is intuitively
obvious.
\begin{proposition}\label{pr:kappa}
For each $x>0$, the value function $v^\dag(x)$ of the optimal
stopping problem \eqref{eq:OSP-general} is a strictly increasing
function of $\kappa$. The same is true for the
problem~\eqref{eq:optimal+utility+exp}.
\end{proposition}
\proof We use the notation $v^\dag(x;\kappa)$ to indicate the
dependence of the value function on the parameter~$\kappa$.
For $\kappa_1<\kappa_2$ and any stopping time
$\tau\not\equiv\infty$, we have
\begin{equation}\label{eq:kappa12}
\EE_x\bigl[\kappa_1\mypp\re^{-\rho\myp\tau\!} +\re^{-\tilde{r}\tau}
g(X_\tau)\bigr]<\EE_x\bigl[\kappa_2\mypp\re^{-\rho\myp\tau\!}
+\re^{-\tilde{r}\tau} g(X_\tau)\bigr]\le v^\dag(x;\kappa_2).
\end{equation}
%
Suppose that $\tau_*$ is a maximizer for the optimal stopping
problem \eqref{eq:OSP-general} with $\kappa=\kappa_1$. Then,
according to \eqref{eq:kappa12},
\begin{equation*}
v^\dag(x;\kappa_1)=\EE_x\bigl[\kappa_1\mypp\re^{-\rho\myp\tau_*\!}
+\re^{-\tilde{r}\tau_*} g(X_{\tau_*})\bigr]<v^\dag(x;\kappa_2),
\end{equation*}
that is, $v^\dag(x;\kappa_1)<v^\dag(x;\kappa_2)$ as claimed. Similar
arguments apply to the problem~\eqref{eq:optimal+utility+exp}.
\endproof

\subsection{Sub-optimal solutions}\label{sec:7.3}

As already mentioned, the optimal stopping problems outlined in
Section~\ref{sec:6.2} are difficult to solve in full generality. To
gain some insight about the qualitative effects of the added
utility-type terms, it may be reasonable to restrict our attention
to solutions in the subclass of hitting times $\tau_b$. Despite such
solutions will only be suboptimal, the advantage is that the reduced
problems can be solved using that all the ingredients are available
explicitly (see Section~\ref{sec:4.1}).

For example, the original problem \eqref{eq:optimal+utility} is
replaced by
\begin{equation}\label{eq:optimal+utility1-sub}
u^\dag(x)=\sup_{b\ge 0}\bigl[\kappa\, \PP_x(\tau_b<\infty) +
\eNPV(x;\tau_b)\bigr].
\end{equation}
Similarly as in Section \ref{sec:4.3}, we only need to maximize the
functional in \eqref{eq:optimal+utility1-sub} over $b\ge x$. Indeed,
if $b\le x$ then $\tau_b=0$ ($\PP_x$-a.s.) and, according to
\eqref{r2d0} and~\eqref{eq:NPV1+},
$$
\sup_{b\le x}\bigl[\kappa\, \PP_x(\tau_b<\infty) +
\eNPV(x;\tau_b)\bigr]=\kappa + \eNPV(x;0)=\kappa+\beta_1x-P,
$$
whereas
\begin{align*}
\sup_{b\ge x}\bigl[\kappa\, \PP_x(\tau_b<\infty) +
\eNPV(x;\tau_b)\bigr]&\ge \bigl[\kappa\, \PP_x(\tau_b<\infty) +
\eNPV(x;\tau_b)\bigr]\bigr|_{b=x}\\
&=\kappa+\beta_1x-P.
\end{align*}

Assume that $\mu-\tfrac12\myp\sigma^2<0$ (for otherwise
$\PP_x(\tau_b<\infty)=1$, thus leading to the same optimal stopping
problem as before). Then, according to \eqref{eq:P(tau<infty)}, the
probability $\PP_x(\tau_{b}<\infty)$ becomes a strictly decreasing
function of $b\in[x,\infty)$, and so the maximum in
\eqref{eq:optimal+utility1-sub} is achieved by a different stopping
strategy, with a lower optimal threshold $b^{\dag\!}$. More
precisely, by virtue of formulas \eqref{eq:P(tau<infty)} and
\eqref{eq:NPV-explicit}, the problem \eqref{eq:optimal+utility1-sub}
is explicitly rewritten as
\begin{equation}\label{eq:optimal+utility1-sub2}
u^\dag(x)=\sup_{b\ge x}\biggl[\kappa\left(\dfrac{x}{b}\right)^{\mynn
1-2\mu/\sigma^2}\! + (\beta_1 b-P)\left(\frac{x}{b }\right)^{\myn
q_*}\biggr],
\end{equation}
where $q_*>1$ is defined in~\eqref{eq:q1*}. Differentiating with
respect to $b$, it is easy to check that the maximizer for the
problem \eqref{eq:optimal+utility1-sub2} is given by
$$
b^\dag=\min\left\{b\ge x\colon
\,a\myp\kappa\left(\frac{b}{x}\right)^{\myn
q_*-a}\!+(q_*-1)\myp\beta_1\myp b\ge Pq_*\right\},
$$
where $a:=1-2\mu/\sigma^2<1<q_*$.

The following (slightly artificial) version of the utility keeps the
spirit of \eqref{eq:optimal+utility1-sub} but is amenable to the
exact analysis:
\begin{equation}\label{eq:optimal+utility+exp-improved1}
u^\dag(x)=\sup_{b\ge 0}\biggl[
\kappa\myp\bigl\{\PP_x(\tau_{b}<\infty)\bigr\}^{q_*/(1-2\mu/\sigma^2)}\!+\eNPV(x;\tau_{b})\biggr].
\end{equation}
Indeed, using the same substitutions \eqref{eq:P(tau<infty)} and
\eqref{eq:NPV-explicit} as before,
\eqref{eq:optimal+utility+exp-improved1} is reduced to
(cf.~\eqref{eq:optimal+utility1-sub2})
\begin{equation}\label{eq:max_amended}
u^\dag(x)=\sup_{b\ge x}\Bigl[(\beta_1 b+\kappa-P)\!\left(\frac{x}{b
}\right)^{\myn q_*}\Bigr],
\end{equation}
which is the same problem as \eqref{eq:r2e**red-again} but with $P$
replaced by $P-\kappa$ (cf.~\eqref{eq:NPV-explicit}). Therefore,
from \eqref{eq:b*new} we immediately obtain the maximizer
\begin{equation}\label{eq:b-dag}
b^\dag=\frac{(P-\kappa)\myp
q_*}{\beta_1(q_*-1)}=b^*\!-\frac{\kappa\myp q_*}{\beta_1(q_*-1)}\le
b^*.
\end{equation}
This is a strictly decreasing (linear) function of $\kappa$; in
particular,  $b^\dag=b^*\mynn$ if $\kappa=0$ and $b^\dag=0$ if
$\kappa=P$. The corresponding value function is given by
(cf.~\eqref{eq:v-new})
\begin{equation}\label{eq:v-new-dag}
u^\dag(x)=
\begin{cases} \displaystyle (\beta_1
b^\dag+\kappa-P)\left(\frac{x}{b^\dag}\right)^{\myn q_*}\!,&x\in[\myp0,b^\dag],\\
\displaystyle \beta_1x+\kappa-P,&x\in[\myp b^\dag,\infty),
\end{cases}
\end{equation}
or more explicitly (cf.~\eqref{eq:v-new1})
\begin{equation}\label{eq:vf_amended}
u^\dag(x)=\begin{cases}
\displaystyle\frac{P-\kappa}{q_*-1}\left(\frac{\beta_1 (q_*-1)\myp
x}{(P-\kappa)\myp q_*}\right)^{q_*}\!,&\displaystyle 0\le x\le
\frac{(P-\kappa)\myp q_*}{\beta_1(q_*-1)},\\[.8pc]
\displaystyle \beta_1x+\kappa-P,&\displaystyle \hphantom{0\le{}}x\ge
\frac{(P-\kappa)\myp q_*}{\beta_1(q_*-1)}.
\end{cases}
\end{equation}
\begin{figure}[h!]
    \centering
\vspace{-1.5pc}    \subfigure[\,$\kappa\mapsto b^\dagger $]{\includegraphics[%
        height=0.31\textheight]{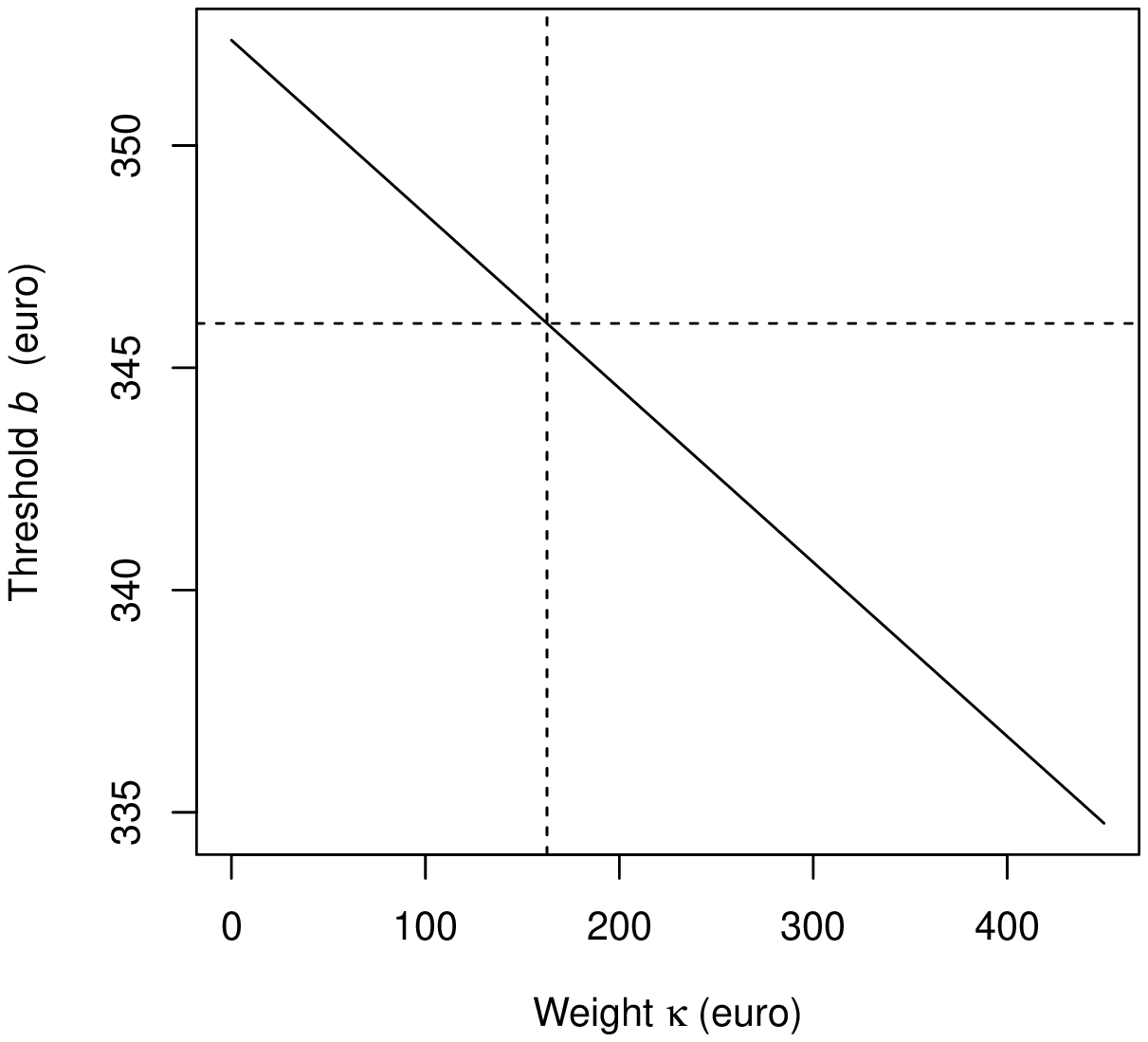}
\put(-199.3,118.3){\begin{rotate}{90}\mbox{\tiny${}^{{}^\dag}$}\end{rotate}}
       \put(-119.5,32.0){\mbox{\scriptsize $\kappa^\dag$}}
        \put(-181.0,128.0){\mbox{\scriptsize $x$}}
}
    \hspace{0.0pc}\subfigure[\,$\kappa\mapsto u^\dag(x)$]{\includegraphics[%
        height=0.31\textheight]{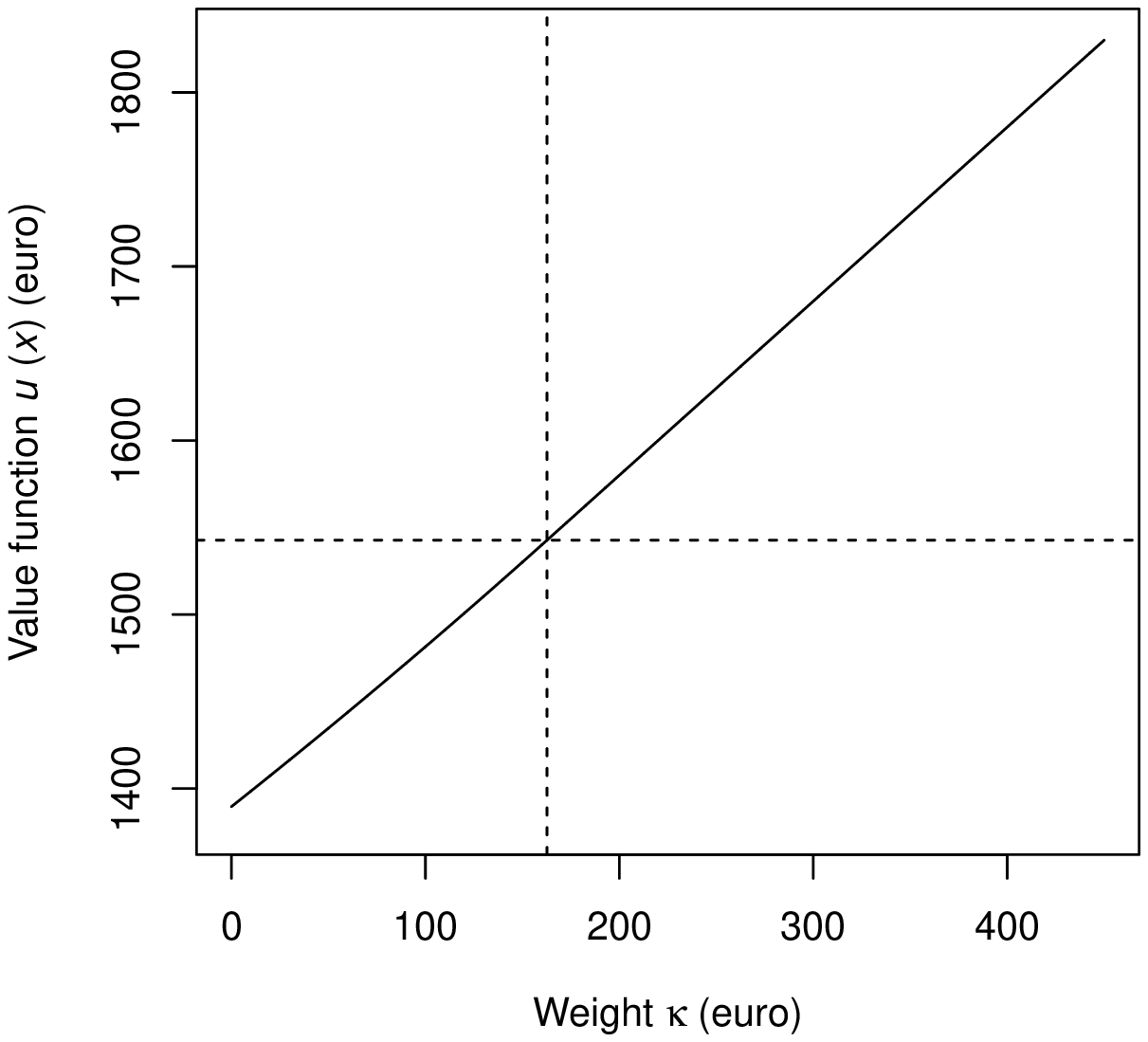}
 \put(-199.8,120.0){\begin{rotate}{90}\mbox{\tiny${}^{{}^\dag}$}\end{rotate}}
        \put(-119.5,32.0){\mbox{\scriptsize $\kappa^\dag$}}
\put(-83.5,97.5){\mbox{\scriptsize
$u^\dag\myn(x)|_{\kappa=\kappa^\dag}$}}
   }
 \vspace{-.5pc}   \caption{
    Functional dependence on
        the preference weight $\kappa$ in the reduced optimal stopping
        problem~\eqref{eq:optimal+utility+exp-improved1}:
        (a) the optimal threshold $b^\dag$ (see~\eqref{eq:b-dag});
        (b) the value function $u^\dag\myn(x)$
        (see~\eqref{eq:vf_amended}).
        Numerical values of the parameters used are as in Example~\ref{ex:5.2}:
        $r=\mu=0.0004$, $P=9\,000$, $\beta=30$, $\sigma=0.02$, and $x=346$. In particular, if $\kappa=0$ then
        $b^\dag\myn$ coincides with $b^*\myn\doteq352.3705$ and $u^\dag\myn(x)$
        coincides with $v(x)\doteq1389.6190$. The dashed vertical
        lines on both plots indicate the value $\kappa^\dag\myn\doteq162.7108$
        (see~\eqref{eq:kappa-dag}) separating different regimes for
        $u^\dag\myn(x)$ according to~\eqref{eq:vf_amended}. When $\kappa=\kappa^\dag\myn$, we have
        $b^\dag\myn=x=346$, shown as a dashed horizontal line in plot~(a); the corresponding
        value function is given by
        $u^\dag\myn(x)=\beta_1x+\kappa^\dag\myn-P\doteq1542.7110$ (see~\eqref{eq:v-new-dag}), shown as a
        dashed horizontal line in plot~(b). Note that the graph of $u^\dag\myn(x)$ in plot (b)
        looks almost linear for $\kappa\in[\myp0,\kappa^\dag]$, because the ratio $\kappa/P$ is quite small, $0\le
        \kappa /P\le \kappa^\dag\myn/P\doteq0.01808$; the slope here is approximately $v(x)(q_*-1)/P\doteq0.88448$,
        as compared to slope $1$ of the
        linear graph for $\kappa\ge \kappa^\dag$.}
    \label{fig4}
\end{figure}%
\begin{figure}[h!]
\vspace{-2.0pc}    \centering
\subfigure[$b\mapsto \PP_x(\tau_b<\infty)$ $\bigl(\mu<\tfrac12\myp\sigma^2\bigr)$]{\includegraphics[%
    height=0.3\textheight]{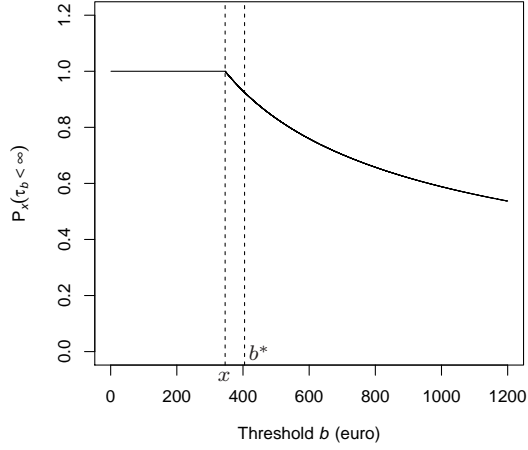}
\put(-130.8,33.3){\mbox{\scriptsize $x$}}
\put(-119.1,41.2){\mbox{\scriptsize $b^*$}} }
    \hspace{0.0pc}\subfigure[$b\mapsto \EE_x(\tau_{b})$ $\bigl(\mu>\tfrac12\myp\sigma^2\bigr)$]{\includegraphics[%
    height=0.3\textheight]{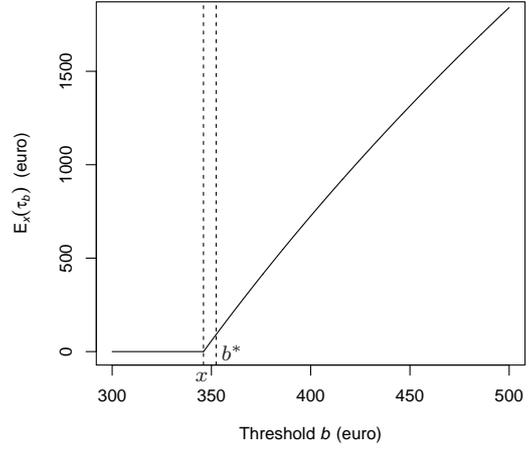}
\put(-139.8,33.3){\mbox{\scriptsize $x$}}
\put(-130.0,41.2){\mbox{\scriptsize $b^*$}} }\\[-.5pc]
    \subfigure[$b\mapsto \eNPV(x;\tau_b)$ $\bigl(\mu<\tfrac12\myp\sigma^2\bigr)$]{\includegraphics[%
    height=0.3\textheight]{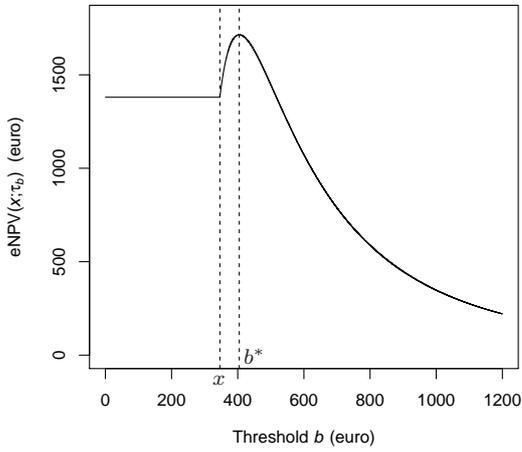}
\put(-130.8,33.3){\mbox{\scriptsize $x$}}
\put(-119.1,41.2){\mbox{\scriptsize $b^*$}} }\hspace{.7pc}\subfigure[$b\mapsto
\eNPV(x;\tau_b)$ $\bigl(\mu>\tfrac12\myp\sigma^2\bigr)$]{\includegraphics[%
    height=0.3\textheight]{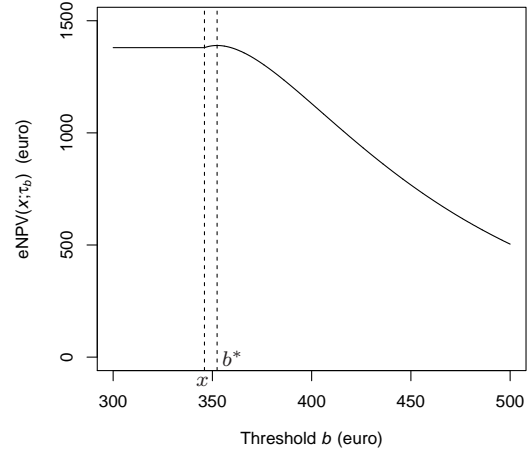}
\put(-139.8,33.3){\mbox{\scriptsize $x$}}
\put(-130.0,41.2){\mbox{\scriptsize $b^*$}}}
 \vspace{-.5pc}   \caption{Theoretical graphs
    for functionals of the hitting time $\tau_b$ versus
threshold $b\ge0$.
    \emph{Upper row:}
    (a) the hitting probability $\PP_x(\tau_b<\infty)$
    (see~\eqref{eq:P(tau<infty)});
    (b) the mean hitting time $\EE_x(\tau_{b})$ (see~\eqref{eq:E(tau)}).
    \emph{Bottom row:}
    the expected net present value
$\eNPV(x;\tau_b)$ (see~\eqref{eq:NPV-explicit}) with $\mu<
\frac12\myp\sigma^2$ (c) or $\mu> \frac12\myp\sigma^2$ (d). The
values of parameters used throughout are as in
Section~\ref{sec:5.4}: $x = 346$, $P=9\,000$, $\beta_1=30$, $\mu=
0.0004$, and $\sigma =0.04$ (left) or $\sigma =0.02$ (right). The
dashed vertical lines on each plot indicate $x$ and the optimal
threshold $b^*\mynn$, respectively; specifically, $b^*\doteq
404.7410$ on the left (see Example~\ref{ex:5.1}) and
$b^*\doteq352.3705$ on the right (see Example~\ref{ex:5.2}).}
    \label{fig5}
\end{figure}%
If $x$ is fixed then the problem value $u^\dag$, as a function of
$\kappa$, is given by the first or the second line in
\eqref{eq:vf_amended} according as $\kappa\in [\myp0,\kappa^\dag]$
or $\kappa\in [\kappa^\dag,\infty)$, respectively, where
\begin{equation}\label{eq:kappa-dag}
\kappa^\dag:=P-\frac{\beta_1(q_*-1)\myp x}{q_*}.
\end{equation}

The dependence of $b^\dag$ and $u^\dag(x)$ upon the utility
parameter $\kappa\in[\myp0,P]$ is illustrated in Fig.~\ref{fig4},
while Fig.~\ref{fig5} demonstrates the functional dependence of the
hitting probability $\PP_x(\tau_b<\infty)$ and the mean hitting time
$\EE_x(\tau_b)$ upon the variable threshold $b\ge0$, along with the
corresponding plots of the expected net present value
$\eNPV(x;\tau_{b})$.

\begin{remark}\label{rm:conversion} Note that
$u^\dag(x)$ is a strictly increasing function of $\kappa\in[0,P]$,
in accord with Proposition~\ref{pr:kappa}. In particular,
$u^\dag(x)$ coincides with the original value function $u(x)$ given
by \eqref{eq:v-new1}, but with the premium $P$ replaced by
$P-\kappa$. This can be interpreted as the individual's consent to
convert additional satisfaction, gained by virtue of pursuing the
optimal stopping problem \eqref{eq:optimal+utility+exp-improved1}
instead of \eqref{eq:r2e1}, into a higher premium,
$P^\dag=P+\kappa$. Such an effect is characteristic of the use of
risk-averse utility functions under the Expected Utility Theory
\cite{Kaas} (see also a discussion below in Section~\ref{sec:6.4}).
\end{remark}

In the case $\mu>\frac12\myp\sigma^2$, instead of
\eqref{eq:optimal+utility+exp} we may consider the simplified
problem
\begin{equation}\label{eq:optimal+utility+exp-b}
u^\dag(x)=\sup_{b\ge0} \bigl[\kappa\myp\exp\{-\EE_x(\tau_b)\} +
\eNPV(x;\tau_b)\bigr].
\end{equation}
Upon the substitution of formulas \eqref{eq:E(tau)} and
\eqref{eq:NPV-explicit}, it is rewritten in the form
(cf.~\eqref{eq:optimal+utility1-sub2})
\begin{equation}\label{eq:optimal+utility1-sub3}
u^\dag(x)=\sup_{b\ge
x}\left[\kappa\,\frac{\ln(b/x)}{\mu-\frac12\myp\sigma^2} + (\beta_1
b-P)\left(\frac{x}{b }\right)^{\myn q_*}\right].
\end{equation}
Again, the maximization problem \eqref{eq:optimal+utility1-sub3} can
be solved (at least, numerically). For an analytic solution, it is
convenient to modify the problem \eqref{eq:optimal+utility+exp-b} as
follows,
$$
u^\dag(x)=\sup_{b\ge 0}
\biggl[\kappa\exp\!\left(-\frac{q_*}{\mu-\frac12\myp\sigma^2}\mypp\EE_x(\tau_{b})\right)
+ \eNPV(x;\tau_{b})\biggr].
$$
Similarly to \eqref{eq:optimal+utility1-sub3}, this leads to the
maximization problem that coincides with \eqref{eq:max_amended} and,
therefore, has the same solution \eqref{eq:b-dag} and
\eqref{eq:v-new-dag} (or, equivalently,~\eqref{eq:vf_amended}).

\subsection{Connections to Expected Utility Theory}\label{sec:7.4}
The considerations above can be linked to the standard Expected
Utility Theory~\cite{Kaas}. In the usual setting, it is assumed that
an individual uses (perhaps, subconsciously) a certain utility
$U(w)$, as a function of financial wealth $w$, to assess losses,
gains and the resulting satisfaction. Generically, given the current
wealth $w$ and some random future loss $Y$, the expected loss
(measured via utility $U(\cdot)$) may be expressed as
$\EE\bigl[U(w-Y)\bigr]$. The individual is inclined to pay a premium
$P$ and buy the insurance policy as long as the expected utility
without insurance is no more than $U(w-P)$,
\begin{equation}\label{eq:EY<P}
\EE\bigl[U(w-Y)\bigr] \le U(w-P).
\end{equation}
The balance condition
\begin{equation}\label{eq:EY=P}
\EE\bigl[U(w-Y)\bigr] = U(w-P)
\end{equation}
determines the maximum premium $P_{\rm max}$ the customer is
prepared to pay (in fact, at this point it makes no difference
whether to buy the insurance or not).

In the baseline case with $U(w)\equiv w$, the conditions
\eqref{eq:EY<P} and \eqref{eq:EY=P} are reduced to
\begin{equation}\label{eq:EY<P*}
P\le P_{\rm max}=\EE(Y).
\end{equation}
However, choosing a different utility function may well change this
threshold. For instance, if the random loss $Y$ has exponential
distribution with parameter $\theta=0.001$, then according to
\eqref{eq:EY<P*} we have $P_{\rm max}=\EE(Y)=1/\theta=1\mypp000$. In
contrast, let the utility function be chosen as
$U(w)=1-\exp\bigl(-\tfrac12\theta w\bigr)$. Here, the utility is
between $0$ and $1$ if the wealth $w$ is positive, but it becomes
increasingly negative for a negative wealth; that is, strong weight
is placed against negative wealth, which may be characteristic of a
risk-averse individual. In this case, it is easy to check that
$$
P_{\rm max}=\frac{2\ln 2}{\theta}=1\mypp386.294>1\mypp000.
$$
Thus, the individual is happy to pay more than before to protect
themselves from the perceived risk of significant losses. That is to
say, an additional amount of satisfaction is convertible into an
extra premium.

In our case, if the UI was to be entered immediately, at time $t=0$,
then the value of this decision would be $\eNPV(x;0)=\beta_1 x-P$
(see \eqref{eq:g-new} and~\eqref{eq:NPV1+}). Clearly, in order for
this to be non-negative, the premium $P$ must satisfy the condition
$$
P\le P_{\rm max}=\beta_1 x.
$$
For instance, in the setting of the numerical example in Section
\ref{sec:5.4}, we get $P_{\rm max}=30\times 346 = 10\,380$, while
the set premium is $P=9\,000$.

Similarly, if the decision was taken at a stopping time $\tau$,
then, conditional on the wage $X_{\tau}$, the maximum premium
payable would be given by $P_{\rm max}=\beta_1 X_{\tau}$. Thus, the
value of $P_{\rm max}$ goes up or down together with the current
wage. However, in our setting the entry time is not decided in
advance, being subject to the stopping rule based on observations
over $(X_t)$. As a result, the value function $v(x)$ ($x>0$) of the
optimal stopping problem is always positive for any premium $P$, no
matter how high (see formula~\eqref{eq:value-b*-exp}). Apparently,
this is manufactured by selecting the threshold $b^*\mynn$ high
enough, which guarantees that, in the (rare) event of hitting it,
the mean value of this strategy will be positive.

This may not be satisfactory from the standpoint of the Expected
Utility Theory; however, there is no contradiction, because in its
standard version this theory does not allow for an optional
stopping. Adding utility terms to the gain function in the spirit of
Sections \ref{sec:6.2} and~\ref{sec:6.3} helps to amend the
situation (see Remark~\ref{rm:conversion}), but the maximum premium
payable still remains indeterminate.

The explanation of this paradox lies in the simple fact that the
gain function in the optimal stopping problems considered so far
does not include any losses. A simple way to account for such losses
is to include \emph{consumption} in the model. Namely, suppose for
simplicity that the consumption rate $c$ is constant; for instance,
the net present value of consumption over time interval $[\myp0,t]$
is given by
$$
\int_{0}^{t}\re^{-rs}c\,\rd{s}=\frac{c\mypp(1-\re^{-r\myp t})}{r}.
$$
It is natural to assume that the wage $X_t$ is sufficient to finance
the consumption, so that $\EE_x(X_t)=x\mypp\re^{\myp\mu\myp t}\ge c$
for all $t\ge0$ (see~\eqref{eq:E(X)}). In turn, for this to hold it
suffices to assume that $X_0=x\ge c$ and $\mu\ge0$. Hence, we need
to take into account consumption only over the unemployment spell
$[\tau_0,\tau_0+\tau_1]$, where the wage is replaced by the UI
benefit. The expected net present value of this consumption is given
by
\begin{align*}
\gamma:=\EE\biggl(\re^{-r\tau_0}\!\int_{0}^{\tau_1}
\!\re^{-rs}c\,\rd{s}\biggr)&=\EE\bigl(\re^{-r\tau_0}\bigr)\cdot
\EE\biggl(\frac{c\mypp(1-\re^{-r\tau_1})}{r}\biggr)\\
&=\frac{\lambda_0\myp c}{(r+\lambda_0)(r+*\lambda_1)},
\end{align*}
using independence of $\tau_0$ and $\tau_1$ and their exponential
distributions (with parameters $\lambda_0$ and $\lambda_1$,
respectively). Thus, our basic optimal stopping problem
\eqref{eq:r2e1} is modified to
\begin{equation*}
v^\ddag(x) =\sup_{\tau} \EE_x \bigl[\re^{-\tilde{r}\tau}
g(X_\tau)-\gamma\bigr],
\end{equation*}
which has the same solution as before (see Section~\ref{sec:2.5})
but with the new value function $v^\ddag(x)=v(x)-\gamma$, that is
(cf.~\eqref{eq:value-b*}),
\begin{equation*}
    v^\ddag(x) = \begin{cases}
    \displaystyle (\beta_1 b^* - P) \left( \frac{x}{\displaystyle b^*}\right)^{\myn q_*}\!-\gamma, &x\in[\myp0,b^*],\\
     \displaystyle\beta_1x - P-\gamma,& x\in[\myp b^*,\infty).
    \end{cases}
\end{equation*}

Now, the inequality $v^\ddag(x)\ge 0$ can be easily solved for $P$
to yield
\begin{equation}\label{eq:Pmax-cons}
P\le P^\ddag_{\rm max}:=\begin{cases} \displaystyle \beta_1b^*-\gamma\left(\frac{b^*}{x}\right)^{\mynn q_*}\!,& x\in[\myp0,b^*],\\
\displaystyle \beta_1x-\gamma,& x\in[\myp b^*,\infty).
\end{cases}
\end{equation}
Note that $P^\ddag_{\rm max}$ in \eqref{eq:Pmax-cons} is a
decreasing function of $\gamma$, but an increasing function of~$x$.
Thus, as could be expected, the maximum affordable premium gets
lower with the increase of consumption, but becomes higher with the
increase of the wage.

\begin{remark}
Of course, consumption can also be incorporated into the optimal
stopping models involving utility (see Sections \ref{sec:6.2}
and~\ref{sec:6.3}), but we omit technical details.
\end{remark}

\section{Concluding remarks}\label{sec:8}
In this paper, we have set up and solved an optimal stopping problem
in a stylized UI model. The model and its solution are useful by
illustrating approaches to optimal strategy of an individual seeking
to get insured. By including consumption in the model, we have also
demonstrated how a fair premium can be calculated, which makes our
UI model usable also from the insurer's perspective.

An explicit closed-form solution of the corresponding optimal
stopping problem was possible due to some simplifying
assumptions\,---\,in particular, exponential distribution of time
$\tau_0$ to loss of job and constant inflation rate~$r$. The
analysis also strongly relied on the simplest model for the wage
process $(X_t)$, that is, geometric Brownian motion with constant
drift $\mu$ and volatility $\sigma^2$.

Let us indicate a few directions of making our UI model more
realistic. Firstly, indefinite term of UI insurance could be
replaced by a finite expiration term for the benefit schedule (akin
to American call option with finite horizon), which would lead to a
harder (time-dependent) optimal stopping problem
(cf.~\cite[\S\mypp25.2]{peskir2006optimal}). Also, the assumption of
exponential distribution of $\tau_0$ needs to be tested on the basis
of real unemployment data. Note, however, that fitting a different
distribution for $\tau_0$ will invalidate the expression
\eqref{eq:NPV} for the expected net present value $\eNPV(x;\tau)$
and, therefore, will change the gain function in the optimal
stopping problem~\eqref{eq:r2e1}, making it more difficult to solve.

The parameters of the model may also need to be made time-dependent,
causing obvious complications to the model. On the other hand, the
implicit assumption of passive waiting for a new job during the
unemployment spell may not be realistic, or at least not desirable
as individuals would rather be expected to seek jobs more
pro-actively. Thus, it may be interesting to combine our UI model
with job-seeking models such as in~\cite{Boshuizen1995}.

The inclusion of utility terms in the optimal setting is novel in
this context, and illuminates significant changes in the
individual's behaviour when driven by utility considerations. In
particular, the value of the optimal stopping problem
\eqref{eq:OSP-general} is an increasing function of the preference
coefficient $\kappa$ (see Proposition~\ref{pr:kappa}). This result
is intuitively appealing, as it conforms with the usual impact of
utility function (under the Expected Utility Theory), allowing one
to convert extra satisfaction into extra premium. This is confirmed
by our analysis of suboptimal solutions in Section \ref{sec:6.3}
(see Fig.~\ref{fig4}). Finally, it would be interesting to study the
optimal stopping problem \eqref{eq:OSP-general} in more detail.

%

\section*{Acknowledgements}
J.S.A.\ was supported by a Leeds Anniversary Research Scholarship
(LARS) from the University of Leeds. Both authors have greatly
benefited from many useful discussions with Tiziano De Angelis, who
has also contributed to the design of this study. J.S.A.\ is
grateful to Elena Issoglio for helpful comments. We thank three
anonymous reviewers for their useful feedback. In particular,
Reviewer \#1 proposed an extension of our insurance model by
inclusion of constant force of mortality and pointed out the
classical paper by \cite{Merton}; Reviewer \#2 commented on positive
truncation in optimal stopping and brought to our attention a paper
by \cite{Villeneuve}; and Reviewer \#3 advised on sensitivity
analysis and economic interpretation.

\end{document}